\documentclass[3p,preprint]{elsarticle}
\usepackage{graphicx}
\usepackage{amsmath}
\usepackage{amsfonts}               
\usepackage[british]{babel}
\usepackage{siunitx}
\usepackage{color}
\usepackage[capitalise]{cleveref}
\usepackage{nicematrix}
\usepackage{placeins}
\usepackage{changes}
\usepackage{float}
\usepackage{makecell}

\usepackage{longtable}
\usepackage{adjustbox}

\def\D{\mathop{}\!\mathrm{d}}
\DeclareSIUnit\Molar{\textsc{m}}
\allowdisplaybreaks                 

\newcommand{\lightgray}{\color[rgb]{0.5,0.5,0.5}}

\begin{document}








\begin{abstract}
Microscale oxygenation plays a prominent role in tumour progression. Spatiotemporal variability of oxygen distribution in the tumour microenvironment contributes to cellular heterogeneity and to the emergence of normoxic and hypoxic populations. Local levels of oxygen strongly affect the response of tumours to the administration of different therapeutic modalities and, more generally, to the phenomenon of resistance to treatments. Several interventions have been proposed to improve tumour oxygenation, being the elevation of the local temperature (hyperthermia) an important one. While other factors such as the metabolic activity have to be considered, the proficiency of the tumour vascular system is a key factor both for the tissue oxygenation and for its temperature maps. Consequently, the interplay of these factors has attracted considerable attention from the mathematical modelling perspective. Here we put forward a transport-based system of partial differential equations aimed at describing the dynamics of healthy and tumour cell subpopulations at the microscale in a region placed between two blood vessels. By using this model with diverse flow conditions, we analyse the oxygen and temperature profiles that arise in different scenarios of vascular status, both during free progression and under thermal therapy. We find that low oxygen levels are associated to elevations of temperature in locations preferentially populated by hypoxic cells, and hyperthermia-induced cell death, being strongly dependent on blood flow, would only appear under highly disrupted conditions of the local vasculature. This results in a noticeable effect of heat on hypoxic cells. Additionally, when pronounced cell death occurs, it is followed by a significant increase in the oxygen levels. Our results provide quantitative insight to the physiological and biological processes taking place at sub-voxel sizes, currently not accessible to standard functional imaging due to spatial resolution limitations.
\end{abstract}


\begin{frontmatter}

\title{Modelling the effect of vascular status on tumour evolution and outcome after thermal therapy}

\author[add1]{Jes\'{u}s J. Bosque}

\author[add1]{Gabriel F. Calvo\corref{cor1}}
\ead{gabriel.fernandez@uclm.es}

\author[add2]{María Cruz Navarro}

\cortext[cor1]{Corresponding author}
	
\address[add1]{Department of Mathematics, Mathematical Oncology Laboratory (MOLAB), University of Castilla-La Mancha, Ciudad~Real, Spain.}	
\address[add2]{Department of Mathematics-IMACI, Facultad de Ciencias y Tecnologías Químicas, University of Castilla-La Mancha, Ciudad Real, Spain}

\end{frontmatter}

\section{Introduction}

Tumour oxygenation is a central issue in the biology and treatment of cancer. As tumour cells proliferate and migrate, they disrupt their microenvironment, inducing changes both in the surrounding healthy tissue and in the local vascular network. These changes produce alterations in the levels of oxygen, which tend to decrease with respect to the levels found in normal tissues. This oxygen deficient state, known as hypoxia, is a hallmark present in most malignant neoplasms~\cite{wilson2011targeting}. Associated to hypoxia there is a cascade of processes that contribute to the transformation of the blood vessels functionality, such as the stabilisation of the hypoxia-inducible factor 1 (HIF-1) and the subsequent release of vascular endothelial growth factors (VEGF) that promote blood vessels formation. The uncontrolled onset of these processes leads to an irregular and highly inefficient vascular network that is often not able to supply the nutrient demand imposed by the persistent growth of the tumour cells~\cite{Bhandari2019}.
\par

\begin{figure}[t]
	\centering
	\includegraphics[width=0.99\textwidth]{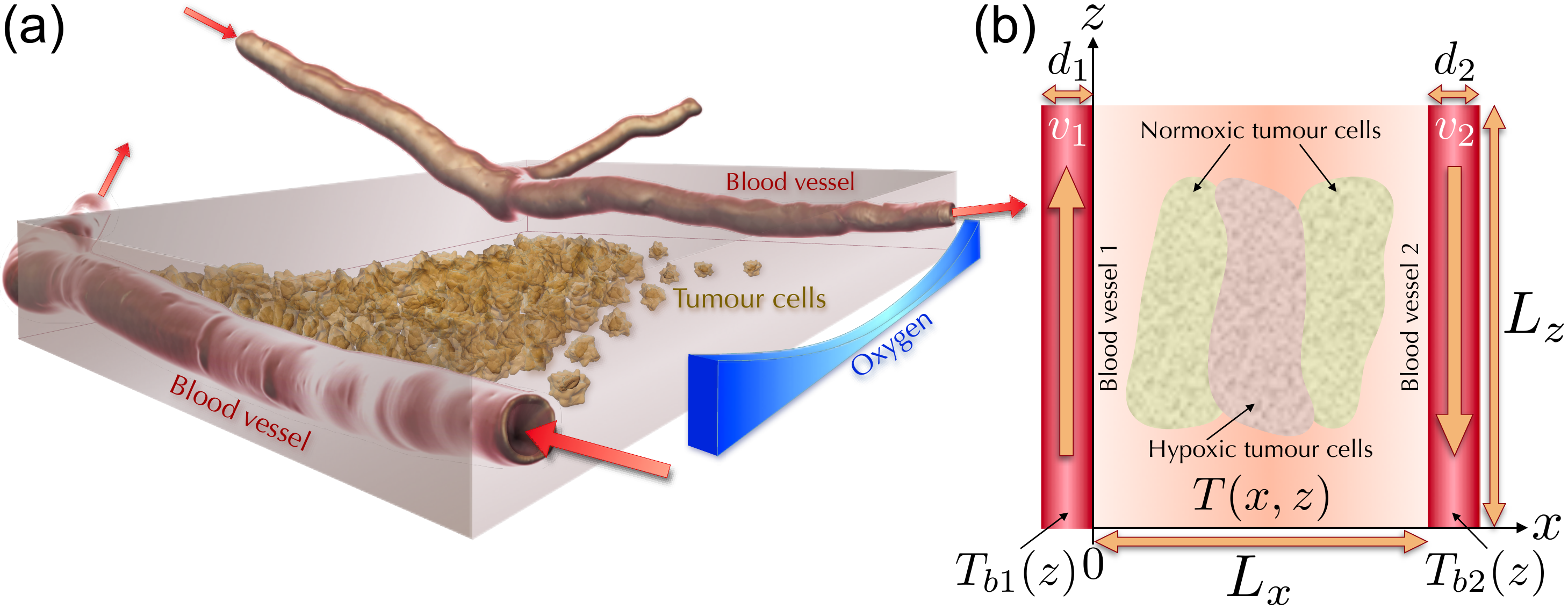}
	\caption{Modelling scenario. \textbf{(a)} Tumour cells proliferate and migrate between two blood vessels, which supply oxygen and nutrients to the surrounding tissue. Depending on the functional status of these vessels and the interplay of the populations, varying normoxia and/or hypoxia conditions may emerge. \textbf{(b)} Section of the tissue located between two blood vessels. Here we address the spatio-temporal evolution of tumour cell populations subjected to different levels of vascular functionality, which affect both the oxygen levels and temperature distribution $T(x,z)$ depending on the thermal therapy course followed. For that we set a partial differential equations-based model which we solve in a rectangular domain having sizes within the typical sub-voxel regime (below one millimetre). The left and right edges communicate the tumour cell region to the blood vessels, with which the tissue exchanges oxygen and thermal energy.}
	\label{Fig:fig1}	
\end{figure}

In this context, hypoxia becomes a determining factor in cancer biology. From the previous depiction two broad hypoxic scenarios have been identified~\cite{michiels2016cycling}. Firstly, \textit{chronic hypoxia} is associated to the limited diffusion experienced by oxygen in a highly oxygen-consuming environment. The oxygen flux provided by arterioles and capillaries is exhausted by cells near the vessels and therefore low oxygen levels appear in areas located far from them, with a typical length range of 100-200 \si{\micro\meter}. Secondly, tumour vessels lack a proper organisation and functionality, and the fact that they are subjected to stasis events leads to fluctuations in blood flow followed by \textit{cycling} (or \textit{acute}) \textit{hypoxia}; during these events the cells are in a state of low oxygenation which lasts for variable periods of time and recovers once the flow is reestablished. These events are modulated both by the constant remodelling of the vascular network, which produces variations with characteristic times in the order of hours, and by the irregular dispersion of red blood cell flux due to the abnormalities of the vasculature, displaying even faster oscillations. A number of previous works have addressed from a mathematical modelling perspective different pathological conditions occurring in tumours that are not easily accessible by experimentation~\cite{McDougall2006,Owen2009,pries2009structural,bernabeu2020abnormal}.
\par

The fluctuating levels of oxygen that may appear in different areas of a tumour, ranging from anoxia to normal oxygen concentrations, affect the evolution of distinct cellular phenotypes that emerge and interact within the tumour contributing to its heterogeneity~\cite{al2019hypoxia,dzyubak2021multi}. Cells that are exposed to low levels of oxygen (henceforth referred to as \textit{hypoxic cells}) adapt to their environmental conditions by expressing a repertoire of molecular signalling pathways. In the case of tumour hypoxic cells, they typically become more motile at the expense of reducing their proliferation rate compared to non-hypoxic tumour cells~\cite{giese2003cost,lewis2016intratumoral}. If the low oxygen conditions persist and become incompatible with the minimum requirements of cells, they undergo an uncontrolled form of death, called necrosis, which involves a cascade of processes that result in the rupture of the cell membrane causing spillage of cell contents into the tumour microenvironment and promoting inflammation and tissue damage~\cite{DArcy2019}. This is often seen in histologic samples of tumours, for instance in glioblastoma~\cite{Rong2006}, in areas located far from the blood vessels. 
\par

Aside from the relevance to tumour evolution, oxygen levels are also key to standard treatment efficacy. Chemotherapy depends on the delivery of the drug to the tumour tissue, which relies on the functionality of the vasculature feeding the tumour, and it is therefore linked to oxygen levels by the status of the vasculature~\cite{dewhirst2017transport}. Additionally, many chemotherapeutic drugs need oxygen for the cytotoxic reactions to take place and, as many of them are directed towards proliferating cells, their action on hypoxic cells may be lower under such conditions~\cite{wouters2007implications}. On the other hand, high oxygen levels are pivotal for radiation therapy efficacy. Few milliseconds after radiotherapy, oxygen participates in the chemical reactions that fix the DNA damage produced by the ionising radiation by forming oxidised forms of the free radicals whose effect on DNA is irreversible~\cite{hall2018radiobiology}. For this reason, cells under very low levels of oxygen are up to three times more resistant to radiotherapy than cells in well oxygenated tissues~\cite{joiner2019basic}. Due to the importance of cancer therapies, the understanding of their resistance and the search for optimisations, there is much ongoing work on the mathematical modelling of these issues \cite{valle2021chemoimmunotherapy,pang2021mathematical}.
\par

Several approaches have been proposed as ways to alleviate tumour hypoxia in order to improve radiotherapy effectiveness~\cite{wigerup2016therapeutic}. One of them is the use of hypoxia-activated prodrugs that are selectively switched-on inside tumour cells that are under oxygen deprivation levels. Other is targeting molecular pathways such as the overexpresion of HIF-1~\cite{Li2021}. Another approach employed to raise oxygenation levels is increasing the temperature in the tumour, a treatment known as hyperthermia. The heating can be induced by several means such as radio frequency antennas~\cite{mattoso2021pointwise}, high intensity focused ultrasound (HIFU)~\cite{ghasemi2020computational}, or other increasingly compelling approaches like nanoparticles, which also allow to exploit novel drug delivery approaches \cite{calvo2020modelling}. In order to optimise the expected benefits, the heating is maintained typically during one hour in the clinical practise. One of the many reported effects of an elevated temperature is the vasodilation of blood vessels, which improves the blood flow allowing the cells to receive more oxygen, however, this might be only one of the factors affecting oxygenation. There is currently an interest in better understanding the contribution of high local temperatures to cancer therapies improvement, being mathematical modelling one of the means used to tackle open questions \cite{li2020thermo}.
\par

Besides many biophysical effects of heat, it is very well known that temperatures in the vicinity of \SI{43}{\degreeCelsius} cause the death of the cells, therefore, thermal therapy is also a cytotoxic therapy \cite{Elming2019}. Although hyperthermia is usually considered as an adjuvant therapy, and many synergistic effects being at play, an important part of its effect comes from the cell death experienced when the tissue is maintained at a high temperature for a sufficient period of time \cite{ahmed2015hyperthermia}. In fact, despite many other effects being triggered, the basis behind thermal dose measuring is always cell death \cite{van2016cem43}. Therefore, this is an important contributing factor to the overall effect that has repercussion in the outcome of patients. Even though the cell death levels are not expected to achieve tumour control only by single-treatment hyperthermia administration, quantifying the interplay of different tumour cell populations thriving under varying physiological conditions in response to thermal therapies could help improve this adjuvant therapeutic modality. A number of authors have tackled the modelling of hyperthermia, with particular emphasis on heat transfer alone~\cite{gupta2013numerical,suleman2020mathematical}. However, to the best of our knowledge, no models have been developed integrating the evolution of multiple cell populations with different vascular conditions and variable oxygen levels under the effect of hyperthermia.
\par

Here we put forward a mathematical model based on partial differential equations (PDEs) to gain insight on the spatio-temporal evolution of normoxic and hypoxic populations thriving in a tissue fed by two blood vessels during free progression and also under hyperthermia administration (see \cref{Fig:fig1}). In addition to the cell populations, we track the oxygen concentration in the tissue as well as the temperature distribution. The blood vessels irrigating the tissue act both as a source for the oxygen and a sink for heat in the tissue. We use this system to analyse the effect of a thermal therapy on normoxic and hypoxic populations under different conditions of blood flow in the vessels. Our approach allows us to elucidate important physiological and biological processes occurring at sub-voxel sizes, which are currently not accessible in patients due to spatial resolution limitations of standard functional imaging used in the clinical setting.
\par

\section{Mathematical model}
\label{Sec:MathModel}
In order to simulate the spatio-temporal evolution of a tumour portion located between two feeding blood vessels (see \cref{Fig:fig1}(a)) and its evolution under thermal treatment, we put forward a system of PDEs that allows us to capture the key variables, i.e. the different cell populations, together with the temperature and oxygen concentration distributions, both within the tissue and in the vessels. Many of the interesting phenomena occurring in the microscopic evolution of cancer cells happen due to the interaction of the cells with their microenvironment, including the vascular tree that feeds the tissue. Since blood vessels play a crucial role in the oxygenation and temperature regulation of the tissue, we focused on the phenomena arising from that interaction. To simplify the problem resolution without losing relevant information, we consider our modelling equations in a rectangular 2D domain $[0,L_{x}]\times[0,L_{z}]$ where the blood vessels run parallel along the left and right sides of the domain (\cref{Fig:fig1}(b)).
\par

Our tumour populations consist of two subpopulations according to their different phenotypic behaviour. These disparate phenotypes are a consequence of the history that the cells have undergone, in particular due to the oxygen levels that they have been exposed to. Tumour cells located in an oxygenated medium are highly proliferative, being the proliferation the preferential process to which they allocate their resources. These cells, which thrive in a well oxygenated medium, are referred to as normoxic tumour cells. In contrast, tumour cells that are persistently subjected to low levels of oxygen adapt to those conditions and acquire specific traits that enable them to survive in a harsh environment \cite{damaghi2021harsh}. This subpopulation of tumour cells are called hypoxic cells and are characterised by a higher motility, as a consequence of their search for resources, and a lower proliferative potential \cite{giese2003cost,lewis2016intratumoral}. In addition to the tumour population, we also consider a population of normal healthy cells that occupy the tissue before the invasion produced by the tumour subpopulations. These are static and slowly proliferative. Their initial occupancy of the tissue is lower than that of the tumour cells since the latter are not subjected to contact inhibition, which allows them to get packed in higher density clusters \cite{mendonsa2018cadherin,pavel2018contact}. They are also affected by the presence of tumour cells which toxify their microenvironment by acidification and release of reactive oxygen species \cite{kumari2018reactive}. Finally, we consider a compartment of necrotic cells which takes into account the debris from cell death in all the other populations, whether they are due to oxygen deprivation, interaction with tumour cells or high temperatures. These necrotic cells occupy a variable space that is expected to be smaller than the corresponding to the other populations.
\par

Besides modelling cell populations, we include in our simulations the main biophysical factor that modifies the tumour dynamics, namely the level of oxygen concentration in the tissue. Since oxygen is necessary for all cells to thrive, its absence affects how cells behave. Oxygen gets delivered to the tissue by the local vasculature so, in order to conform with a minimal model, we set a simple approach for oxygen transport through the vessels as well as the diffusion-reaction within the tissue. Another biophysical element which is mediated by transport to the blood flow is temperature thus, to complete our description, and with the aim of further understanding the effects of thermal therapy, we modelled the transport of thermal energy in the vessels and its diffusion in the tissue.
\par

Our model intends to represent the interaction of all the significant variables in a tissue located between two blood vessels. To do so, we implement all the agents by their corresponding modelling reaction-diffusion PDEs. These will be solved in a 2D rectangular domain isolated at the top and bottom sides and interacting with two countercurrent blood vessels at the left and right edges. The blood flow attributes are oxygen concentration and temperature, which are allowed to change in every vessel's transversal section according to the flow of the fluid. Therefore, for oxygen and temperature we set 1D transport equations for each vessel. 
\par

\subsection{Evolution of cell populations model}
Many PDE-based models have been proposed to describe different aspects of tumour growth~\cite{Enderling2014,Michor2015,Swanson2017,Rockne2019}. Among these, reaction-diffusion equations have been usually employed, specially those assuming Fickian diffusion to account for cell migration and a reaction term consisting of a local logistic growth for the net proliferation, resulting in Fisher-Kolmogorov-type equations~\cite{Belmonte2014,el2019revisiting,elazab2018macroscopic,Badoual2021}. Several works have resorted to adaptations of this basic model to capture key physio-pathological hallmarks found for instance in malignant gliomas~\cite{perez2011bright}, and have successfully led to the identification of image-based biomarkers in the clinic~\cite{perez2017glioblastoma}. Here we set one PDE for each one of the tumour populations cell densities, that is, the normoxic population $n(x,z,t)$ and the hypoxic population $h(x,z,t)$. Their motility is parameterised by the diffusion coefficients $D_n$ and $D_h$, being the former smaller than the latter, and their proliferation coefficients $\rho_n$ and $\rho_h$, with the proliferation of the normoxic population greater than the hypoxic one~\cite{Alicia2012}. Thus, migration is modelled via standard Fickian diffusion, and the proliferation follows a logistic term, where $( 1 - n - h - w - \xi c)$ represents the available space in each point, with $\xi$ being the relative fitness of necrotic cells---with density $c(x,z,t)$---and $w(x,z,t)$, the density of healthy cells. The equations read as

\begin{equation}
\label{Eq:normoxic}
	\frac{\partial n}{\partial t} = D_n \nabla^2 n + \rho_n \left( 1 - n - h - w - \xi c \right) n - \sigma_{nh}(s) n + \sigma_{hn}(s) h - \sigma_t(T,t_{43}) n,
\end{equation}

\begin{equation}
\label{Eq:hypoxic}
	\frac{\partial h}{\partial t} = D_h \nabla^2 h + \rho_h \left( 1 - n - h - w - \xi c \right) h + \sigma_{nh}(s) n - \sigma_{hn}(s) h - \sigma_{hc}(s) h - \sigma_t(T,t_{43}) h.
\end{equation}
The terms $\sigma_{nh}$ and $\sigma_{hn}$, detailed in \ref{Sec:App_PhenotypicSwitch}, account for the phenotypic switch between normoxic and hypoxic cells as a function of the oxygen concentration $s$, while $\sigma_{hc}$ and $\sigma_{t}$ represent respectively the cell death due to low oxygen, and to high temperature $T$ and/or thermal accumulated dose $t_{43}$ (see \ref{Sec:App_ThermalDeath}).
\par

The equation for the healthy tissue consists of a logistic proliferation term with a (relatively small) proliferation rate $\rho_w$ and three terms for different sources of cell death: contact interaction with tumour cells (parameterised by $\lambda$), lack of oxygen $\sigma_{nh}(s)$ (identical to the transition from the normoxic population to the hypoxic one), and thermally-induced death $\sigma_{t}(T,t_{43})$

\begin{equation}
\label{Eq:healthy}
	\frac{\partial w}{\partial t} = \rho_w \left( 1 - n - h - w - \xi c \right) w - \lambda (n+h) w - \sigma_{nh}(s) w - \sigma_t(T,t_{43}) w.
\end{equation}

All cell death terms in the above equations contribute to the expansion of a necrotic compartment leading to the following equation

\begin{equation}
\label{Eq:necrotic}
	\frac{\partial c}{\partial t} = \lambda (n+h) w + \sigma_{nh}(s) w + \sigma_{hc} h + \sigma_t(T,t_{43}) (n+h+w).
\end{equation}

The set of nonnegative cell density functions $n, h, w$ and $c$ satisfy the following no-flux (i.e., homogeneous Neumann) boundary conditions 
\begin{align}
	\left. \frac{\partial P}{\partial x} \right|_{(0,z,t)} = \left. \frac{\partial P}{\partial x} \right|_{(L_x,z,t)} = \left. \frac{\partial P}{\partial z} \right|_{(x,0,t)} = \left. \frac{\partial P}{\partial z} \right|_{(x,L_z,t)} = 0,  
\end{align}
and initial conditions $P(x,z,0) = P_{0}(x,z)$, where $P = \{ n,h,w,c \}$.

\subsection{Thermal evolution}
\label{SubSec:ThermalEvo}
The evolution of temperature $T(x,z,t)$ in the tissue is given by the heat equation
\begin{equation}
\label{Eq:Heat}
	\delta C \frac{\partial T}{\partial t} = \kappa \nabla^2 T + Q_n n + Q_h h + Q_w w + F(t),
\end{equation}
where $\delta$, $C$ and $\kappa$ are the tissue density, its heat capacity, and its thermal conductivity, respectively. The source terms $Q_n$, $Q_h$ and $Q_w$ correspond to the heat metabolic contributions of the different types of cells and $F(t)$ is the externally applied heat power. 
\par
Regarding the boundary conditions, we assume no fluxes across the top and bottom boundaries, 

\begin{equation}
	\left. \frac{\partial T}{\partial z} \right|_{(x,0,t)} = \left. \frac{\partial T}{\partial z} \right|_{(x,L_z,t)} = 0,
\end{equation}
whereas for the left and right boundaries, where the heat is dissipated through the blood vessels, we model the heat transfer by means of the Newton cooling law. Therefore, we have
\begin{align}
	\left. \frac{\partial T}{\partial x} \right|_{(0,z,t)} &= \frac{\eta}{\kappa} \left( T( 0,z,t) - T_{b1}(z,t) \right), & \left. \frac{\partial T}{\partial x} \right|_{(L_x,z,t)} &= -\frac{\eta}{\kappa} \left( T(L_x,z,t) - T_{b2}(z,t) \right),
\end{align}
where $\eta$ denotes the heat transfer coefficient in the blood-tissue interface, and $T_{b1}(z,t)$, $T_{b2}(z,t)$, the temperature corresponding to the blood in the left and right vessels, respectively. From now on we will refer to the left vessel with subscript 1 and to the right one with subscript 2.
\par

In each of the blood vessels, whose diameters are $d_1$ and $d_2$, the blood circulates in opposite directions with a velocity magnitude $v_1$ and $v_2$, with a flow assumed to be incompressible ($\partial_z v_1 = \partial_z v_2 = 0$). The conservation of energy leads to the following transport equations which account for the temperature in each section of the vessel

\begin{equation}
\label{Eq:Tb1}
	\frac{\partial T_{b1}}{\partial t} + v_1 \frac{\partial T_{b1}}{\partial z} = \frac{\eta}{d_1 \delta_{b} C_b} \left( T(0,z,t) - T_{b1}(z,t) \right),
\end{equation}

\begin{equation}
\label{Eq:Tb2}
	\frac{\partial T_{b2}}{\partial t} - v_2 \frac{\partial T_{b2}}{\partial z} = \frac{\eta}{d_2 \delta_{b} C_b} \left( T(L_x,z,t) - T_{b2}(z,t) \right),
\end{equation}
with $\delta_{b}$ being the blood density and $C_b$ its heat capacity. Notice that each of these equations is coupled with the temperature at the left and right boundaries of the tissue, respectively. The boundary conditions for \cref{Eq:Tb1,Eq:Tb2} give the entrance temperature to the vessels, which is set to a value $T_{b0}$
\begin{align}
    T_{b1}(0,t)&= T_{b0}, & T_{b2}(L_z,t)&= T_{b0}.
\end{align}

Following a classical assumption in the modelling of circulation within blood vessels, we model the variations of flow by means of a Poiseuille law that relates the radius of the vessel $r$ with the amount of blood flow rate $\dot{V}$. Given a fixed pressure difference $\Delta p$ between the entry and the outlet of the vessel and a viscosity $\mu$, the blood flow rate is ruled by
\begin{equation}
\label{Eq:Poiseuille}
    \dot{V} = \frac{\pi\Delta p}{8 \mu L_z} r^4. 
\end{equation}

\noindent In our model, the blood vessels are considered to be cylindrical with nominal radii $r_{01}$ and $r_{02}$ for the left and right vessel respectively. The application of hyperthermia is known to induce vasodilation of the vessels~\cite{Berezhnoi2018}. Thus, during the simulation of the therapy, the radii of the vessels are allowed to vary linearly with temperature change and a expansion coefficient $\chi$. This assumption is justified by the relatively narrow temperature range involved (about $10\%$). These linear relationships for the radii $r_1$ and $r_2$ of each vessel during the treatment are given by 
\begin{equation}
\label{Eq:r1_variation}
	r_1 = r_{01} \left( 1 + \chi \left( T_{bm1} - T_{01} \right) \right),
\end{equation}
\begin{equation}
\label{Eq:r2_variation}
	r_2 = r_{02} \left( 1 + \chi \left( T_{bm2} - T_{02} \right) \right),
\end{equation}

\noindent where $T_{bm1}$ and $T_{bm2}$ are the temperatures of the blood in the midpoint of each vessel provided that they do not exceed 41 \si{\degreeCelsius}
\begin{equation}
    T_{bm1}(t) = \min{\left( T_{b1}\!\!\left( \frac{L_z}{2},t \right)\!, \; 41\right)},
\end{equation}
\begin{equation}
    T_{bm2}(t) = \min{\left( T_{b2}\!\!\left( \frac{L_z}{2},t \right)\!, \; 41\right)},
\end{equation}

\noindent and $T_{01},\,T_{02}$ are their reference temperatures, i.e. the temperatures at the same midpoint and instant before the treatment start. As the blood flow rate $\dot{V}$ in cylindrical vessels of radius $r$ and velocity $v$ is given by $\dot{V}=v \pi r^2$, the Poiseuille law \cref{Eq:Poiseuille} can be rewritten to specify the dependency of velocity with the square of the radius as
\begin{equation}
    v = \frac{\Delta p}{8 \mu L_z} r^2,
\end{equation}

\noindent that allows us to relate the velocity for a variable radius $r$ to the velocity $v_0$ at a reference radius $r_0$, as $v=v_0 \frac{r^2}{r_0^2}$. Thus, the thermal dilation of the vessels during the application of hyperthermia leads to a variation of the velocities in both vessels whose values are given by
\begin{equation}
\label{Eq:v1_variation}
	v_1 = v_{01} \left( 1 + \chi \left( T_{bm1} - T_{01} \right)   \right)^2,
\end{equation}
\begin{equation}
\label{Eq:v2_variation}
	v_2 = v_{02} \left( 1 + \chi \left( T_{bm2} - T_{02} \right)   \right)^2.
\end{equation}

The initial conditions considered for the temperature, both in the tissue and the vessels, are
\begin{align}
\label{Eq:icT}
T(x,z,0)&=T_0,  \\	
T_{b1}(z,0)&=T_{b0},  \\	
T_{b2}(z,0)&=T_{b0}. 
\end{align}

\subsection{Oxygen evolution}
The concentration of oxygen in the tissue, $s(x,z,t)$, is modelled via a diffusion-reaction PDE that has been extensively used in the literature \cite{Alicia2012,kingsley2021bridging}. It consists of a diffusion term with diffusion coefficient $D_s$ and a reaction term that accounts for consumption by the cells given by a Michaelis-Menten term having a saturation constant $K_M$. It reads as
\begin{equation}
	\frac{\partial s}{\partial t} = D_s \nabla^2 s - \left( \alpha_n n + \alpha_h h + \alpha_w w \right) \frac{s}{K_M+s},
	\label{eq:PDEOxygen}
\end{equation}
where the constants $\alpha_n,\,\alpha_h$ and $\alpha_w$ are the oxygen consumption rates per cell of each of the populations. Similarly as in previous equations, we impose no-flux boundary conditions at the top and bottom boundaries
\begin{equation}
	\left. \frac{\partial s}{\partial z} \right|_{(x,0,t)} = \left. \frac{\partial s}{\partial z} \right|_{(x,L_z,t)} = 0,
\end{equation}

\noindent and, as in the thermal case, we have boundary conditions expressing the conservation of energy in the solid-liquid interface 
\begin{align}
	\left. \frac{\partial s}{\partial x} \right|_{(0,z,t)} &= \frac{\gamma}{D_s} \left( s(0,z,t) - s_{b1}(z,t) \right), & \left. \frac{\partial s}{\partial x} \right|_{(L_x,z,t)} &= -\frac{\gamma}{D_s} \left( s(L_x,z,t) - s_{b2}(z,t) \right).
\end{align}

\noindent The transfer of energy in that interface is as a flux proportional to the differences in concentration ruled by the permeability parameter $\gamma$. Finally, the concentration of oxygen $s_{b1}(z,t),\,s_{b2}(z,t)$ through the blood vessels is given by the transport equations
\begin{equation}
	\frac{\partial s_{b1}}{\partial t} + v_1 \frac{\partial s_{b1}}{\partial z} = \frac{\gamma}{d_1} \left( s{(0,z,t)} - s_{b1}(z,t) \right),
	\label{eq:PDEOxyTransport1}
\end{equation}
\begin{equation}
	\frac{\partial s_{b2}}{\partial t} - v_2 \frac{\partial s_{b2}}{\partial z} = \frac{\gamma}{d_2} \left( s{(L_x,z,t)} - s_{b2}(z,t) \right),
	\label{eq:PDEOxyTransport2}
\end{equation}
with boundary conditions
\begin{align}
    s_{b1}(0,t)&= s_{b0}, & s_{b2}(L_z,t)&= s_{b0}.
\end{align}

The initial conditions considered for the oxygen concentration, both in the tissue and the vessels, are
\begin{align}
\label{Eq:icS}
s(x,z,0)&=s_0,  \\	
s_{b1}(z,0)&=s_{b0},  \\	
s_{b2}(z,0)&=s_{b0}. 
\end{align}

\subsection{Thermal damage accumulation}
The application of hyperthermia treatment causes the accumulation of a thermal dose in the tissue that may end up being lethal for the cells. The measure typically used for the accumulated dose is the number of equivalent minutes at 43 \si{\degreeCelsius}. Thus, we use a function $t_{43}(x,z,t)$ for the thermal dose in the tissue that represents the accumulated dose during hyperthermia treatment in each point $(x,z)$ at a given time $t$. As detailed in \ref{Sec:App_ThermalDeath}, the equation that gives the time dependence of this function is
\begin{equation}
	\frac{\partial t_{43}}{\partial t} (x,z,t) = R^{43-T(x,z,t)} / 60,
\end{equation}
with an initial condition $t_{43}(x,z,0)=0$.

\section{Numerical implementation}
\label{Sec:Calculation}
To numerically solve the above systems of PDEs, \crefrange{Eq:normoxic}{Eq:Tb2} and \crefrange{eq:PDEOxygen}{eq:PDEOxyTransport2}, we employed the method of lines~\cite{schiesser2012partial}. This method allows for a simple implementation and it is also efficient for time integration. The method consists in using a finite difference scheme to discretise the spatial variables of the problem, while the temporal part remains treated as a continuous problem. In this way, an ordinary differential equation (ODE) is associated to each spatial point and variable. The resulting system of ODEs is then solved by standard routines, such as via Runge-Kutta methods. 
\par

The spatial domain for the tissue  $[0, L_x]\times[0, L_z]$ is divided in $M$ subintervals in the $x$-direction and $N$ in the $z$-direction, so we have an equispaced rectangular grid of $(M+1)\times (N+1)$ nodes, with $\Delta x=\frac{L_x}{M}$ and $\Delta z=\frac{L_z}{N}$ being the discretisation steps. The spatial domain for each vessel $[0, L_z]$ is divided in $N_V=N\times m$ subintervals, with $m\in \mathbb{N}$, so in the resulting system, there are contact nodes between the blood and the tissue. Each vessel is thus discretised as a vector of dimension $N_V +1$. \Cref{Fig:figC1} shows a scheme of the distribution of the different spatial nodes, both for the computational domains of the tissue and for the vessels. Further details on the spatial discretisation can be found in \ref{Sec:App_Grid}.
\par

We will first consider the internal points of the tissue's grid. We define $N_\textrm{int}=(M-1)\times(N-1)$ and regard each unknown in the internal points of the tissue as a vector ${\boldsymbol \Psi}$ having components $\psi_{k}$ with $k=1,2,\ldots,N_\textrm{int}$ starting from the point $(\Delta x,\Delta z)$ (see \ref{Sec:App_Grid} and \cref{Fig:figC1}). For these arrays associated to that arrangement of interior points we have the following system of ODEs, which are the result of using a fourth-order finite difference discretisation in the spatial terms in the corresponding PDEs:
\begin{align}
\begin{split}
\label{Eq:ODE1}
    \frac{\D n_k}{\D t} =&\; D_n \sum_{l=1}^{(M+1)\times (N+1)} \left( G_{kl}+H_{kl} \right) n_l + \rho_n \left( 1-n_k-h_k-w_k-\xi c_k \right) n_k \\ 
    &- \sigma_{nh}(s_k)n_k + \sigma_{hn}(s_k)h_k - \sigma_{t}(T_k,(t_{43})_k)n_k,
\end{split}
\\
\begin{split}
\label{Eq:ODE2}
    \frac{\D h_k}{\D t} =&\; D_h \sum_{l=1}^{(M+1)\times (N+1)} \left( G_{kl}+H_{kl} \right) h_l + \rho_h \left( 1-n_k-h_k-w_k-\xi c_k \right) h_k \\ 
    &+ \sigma_{nh}(s_k)n_k - \sigma_{hn}(s_k)h_k - \sigma_{hc}(s_k)h_k - \sigma_{t}(T_k,(t_{43})_k)h_k,
\end{split}
\\ 
\begin{split}
    \frac{\D w_k}{\D t} =&\; \rho_w \left( 1-n_k-h_k-w_k-\xi c_k \right) w_k - \lambda \left( n_k+h_k \right) w_k \\
    &- \sigma_{nh}(s_k)w_k - \sigma_{t}(T_k,(t_{43})_k)w_k, 
\end{split}
\\
\begin{split}
\label{Eq:ODE4}
    \frac{\D c_k}{\D t} =&\; \lambda \left( n_k+h_k \right) w_k + \sigma_{nh}(s_k)w_k + \sigma_{hc}(s_k)h_k \\
    &+ \sigma_{t}(T_k,(t_{43})_k)\left( n_k+h_k+w_k \right), 
\end{split}
\\
\label{Eq:ODE5}
    \frac{\D T_k}{\D t} =&\; \frac{1}{\delta C} \left( \kappa \sum_{l=1}^{(M+1)\times (N+1)} \left( G_{kl}+H_{kl} \right) T_l + Q_n n_k + Q_h h_k + Q_w w_k + F_k(t) \right), 
    \\
\begin{split}
\label{Eq:ODE6}
    \frac{\D s_k}{\D t} =&\; D_s \sum_{l=1}^{(M+1)\times (N+1)} \left( G_{kl}+H_{kl} \right) s_l - \left( \alpha_n n_k + \alpha_h h_k + \alpha_w w_k \right)\frac{s_k}{K_M+s_k}, 
\end{split}
\\
\label{Eq:ODE7}
    \frac{\D (t_{43})_k}{\D t} =&\; \frac{1}{60} \left( \frac{1}{4} + \frac{1}{4} \tanh\left( \frac{T_k - 42}{1.5} \right) \right)^{43-T_i},
\end{align}
\par
\noindent where $G$ and $H$ are the two differentiation matrices representing the numerical second-order derivatives along the $x$ and $z$ directions, respectively. The details on how these matrices are built are given in~\ref{Sec:App_DerMat}. The intervening variables in \crefrange{Eq:ODE1}{Eq:ODE7} are supplemented with suitable initial conditions at each node $k=1,2,\ldots,N_\textrm{int}$.

To solve the equations that model the evolution of thermal energy and oxygen concentrations in the blood flow, we define for each vessel one equispaced grid with $N_V$ divisions. To numerate the points, we set one index $q=1,\dotsc,N_V+1$, where one of the ends is always imposed by the boundary conditions. The spatially discretised equations in these grids give rise to the following ODEs to be solved together with \crefrange{Eq:ODE1}{Eq:ODE7}
\begin{align}
    \label{Eq:ODE8}
    \frac{\D (T_{b1})_q}{\D t} &= \frac{\eta}{d_1 \delta_b C_b} \left( T_{1q} - (T_{b1})_q \right) - v_1 \sum_{r=1}^{N_V + 1} A_{qr} (T_{b1})_r. \\
    \frac{\D (T_{b2})_q}{\D t} &= \frac{\eta}{d_2 \delta_b C_b} \left( T_{2q} - (T_{b2})_q \right) + v_2 \sum_{r=1}^{N_V + 1} B_{qr} (T_{b2})_q. \\
    \frac{\D (s_{b1})_q}{\D t} &= \frac{\gamma}{d_1} \left( s_{1q} - (s_{b1})_q \right) - v_1 \sum_{r=1}^{N_V + 1} A_{qr} (s_{b1})_q. \\
    \label{Eq:ODE11}
    \frac{\D (s_{b2})_q}{\D t} &= \frac{\gamma}{d_2} \left( s_{2q} - (s_{b2})_q \right) + v_2 \sum_{r=1}^{N_V + 1} B_{qr} (s_{b1})_q.
\end{align}
The intervening variables in \crefrange{Eq:ODE8}{Eq:ODE11} are supplemented with suitable initial conditions at nodes $q=2,\ldots,N_{V}+1$ for variables $(T_{b1})_q$ and $(s_{b1})_q$, and $q=1,\ldots,N_{V}$ for variables $(T_{b2})_q$ and $(s_{b2})_q$. Note that from the entry points in \crefrange{Eq:ODE8}{Eq:ODE11} we have
\begin{align}
    (T_{b1})_1 =& T_{b0}, & (T_{b2})_{N_V+1} =& T_{b0}, \\
    (s_{b1})_1 =& s_{b0}, & (s_{b2})_{N_V+1} =& s_{b0}.
\end{align}
Therefore these values are used in the calculations of the ODEs of the adjacent points, but do not have an ODE themselves as their temporal variation is known. \Crefrange{Eq:ODE8}{Eq:ODE11} are coupled to the oxygen concentration and temperature in the tissue by means of the exchange of energy and mass through the boundary, which is proportional to the differences in the solid-liquid phases. Since the number of divisions in vessels' grids are $m$ times that of the tissue boundary, the variables $T_{1q},\,T_{2q},\,s_{1q}$ and $s_{2q}$ are computed through a linear interpolation of the corresponding $T_{1,j}$, $T_{M+1,j}$, $s_{1,j}$, $s_{M+1,j}$.
\par

Matrices $A$ and $B$ in \crefrange{Eq:ODE8}{Eq:ODE11} are the differentiation matrices for a first-order spatial derivative in the $z$ direction along the vessel path. To eliminate perturbations that typically affect transport equations, we use a five point biased upwind approximation in them, what forces us to use two different derivatives, one for the left vessel (matrix $A$), where blood flows in the direction of the axis, and another for the right vessel (matrix $B$), where blood flows opposite to the direction of the axis. Details on how these matrices are built can be found in \ref{Sec:App_DerMat}.
\par

In the resolution of \crefrange{Eq:ODE1}{Eq:ODE4} for cellular populations, the Neumann boundary conditions are used to obtain the relations that set the values of points in the boundary. We employ the following matrix form for each variable in the tissue $P_{i,j}$, $i=1,...,M+1$, $j=1,...,N+1$. Via a fourth-order discretisation we get 
\begin{align}
    \label{Eq:P_i1}
    P_{i,1} =& \frac{1}{25} \left( 48P_{i,2} - 36P_{i,3} + 16P_{i,4} - 3P_{i,5} \right), \\
    \label{Eq:P_iN+1}
    P_{i,N+1} =& \frac{1}{25} \left( 48P_{i,N} - 36P_{i,N-1} + 16P_{i,N-2} - 3P_{i,N-3} \right), \\
    P_{1,j} =& \frac{1}{25} \left( 48P_{2,j} - 36P_{3,j} + 16P_{4,j} - 3P_{5,j} \right), \\
    \label{Eq:P_M+1j}
    P_{M+1,j} =& \frac{1}{25} \left( 48P_{M,j} - 36P_{M-2,j} + 16P_{M-3,j} - 3P_{M-4,j} \right). 
\end{align}
for $P=\{n,h,w,c\}$. On the other hand, for tissue variables that are coupled to the vessel ones, i.e. $U=\{s,T\}$, the same \cref{Eq:P_i1,Eq:P_iN+1} are applied for the top and bottom boundaries as the same homogeneous Neumann conditions apply, whereas the points in the liquid-solid interface are ruled by the schemes

\begin{align}
    \label{Eq:U_1j}
    U_{1,j} =& \frac{1}{25+12a\Delta x} \left( 48U_{2,j} - 36U_{3,j} + 16U_{4,j} - 3U_{5,j} + 12a \Delta x U_{b1,j} \right), \\
    \label{Eq:U_M+1j}
    U_{M+1,j} =& \frac{1}{25+12a\Delta x} \left( 48U_{M,j} - 36U_{M-1,j} + 16U_{M-2,j} - 3U_{M-3,j} + 12a \Delta x U_{b2,j} \right),
\end{align}
where $a=\gamma/D_s$ for the temperature and $a=\eta/\kappa$ for the oxygen concentration. Furthermore, $U_{b1,j}, U_{b2,j}$ stand for the discretised vectors of $s_{b1}, s_{b2}$ when computing $s$, and for $T_{b1}, T_{b2}$ when computing $T$.
\par

Notice that when the simulation involves the application of an external heat power, that is, when simulating the administration of hyperthermia treatment, the radii of the vessels experiments a vasodilation due to the elevated temperature, affecting also the velocity of blood in the vessels as indicated in \cref{SubSec:ThermalEvo}. In this case, the diameters $d_1,d_2$ and velocities $v_1,v_2$ that appear in the above equations change with the temperature according to the expressions \cref{Eq:r1_variation,Eq:r2_variation,Eq:v1_variation,Eq:v2_variation}.
\par

Once we have discretised the spatial terms of our original PDEs by means of a finite difference scheme, we obtain a system of $7N_i+4N_V$ coupled ODEs formed by \crefrange{Eq:ODE1}{Eq:ODE7} and \crefrange{Eq:ODE8}{Eq:ODE11}, which only depend on time, together with the algebraic relations that enforce the boundary conditions. We use in the simulations the descriptive parameters that are shown in \cref{Tab:Tabla} from \ref{Sec:App_Tabla}, which characterise the problem. For the solution of the resulting system we employ standard Runge-Kutta methods for stiff problems. In our case, we implement the ODEs in the MATLAB language and use the solver function {\fontfamily{cmtt}\selectfont ode15s} to perform the integration in time. To accelerate the execution of this kind of built-in function it is convenient to provide the pattern of sparcity of the Jacobian, which expresses the coupling between the different equations. We provide the Jacobian pattern as explained in \ref{Sec:App_Jacobian}. Henceforth, all the system results have been obtained with MATLAB (R2020a, The MathWorks, Inc., Natick, MA, USA), run in a 24-core 192 GB RAM 2.7 GHz Mac Pro (2019).
\par

\section{Results}
\subsection{Initial tumour development}
\label{SubSec:Results_FREE}
Firstly, to better understand the progression during free proliferative growth, we simulate the evolution of populations of tumour cells, oxygen concentration and temperature. We describe next the initial conditions considered. For cell populations, a small and highly localised population of tumour normoxic cells is initially centred in the tissue, which is otherwise occupied by healthy cells at a constant density $w_{00}$, with no hypoxic tumour cells nor necrotic cells present. The densities are

\begin{align}
    n(x,z,0) &= n_{00} \exp \left( - \frac{(x-L_x/2)^2}{\sigma_x^2} - \frac{(z-L_z/2)^2}{\sigma_z^2} \right), \\
    w(x,z,0) &= w_{00}, \\
    h(x,z,0) &= 0, \\
    c(x,z,0) &= 0,
\end{align}
where $\sigma_x$ and $\sigma_z$ denote characteristic widths of the initial tumour normoxic cell distribution.
\par
The initial conditions for oxygen and temperature in the tissue are set to constant values $s_0$ and $T_0$, which rapidly attain the steady-state values consistent with the initially small population. The variables in the vessels are ruled by Dirichlet conditions at their entrance, which are located at bottom side ($z=0$) for the left vessel (numbered as 1), and at the top size ($z=L_z$) for the right one (numbered as 2), respectively. The initial conditions for the oxygen and temperature within the vessel are set to the same value as the inflow oxygen and temperature imposed by the Dirichlet conditions, i.e. $s_{b0}$ and $T_{b0}$. Again, they rapidly tend to the values imposed by the transfer to and from the tissue. Therefore we have

\begin{align}
    s(x,z,0) &= s_0, \\
    T(x,z,0) &= T_0, \\
    s_{b1}(z,0) &= s_{b2}(z,0) = s_{b0} = s_{b1}(0,t) = s_{b2}(L_z,t), \\
    T_{b1}(z,0) &= T_{b2}(z,0) = T_{b0} = T_{b1}(0,t) = T_{b2}(L_z,t).
\end{align}

\subsubsection{Case of functional vasculature}

We first performed simulations with a value of the velocity in both vessels equal to $v_1=v_2=1600$ \si{\micro\meter\per\second}. This represents a scenario in which the vasculature is functional and thus can provide a sufficient flow to maintain the demands of the tissue. From the cell densities $n(x,z,t)$, $h(x,z,t)$, $w(x,z,t)$ and $c(x,z,t)$, we calculate the number of cells $n_{\mathrm{num}}(t)$, $h_{\mathrm{num}}(t)$, $w_{\mathrm{num}}(t)$ and $c_{\mathrm{num}}(t)$, contained in the region of interest, by spatially integrating the solutions of \cref{Eq:normoxic,Eq:hypoxic,Eq:healthy,Eq:necrotic}, and apply a conversion factor $\phi$ that accounts for all the cells surrounding the blood vessels in tumour cords in a 3D biological scenario (see \ref{Sec:App_Geometrical})

\begin{align}
\label{Eq:n_num}
    n_{\mathrm{num}}(t) &= \phi \int_0^{L_x} \!\!\! \int_0^{L_z} n(x,z,t) \D x \D z, \\
    h_{\mathrm{num}}(t) &= \phi \int_0^{L_x} \!\!\! \int_0^{L_z} h(x,z,t) \D x \D z, \\
    w_{\mathrm{num}}(t) &= \phi \int_0^{L_x} \!\!\! \int_0^{L_z} w(x,z,t) \D x \D z, \\
\label{Eq:c_num}
    c_{\mathrm{num}}(t) &= \phi \int_0^{L_x} \!\!\! \int_0^{L_z} c(x,z,t) \D x \D z. \\
\end{align}

To study the evolution of oxygen and temperature in the region of interest we used aggregate spatial values by averaging over the computational domain 

\begin{align}
\label{Eq:s_mean}
    \overline{s}(t) &= \frac{1}{L_x L_z} \int_0^{L_x} \!\!\! \int_0^{L_z} s(x,z,t) \D x \D z, \\
    \overline{T}(t) &= \frac{1}{L_x L_z} \int_0^{L_x} \!\!\! \int_0^{L_z} T(x,z,t) \D x \D z.
\end{align}

\begin{figure}[!t]
    \centering
    \includegraphics[width=1 \textwidth]{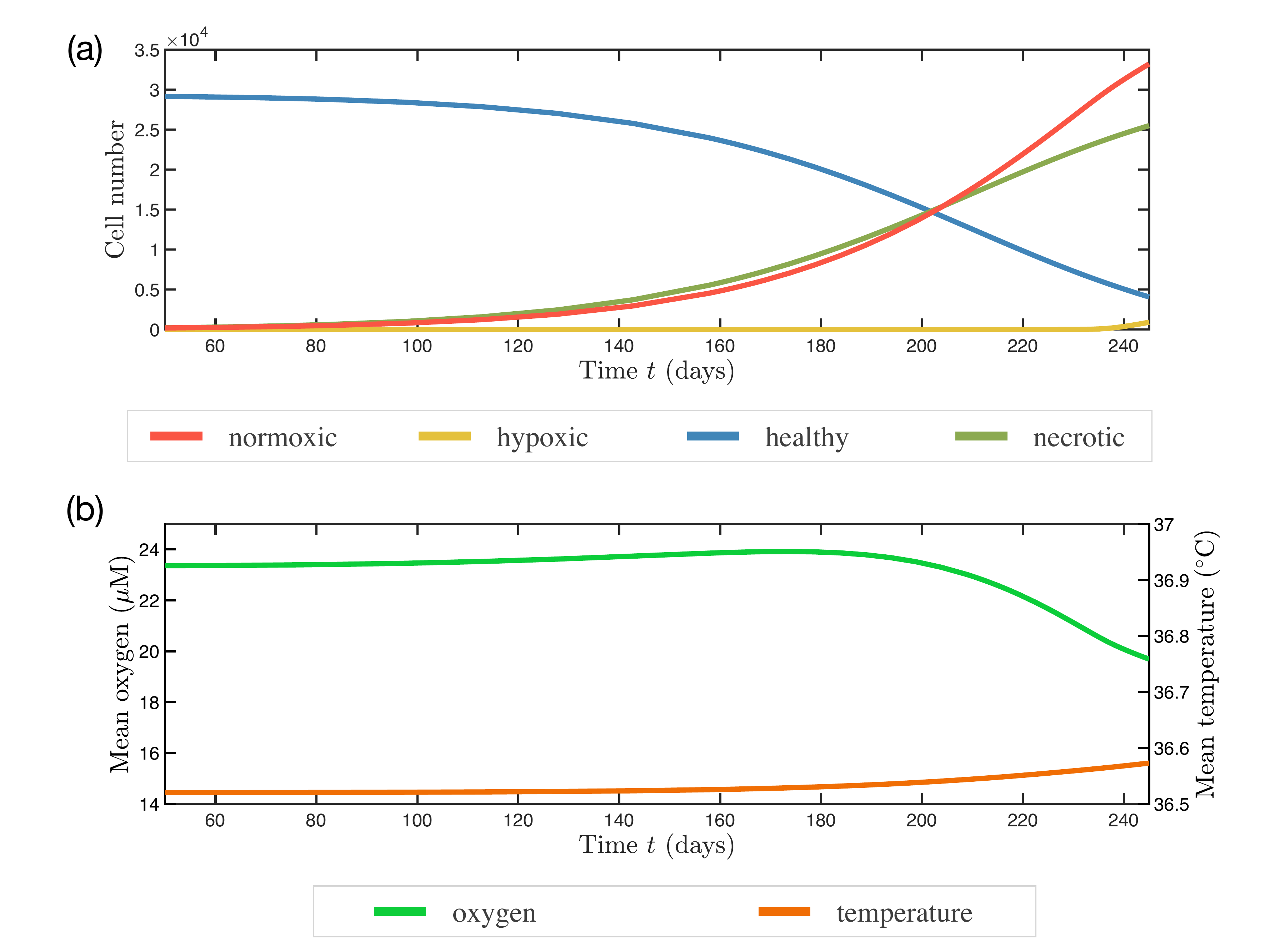}
    \caption{Time evolution of the spatially aggregated variables from the PDEs solution is a case study where the blood velocity in the vessels is $v_1=v_2=1600$ \si{\micro\meter\per\second}. The top figure \textbf{(a)} shows the total number of normoxic tumour cells ($n_{\mathrm{num}}(t)$), hypoxic tumour cells ($h_{\mathrm{num}}(t)$), healthy cells ($w_{\mathrm{num}}(t)$) and necrotic cells ($c_{\mathrm{num}}(t)$); while the healthy cells decay due to the increasing presence of tumour cells contributing to the development of a necrotic compartment, the normoxic tumour cells proliferate to be the predominant population. The bottom figure \textbf{(b)} shows the evolution of the mean value of oxygen ($\overline{s}(t)$), which decreases to medium values when the normoxic populations takes over, and the mean temperature ($\overline{T}(t)$), which experiments a very light raise. Overall, the correct functionality of blood vessels maintains the physical variables within the normal values.}
    \label{Fig:fig2}
\end{figure}

The results of the temporal evolution of these variables between 0 and 245 days ($t_{\mathrm{end}}=245$ days) are shown in \cref{Fig:fig2}. As the time progresses, the healthy cell population ($w_{\mathrm{num}}$) that initially dominates the tissue steadily decreases due to the presence of tumour cells proliferating and infiltrating the region. Cell death from this compartment contributed to the population of necrotic cells ($c_{\mathrm{num}}$). Meanwhile, the normoxic tumour cells ($n_{\mathrm{num}}$), located at the centre of the region, proliferate and start taking over the region. After 200 days, the cancerous cells are the most numerous population. The associated raise in oxygen consuming cells above the capacity of the tissue makes the mean levels of oxygen ($\overline{s}$) decrease by a 16 \% at the end of the period. However, the mean values of oxygen concentration do not plunge to hypoxic levels since the blood vessels are able to supply enough oxygen to keep the cells alive. At the end of the studied time window, the oxygen reaches its lowest level and, as a consequence, a small number of hypoxic tumour cells appear in the system, having $h_{\mathrm{num}}(t_{\mathrm{end}})> 0$. The progressive stress put on the vessel resources makes the refrigeration provided by blood flow more difficult, thus leading to a slight increase of 0.05 \si{\degreeCelsius} in the mean temperature ($\overline{T}$).
\par

\begin{figure}
    \centering
    \includegraphics[width=0.95 \textwidth]{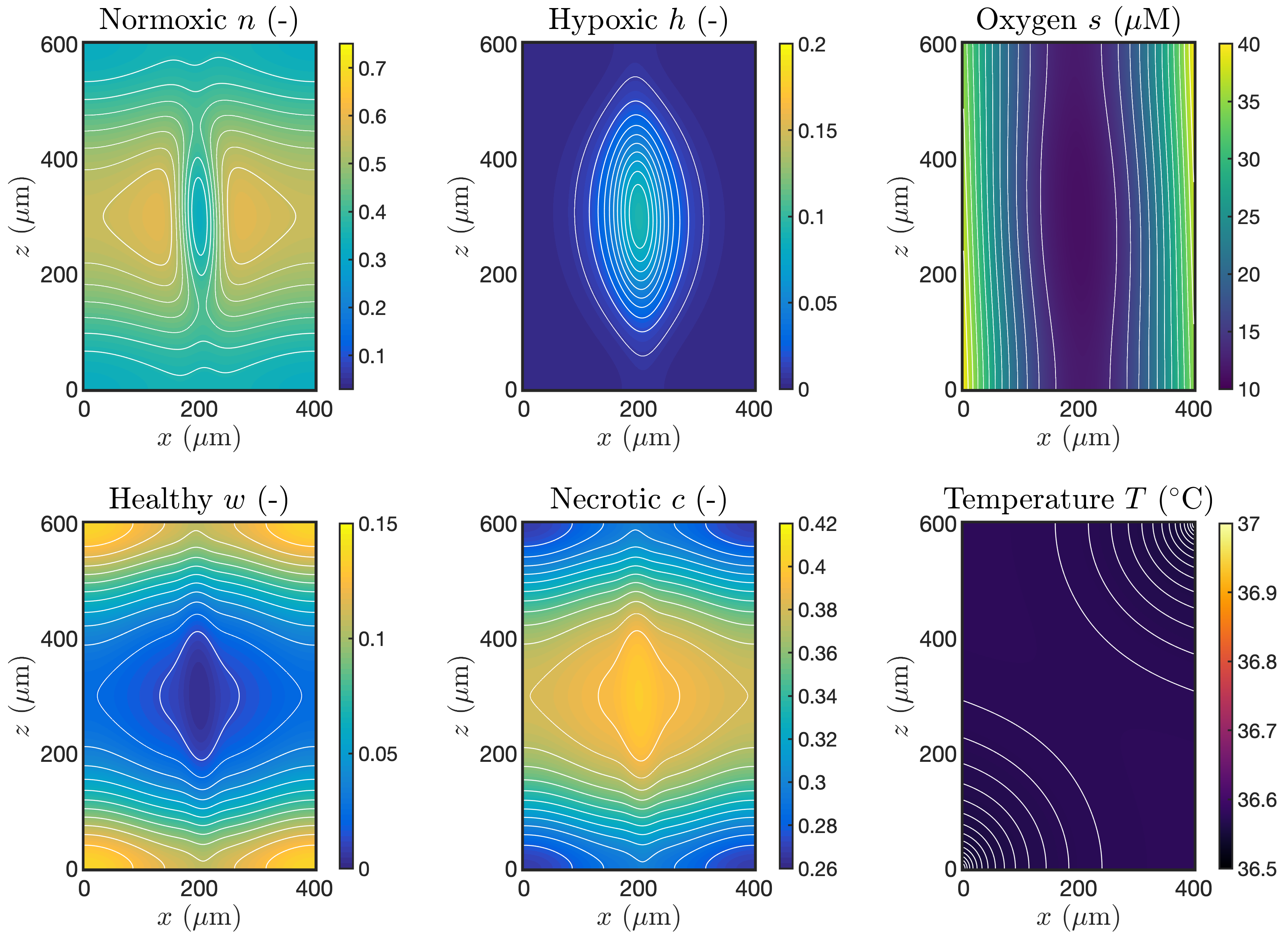}
    \caption{Maps of the model functions at the final time ($t_{\mathrm{end}}=245$ days) in a case simulation with a functional vasculature characterised by a blood velocity of $v_1=v_2=1600$~\si{\micro\meter\per\second}. The normoxic population ($n(x,z,t_{\mathrm{end}})$) occupies most of the tissue, specially in the transversal central areas. The oxygen concentration ($s(x,z,t_{\mathrm{end}})$) leads to low values at the centre of the region, far from the blood vessels, causing the appearance of a small number of hypoxic cells ($h(x,z,t_{\mathrm{end}})$) in the low oxygen area. Healthy cells ($w(x,z,t_{\mathrm{end}})$) cannot survive in those places where the oxygen is low and the tumour cells take over, thus resulting in the emergence of a central necrotic core ($c(x,z,t_{\mathrm{end}})$). There is a uniform low temperature distribution ($T(x,z,t_{\mathrm{end}})$) due to the effect of an adequate blood flow.}
    \label{Fig:fig3}
\end{figure}

The geometric setting of the tissue, where different cells are located at different distances from the feeding vessels, gives rise to heterogeneous spatial configurations. The solutions of the PDEs provide the complete spatio-temporal information and, after having explored the temporal change of the bulk variables, it is interesting to study the spatial configuration that remains at the final time. \Cref{Fig:fig3} shows the final states of the tissue related unknowns at the end of the simulation. Normoxic tumour cells ($n$) have dominated the tissue and appear at high densities in all the domain, being their mean density 0.452 in all the region. Some parts are especially populated by these cells, particularly the central transversal area at intermediate distances (60 to 150 \si{\micro\meter}) from the vessels. The original healthy cells ($w$) have left the central part of the tissue and have being relegated to well-oxygenated positions distant from the point of initiation of the tumour proliferation. This death of healthy cells leaves a core of necrotic cells ($c$) which is located at the centre of the area. The higher pressure put on oxygen delivery by the new tumour cells produces a notable drop in the oxygen concentration ($s$) parallel to the blood vessels path. The lowest oxygen levels in the centre reach 11.22~\si{\micro\Molar}, what is low enough to trigger the transformation of some normoxic tumour cells to their hypoxic counterparts ($h$), which have an altered metabolism and are characterised by a lower proliferation capacity but a higher motility. This population appears in the centre part of the tissue, parallel to the blood vessels and in the farthest distances from them, but only in low densities with a maximum of 0.092. Temperatures $T$ in the tissue remain low, with minimum values located at the entrance of fresh blood and extending their refrigerating effect to nearby areas. As to the levels of temperature ($T_{b1},\,T_{b2}$) and oxygen ($s_{b1},\,s_{b2}$) in the blood contained within the vessels, their evolutions along their paths at the last time of the simulation are depicted in \cref{Fig:figH1}(a) and \cref{Fig:figH1}(b) respectively. In the latter, a drop in oxygen levels from the inlet to the outlet of the vessels reflects the tissue consumption, while the elevation of the temperature in the former reflects the refrigerating effect of the flow. Nevertheless, both the drops in oxygen and the elevation in temperature are small under these conditions of blood flow.
\par

\begin{figure}[!ht]
    \centering
    \includegraphics[width=1 \textwidth]{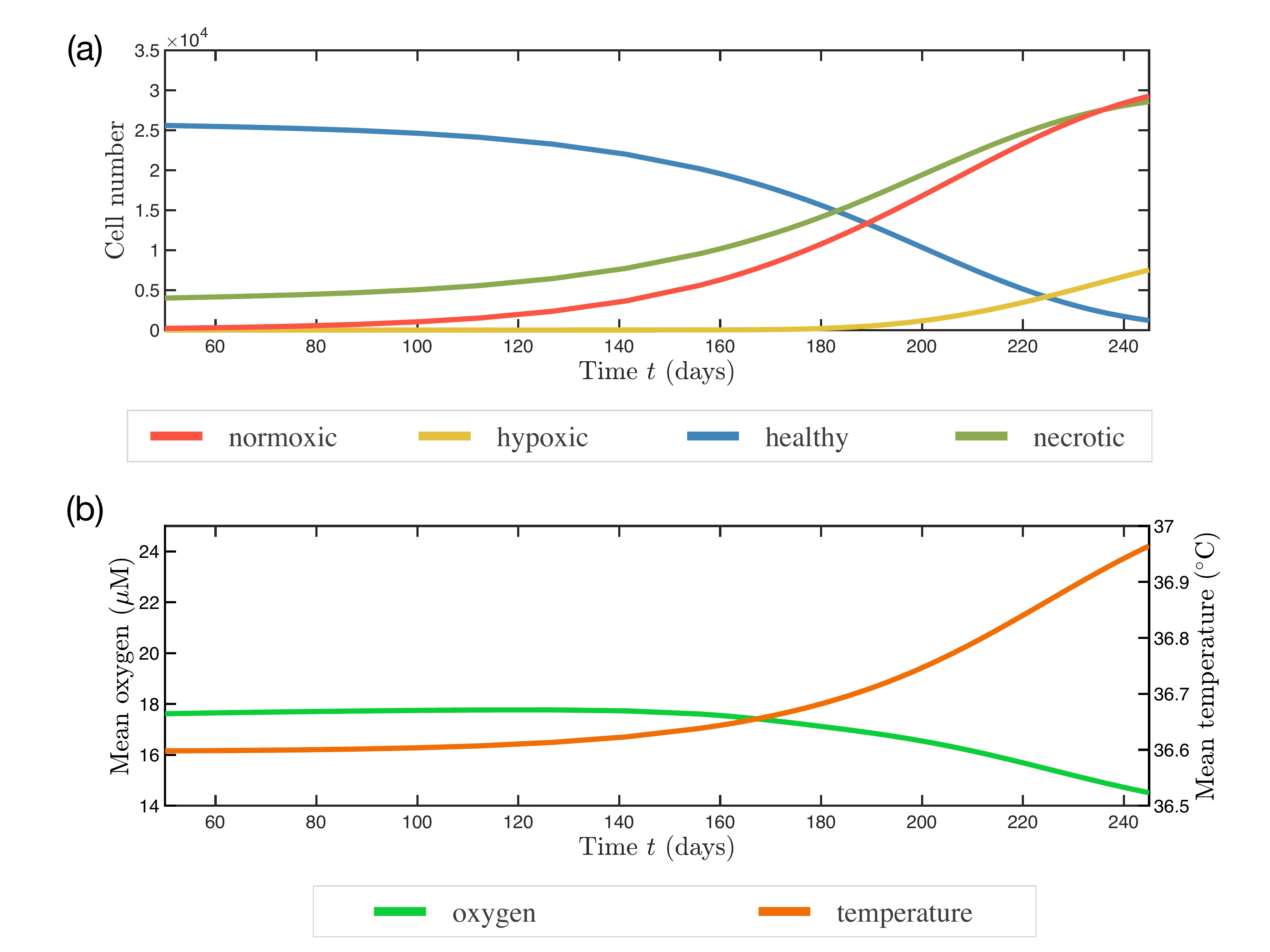}
    \caption{\textbf{(a)} Time evolution of the aggregated variables $n_{\mathrm{num}}(t),\;h_{\mathrm{num}}(t),\;w_{\mathrm{num}}(t)$ and $c_{\mathrm{num}}(t)$ from the solution to the model PDEs in a case with a dysfunctional vasculature where the blood velocity is $v_1=v_2=300$ \si{\micro\meter\per\second}. \textbf{(b)} Since the vasculature is not able to provide enough oxygen, the mean oxygen values in the tissue ($\overline{s}(t)$) become relatively low, giving rise to the development of hypoxic cells. Due to a less efficient vasculature capable to extract the metabolic heat generated, the temperature ($\overline{T}(t)$) steadily increases to values near 37 \si{\degreeCelsius} at the end of the simulation.}
    \label{Fig:fig4}
\end{figure}

\subsubsection{Case of impaired vasculature}

After having explored the evolution of the system, we wondered what differences there were for the same configuration evolving under an impaired blood flow. This situation appears indeed during cancer progression, due to the pathological characteristics of tumour vasculature, whose unregulated genesis leads to and erratic behaviour. To determine the conditions that may develop when the blood flow is insufficient, we solved the equations from \cref{Sec:MathModel} with a blood velocity in the vessels of $v_1=v_2=300$ \si{\micro\meter\per\second}. The time evolution of the spatially aggregated variables from the solutions is shown in \cref{Fig:fig4}. In the same way that the previous case, the initial population of healthy cells decreases due to the contact with the blooming tumour population, which progresses steadily and until the day 180 is predominantly constituted by normoxic cells. From that point, the mean level of oxygen ($\overline{s}$), which was already lower than in the previous case, starts declining as a consequence of the high demand imposed by the tissue. The lack of oxygen unleashes the transformation of normoxic cells in hypoxic cells which make up an important part of the tumour cells at the end of the simulation. The final levels of oxygen are low for this case, what can also be seen in the final state of the oxygen concentration in the blood vessels ($s_{b1},\,s_{b2}$) which show a strong reduction along its path (\cref{Fig:figH1}(d)). The mean temperature, for his part, rises with the saturation of the tissue by tumour cells, reaching values near to 37 \si{\degreeCelsius}.    
\par

\begin{figure}[!ht]
    \centering
    \includegraphics[width=0.95 \textwidth]{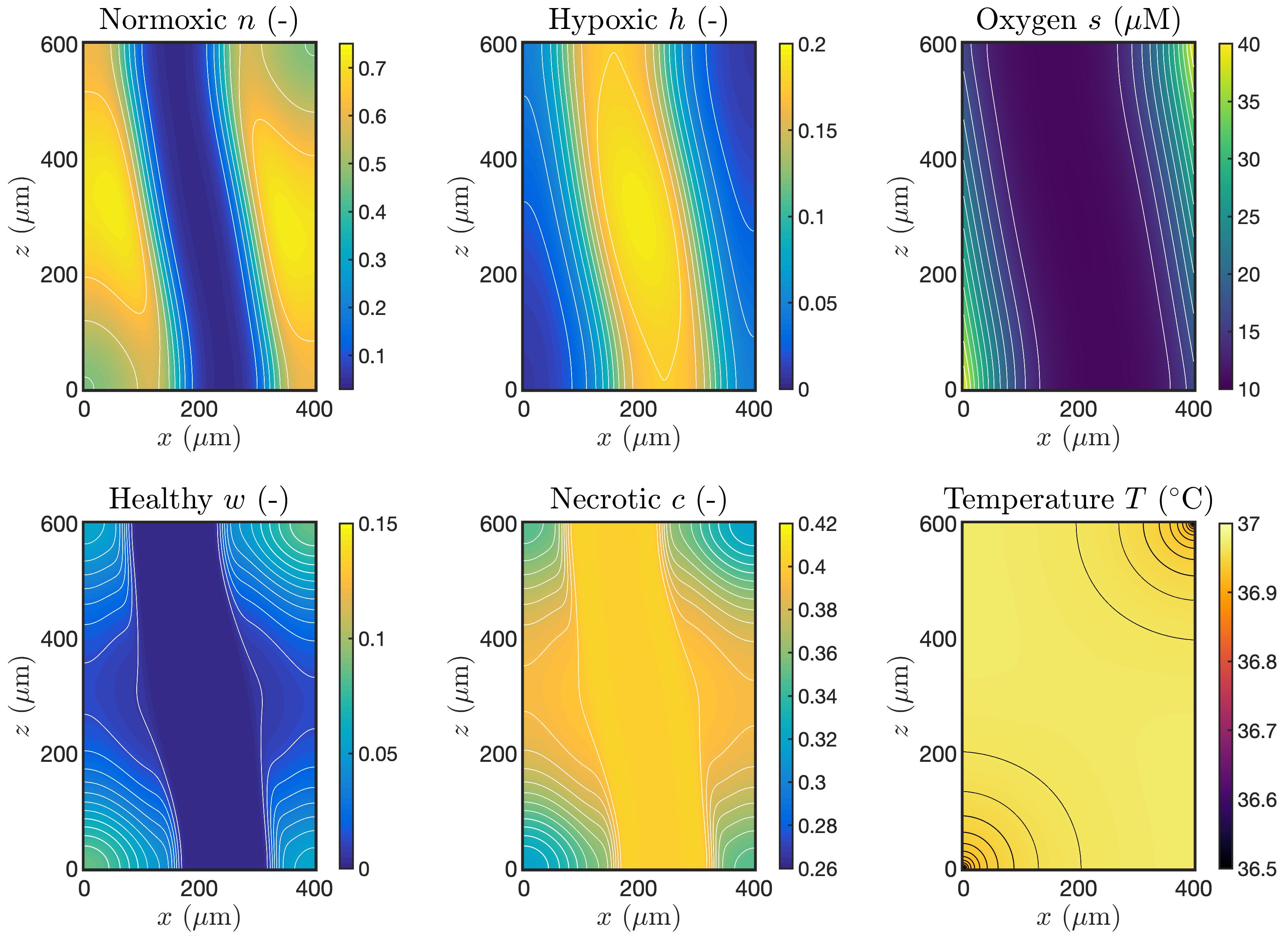}
    \caption{Maps of the solutions of the PDEs at the final time ($t_{\mathrm{end}}=245$ days) in a system with a velocity of $v_1=v_2=300$ \si{\micro\meter\per\second} which is a representation of an abnormal vasculature with poor performance. The low capability of the vessels to provide oxygen causes a wide area of low oxygen concentration ($s(x,z,t_{\mathrm{end}})$). This leads to a strip where no healthy cells ($w(x,z,t_{\mathrm{end}})$) nor normoxic tumour cells ($n(x,z,t_{\mathrm{end}})$) are present, but a hypoxic population ($h(x,z,t_{\mathrm{end}})$) with higher tolerance to harsh conditions ensues. Likewise, the low levels of blood flow hinders the extraction of metabolic heat, leading to an elevation of the temperature ($T(x,z,t_{\mathrm{end}})$).}
    \label{Fig:fig5}
\end{figure}

It is interesting to observe the spatial maps of these solution at the final time that are shown in \cref{Fig:fig5} for the case of $v_1=v_2=300$ \si{\micro\meter\per\second}. We can see now that the high densities of normoxic tumour cells are located adjacent to the vessels, while there is a very important number of hypoxic tumour cells that are occupying the centre of the region. That centre part is distinguished by a wide strip of low oxygen concentration. In that strip, healthy cells cannot thrive and, as a consequence, there is no healthy population there. In exchange, the healthy cells that have died leave a necrotic core that occupy a large part of the area. As to temperature, we can see an almost uniform high temperature resulting from the metabolic heat generation and the low capability of blood vessels to remove heat. Therefore, for tumour cells thriving among damaged blood vessels, higher levels or hypoxic cells are expected, at the same time that local temperatures will be high with low oxygen levels which are only practicable at locations near the vessels. 
\par

\begin{figure}[t]
    \centering
    \includegraphics[width=1 \textwidth]{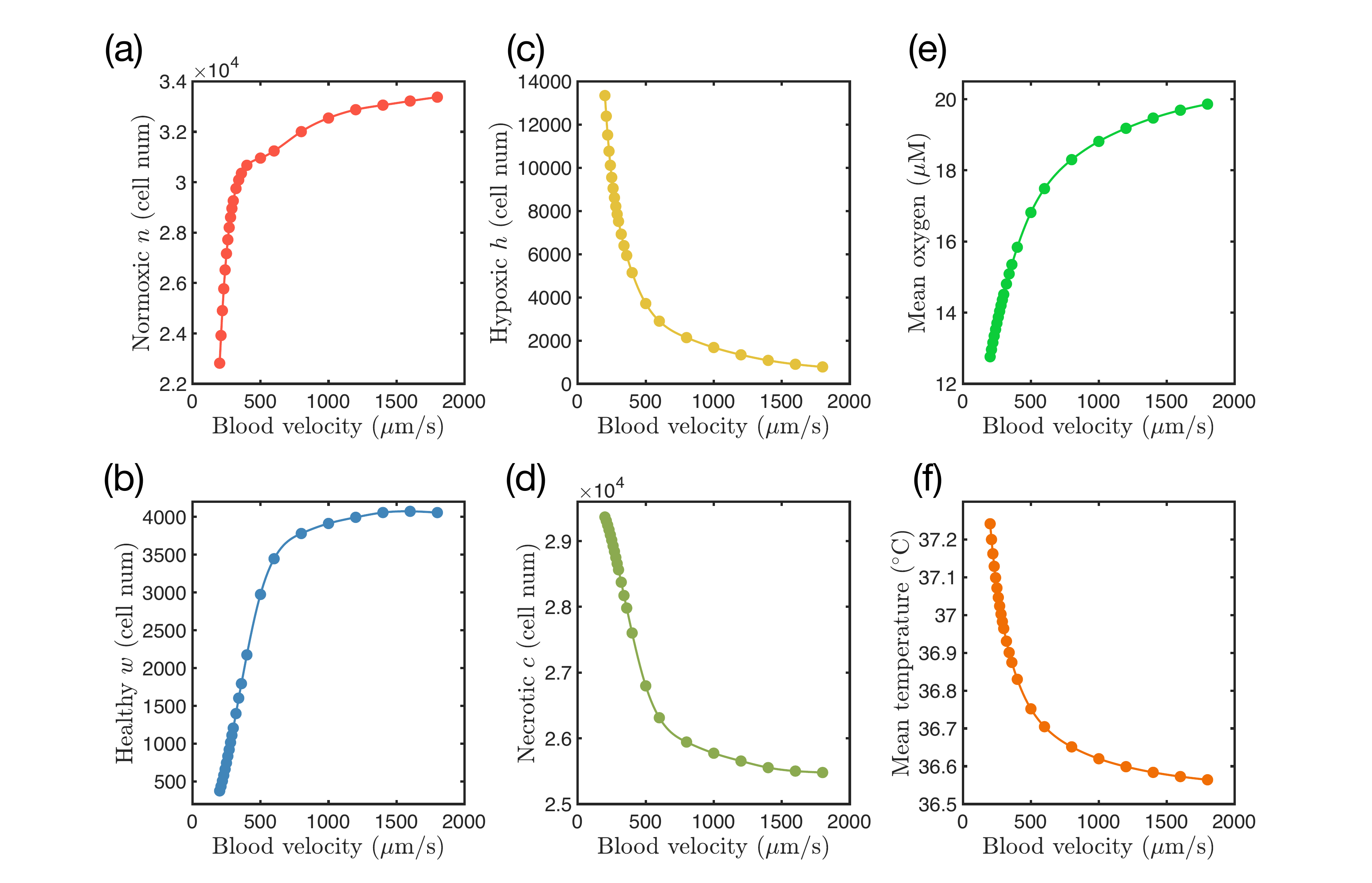}
    \caption{Results of the bulk variables: \textbf{(a)} $n_{\mathrm{num}}(t_{\mathrm{end}})$, \textbf{(b)} $w_{\mathrm{num}}(t_{\mathrm{end}})$, \textbf{(c)} $h_{\mathrm{num}}(t_{\mathrm{end}})$, \textbf{(d)} $c_{\mathrm{num}}(t_{\mathrm{end}})$, \textbf{(e)} $\overline{s}(t_{\mathrm{end}})$ and \textbf{(f)} $\overline{T}(t_{\mathrm{end}})$ at the last time point ($t_{\mathrm{end}}=245$ days) from simulations of the system with different velocities indicated in the x-axis. As the velocity of blood in the vessels increases, the level of oxygen increases, leading to a lower presence of hypoxic tumour cells and an increase in both normoxic tumour cells and healthy cells. A greater blood velocity involves a greater cooling ability and, therefore, the mean temperature in the tissue decreases.}
    \label{Fig:fig6}
\end{figure}

\subsubsection{Influence of vascular functionality}

We can see that the flow in the blood vessels, which is used here as a proxy for the vascular level of functionality, has a strong influence on the evolution of the main variables of the problem. Therefore we wish to investigate how the solutions change for a range of levels of blood flow in the vessels. With that aim, we perform simulations of the system with equal velocities in the vessels ranging from $v_1=v_2=200$ \si{\micro\meter\per\second} to $v_1=v_2=1800$ \si{\micro\meter\per\second}. The results are shown in \cref{Fig:fig6}, where we have depicted the total number of cells of each population at the time $t_{\mathrm{end}}=245$ days of simulations ($n_{\mathrm{num}}(t_{\mathrm{end}})$, $w_{\mathrm{num}}(t_{\mathrm{end}})$, $h_{\mathrm{num}}(t_{\mathrm{end}})$ and $c_{\mathrm{num}}(t_{\mathrm{end}})$ in panels (a) to (d)) and the mean values of oxygen and temperature ($\overline{s}(t_{\mathrm{end}})$, $\overline{T}(t_{\mathrm{end}})$ in panels (e) and (f), respectively). In the figures we can distinguish two different regimes, below and above $v_1=v_2=500$ \si{\micro\meter\per\second}. Below that point, the specific level of the velocity becomes critical. Very low levels of blood velocity are linked to small number of normoxic cell populations in favour of hypoxic cells which appear in their place. This is a consequence of the mean oxygen concentration evolution that is well reduced when the blood flow is low. Likewise, the number of healthy cells is reduced to almost zero for very small blood velocities, what results in the accumulation of necrotic cells. Regarding mean temperatures, their values get increased when the flow is reduced, therefore, the accumulation of tumour cells in areas marked by an impaired local blood flow are expected to get hotter that the homeostatic temperatures of the tissue. Above $v_1=v_2=500$ \si{\micro\meter\per\second} the same trends remain, however, the relative importance of blood velocity at high values decline and the solutions get more similar among each other.
\par

\begin{figure}[ht]
    \centering
    \includegraphics[width=0.45 \textwidth]{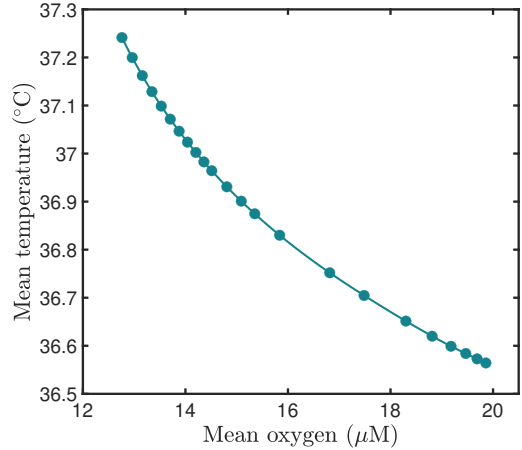}
    \caption{From the simulations carried out in \cref{SubSec:Results_FREE} for a range of variation of the velocities in the vessels whose overall results are depicted in \cref{Fig:fig6}, we take the final mean values of oxygen $\overline{s}(t_{\mathrm{end}})$ and temperature $\overline{T}(t_{\mathrm{end}})$. Under the assumptions followed here we encounter a strong inverse relationship between oxygen and temperature in a microscopic irrigated portion of tissue.}
    \label{Fig:fig7}
\end{figure}

Interestingly, the two biophysical variables from the tissue follow opposite behaviours for different conditions of the vascular system. From the results of our model, when the blood flow is low, the temperature tends to increase, while the oxygen concentration follows a declining trend. The status of the vasculature plays an important role in both, therefore one would expect a linked behaviour among the two. From the multiple simulations carried out for different flow conditions we obtain mean values of oxygen concentration and mean temperature in a microscopic region that surround two blood vessels. In \cref{Fig:fig7} we represent the calculated values of temperature against the levels of oxygen present in that system. It is observed that both of them are intimately ligated by an inverse relation. According to it, portions of tissue under the condition of hypoxia would display a higher temperature than well oxygenated tissues. However, it must be noted that the range of variation of oxygen is much higher than the variation that appears in the temperature.
\par

\FloatBarrier
\subsection{Thermal therapy response}
We have shown that the differences in blood flow lead to different ways in which the system evolves resulting in disparate final states that will have diverse responses to therapies, for instance, radiotherapy. We studied next how these distinct states behave differently under the application of thermal therapy with the goal of discriminating effects that appear at the microscale. In order to investigate that, we built up from the results of free cancer evolution up to 245 days that had been calculated in the previous section, which we used as the initial condition for the new simulations. Starting from them, we performed simulations of the system under the application of an external heat power of $P=40$~\si{\watt\per\liter}, value that is a standard in hyperthermia treatment.
\par

\subsubsection{Effect of vascular impairment}
\begin{figure}[!b]
    \centering
    \includegraphics[width=1 \textwidth]{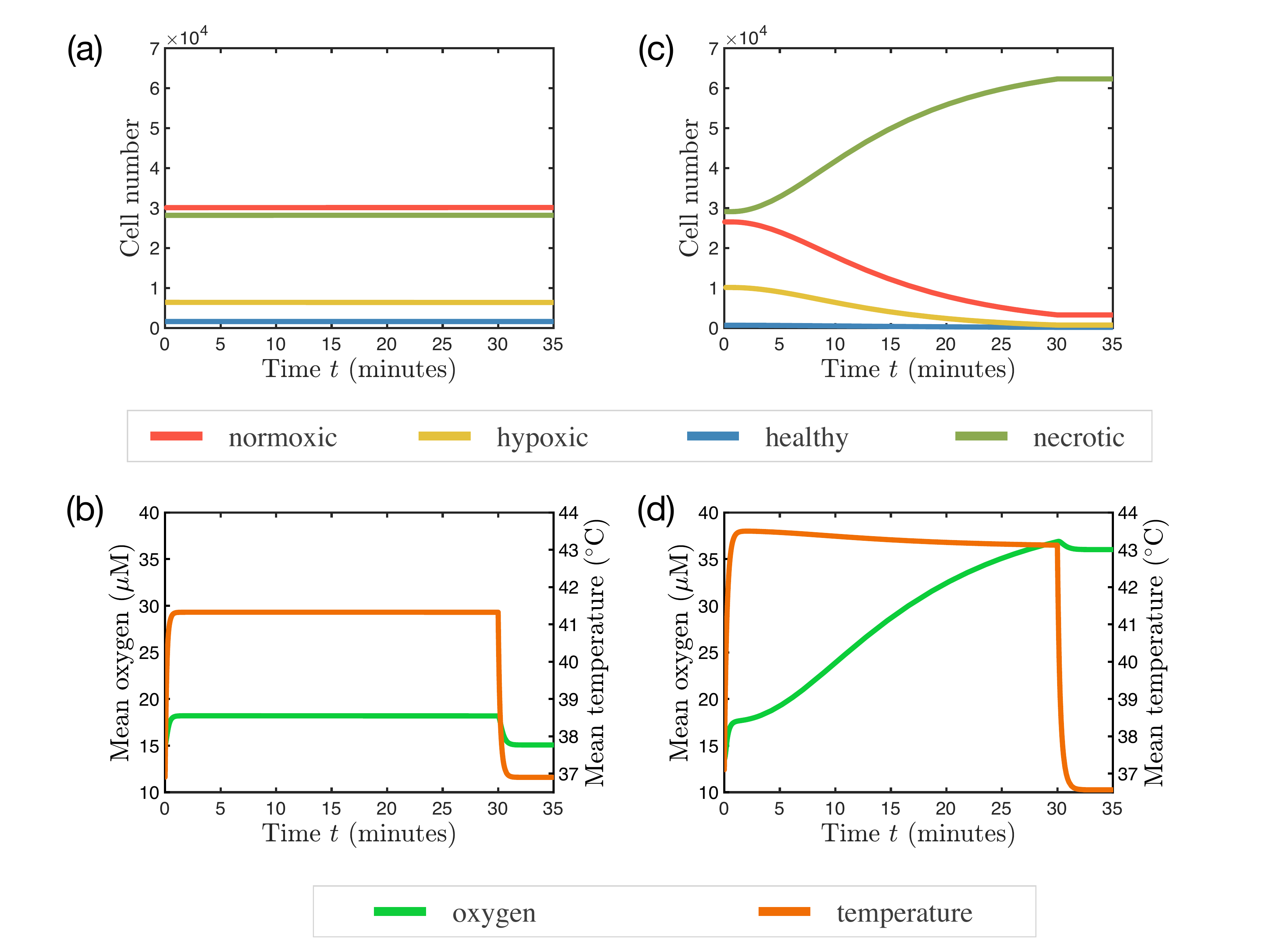}
    \caption{Evolution of the system under thermal therapy applied between $t=0$ minutes and $t=30$ minutes with a power of $P=40$ \si{\watt\per\liter} for two different cases of blood vessels functionality. \textbf{(a)} and \textbf{(b)} Simulation with a nominal---prior to the treatment---blood velocity of $v_{01}=v_{02}=340$~\si{\micro\meter\per\second}. \textbf{(c)} and \textbf{(d)} Simulation with a nominal blood velocity of $v_{01}=v_{02}=240$~\si{\micro\meter\per\second}. Both simulations start from an initial state calculated by the free evolution of the system under the same blood flow nominal values during a time of 245 days. \textbf{(a)} and \textbf{(c)} show the evolution of the number of cells in each population ($n_{\mathrm{num}}(t)$, $h_{\mathrm{num}}(t)$, $w_{\mathrm{num}}(t)$ and $c_{\mathrm{num}}(t)$) during therapy and \textbf{(b)} and \textbf{(d)}, the corresponding mean oxygen concentration ($\overline{s}(t)$) and mean temperature ($\overline{T}(t)$). Since the first case (left column) does not reach temperatures over 43 \si{\degreeCelsius}, cell death is negligible. For the second case (right column) the impaired blood flow is not able to remove the heat excess and the temperature reaches higher values. This leads to important levels of thermal cell death resulting in a reduction of tumour cells. This decrease in the cell populations gives rise to a reoxygenation effect.} 
    \label{Fig:fig8}
\end{figure}

As it was detailed in \cref{Sec:Calculation}, the simulations under thermal therapy allow for the variation of the conditions of blood flow that come from the thermal effect of vasodilation. These simulations also include in the calculation the accumulation of thermal dose $t_{43}$ that is received in each point of the tissue. We apply the treatment during 30 minutes between $t=0$ min and $t=30$ min and observe the evolution for two different cases with different levels of blood flow, namely $v_{01}=v_{02}=340$~\si{\micro\meter\per\second} and $v_{01}=v_{02}=240$~\si{\micro\meter\per\second}. The results of the time evolution of the cell numbers of each of the populations for the first case are shown in \cref{Fig:fig8}(a), while the corresponding mean levels of oxygen and temperature are depicted in \cref{Fig:fig8}(b). As it can be seen, the temperature reached a maximum value of $\overline{T}=41.33$~\si{\degreeCelsius}, what makes any contribution to cell death almost inappreciable. Therefore, the evolution of all the cell populations remained almost constant between the beginning and the end of the treatment. The mean levels of oxygen experimented a rise of 20.6\% due to the increase in blood flow provided by thermal induced vasodilation, however, the levels returned to the nominal values shortly after the cessation of the heating.
\par

The case with a severely impaired blood flow (i.e., $v_{01}=v_{02}=240$~\si{\micro\meter\per\second}), shows a different outcome. In this case the temperature reaches values over 43~\si{\degreeCelsius}, inducing an important level of cell death along the 30~\si{\minute} of the therapy application. Both normoxic tumour cells and hypoxic tumour cells experiment a steady reduction in their numbers, while its rate gets increased with the accumulation of thermal dose $t_{43}$ (note the decrease in the slope of the cell number as time progresses from $t=0$ min). The number of normoxic tumour cells present at the end of the treatment is 12.2\% of what it was at the beginning, while for hypoxic cells it is 7\%. As a consequence of the reduction of consumers, the level of oxygen experiments a big increase during the treatment, being its mean value at the end of the treatment 2.7 times bigger than at the beginning. In this case as in the previous one, the effect of vasodilation is also noticeable in the elevation of the oxygen concentration, and again this effect ceases rapidly once the treatment is over.  
\par

\begin{figure}[!b]
    \centering
    \includegraphics[width=0.95 \textwidth]{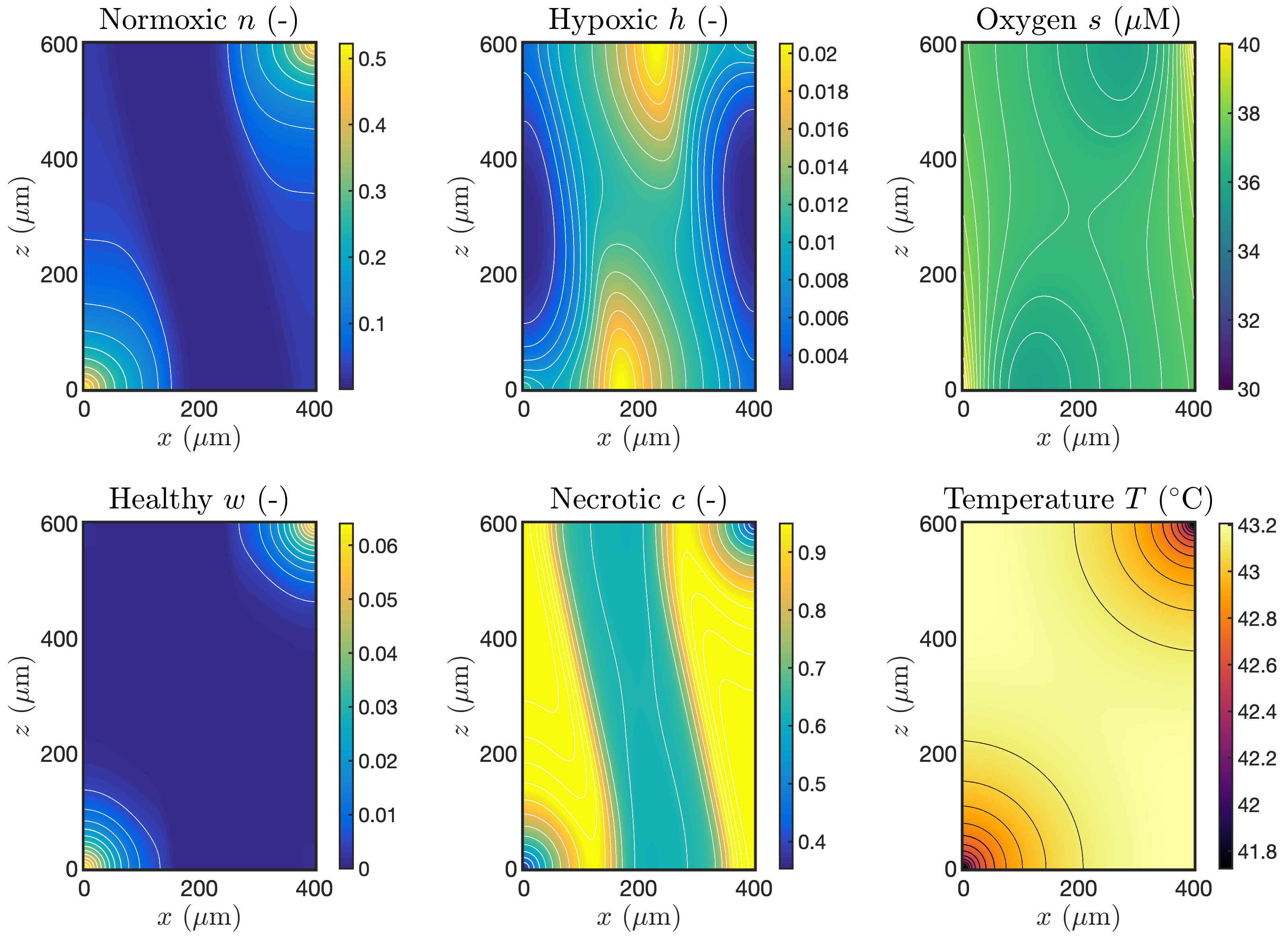}
    \caption{Maps of the final state of the cell population densities ($n(x,z,t_{\mathrm{end}})$, $h(x,z,t_{\mathrm{end}})$, $w(x,z,t_{\mathrm{end}})$, $c(x,z,t_{\mathrm{end}})$), oxygen concentration ($s(x,z,t_{\mathrm{end}})$) and temperature distribution ($T(x,z,t_{\mathrm{end}})$) at the end of 30 minutes of thermal therapy ($t_{\mathrm{end}}=30$ min) applying a power of $P=40$ \si{\watt\per\liter} in a tissue irrigated at the left and right sides by two blood vessels where blood circulates previously to vasodilation at a velocity of $v_{01}=v_{02}=240$ \si{\micro\meter\per\second}. The effect of the applied power elevates the temperature above $T=43$~\si{\degreeCelsius}, causing the death of many of the cells present in the tissue. In this case, there remain normoxic tumour cells in the areas near the entrance of fresh blood, and some hypoxic tumour cells in the vicinity of those entrances, but far enough from the vessels for hypoxic cells to have developed. Those normoxic niches are precisely the areas of lower temperature during the therapy. Moreover, the death of many oxygen consuming cells leads to an uprise in the oxygen levels of the overall region of interest.} 
    \label{Fig:fig9}
\end{figure}

In \cref{Fig:fig9} we show the spatial maps at the end of the 30~minutes of hyperthermia application for the case of $v_{01}=v_{02}=240$ \si{\micro\meter\per\second}, i.e. the case of impaired blood flow, which is the only of the two that shows a change in the populations induced by the treatment. The normoxic population after the treatment is only present at low densities and it is preferentially found in cold niches near the entrance of fresh blood from the vessels. The highest densities of hypoxic tumour cells can be mostly found far from the oxygen source of the blood vessels, restricted to small areas that are sufficiently far from the vessel to maintain the lowest levels of oxygen, but sufficiently near to it as to have a cooler temperature and therefore a smaller effect of the treatment. The general levels of oxygen are now much higher that previously to the treatment, with two zones with minimum values (35.7~\si{\micro\Molar}) where the hypoxic cells are located. As we see in the map, the temperature levels are fairly constant in the studied region, with the exception of the corners that are very near to the entry of the vessels and are thus very strongly influenced by the incoming fresh blood.
\par

\begin{figure}[!t]
    \centering
    \includegraphics[width=1 \textwidth]{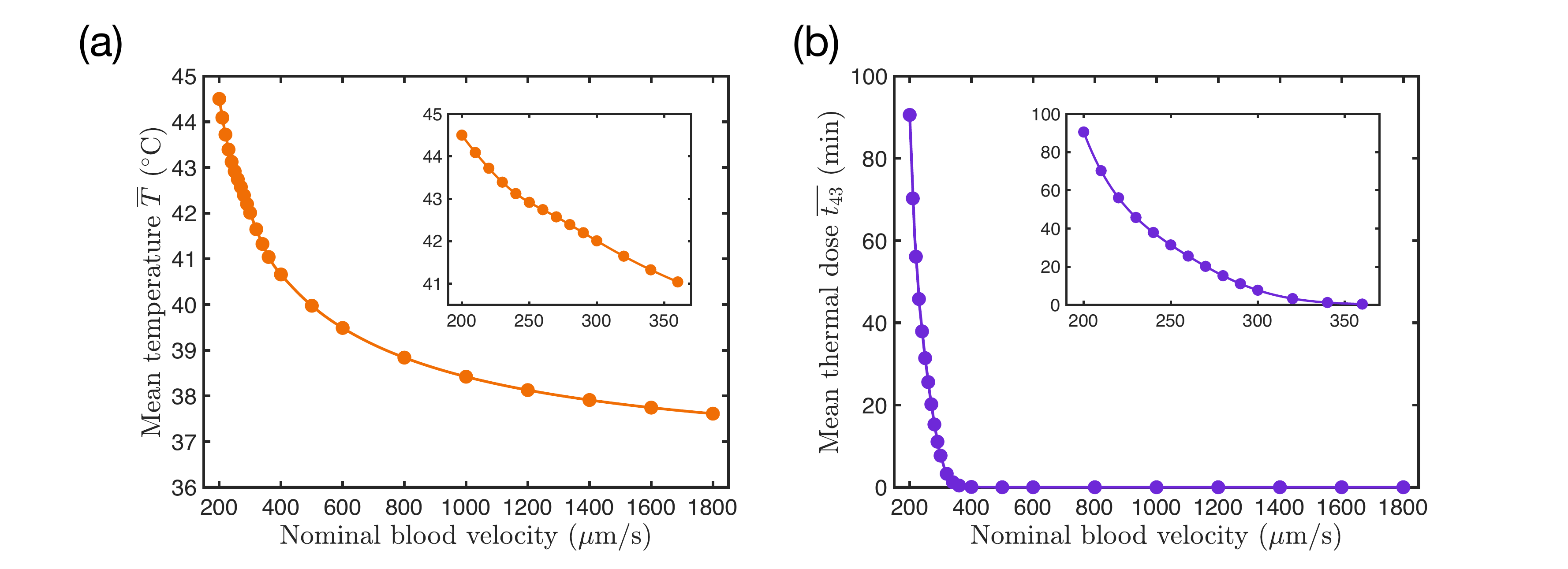}
    \caption{Results from the final instant of a 30-minute hyperthermia treatment performed over our \textit{in silico} model with a power of $P=40$~\si{\watt\per\liter}. Several simulations are carried out for a range of values of the nominal velocity in the blood vessels. The figures show the results for each of these values. \textbf{(a)} Mean temperature $\overline{T}(t_{\mathrm{end}})$ in the tissue at the end of the treatment. \textbf{(b)} Mean accumulated thermal dose at the end of the treatment $\overline{t_{43}}(t_{\mathrm{end}})$.  Insets in both panels provide a zoomed view of the initial time frames.} 
    \label{Fig:fig10}
\end{figure}

Therefore, the temperatures reached during the treatment are highly dependent of the blood flow that supplies the tissue, which in this work is synthesised by the blood velocity. We then wondered what levels of mean temperature in the tissue would appear for other different values of blood velocity and to research that we simulated the evolution of all the configurations that we had obtained previously from the free evolution for 245 days (\cref{SubSec:Results_FREE}) under a treatment of hyperthermia lasting for 30 minutes with a power $P=40$~\si{\watt\per\liter}. At the final time of the treatment, $t_{\mathrm{end}}=30$~min, we measured the mean temperature in the tissue ($\overline{T}(t_{\mathrm{end}})$). In \cref{Fig:fig10}(a) we represent these data as a function of the different nominal blood velocities characteristic of each simulation. We show that the highest temperatures during the treatment are obtained for low blood velocities, reaching a value of 44.5 \si{\degreeCelsius} for a nominal blood velocity of $v_{01}=v_{02}=200$~\si{\micro\meter\per\second}. Increasing values of velocity steeply decreased the attained temperature values, which are under 39.5 \si{\degreeCelsius} for velocities equal or higher than 600 \si{\micro\meter\per\second}. For the highest nominal values of velocity considered here, that is $v_{01}=v_{02}=1600$ and $v_{01}=v_{02}=1800$ \si{\micro\meter\per\second}, the mean temperature does not exceeds 37.8~\si{\degreeCelsius}. Consequently, the level of blood flow supply in the tissue, which is related to the local vasculature functionality, is a crucial factor for the attainable temperatures during thermal treatment. Beyond this, we plot the final mean thermal dose $\overline{t_{43}}(t_{\mathrm{end}})$ that the tissue has accumulated at the end of the treatment (obtained by an average analogous to \cref{Eq:s_mean}) against the corresponding nominal blood velocity in the vessels that was used in each simulation (\cref{Fig:fig10}(b)). For a thermal treatment of 30 minutes, thermal doses greater than 50 equivalent minutes can be found under low blood flow conditions. The accumulated dose decreases severely with increases in blood velocity and it becomes nearly zero for velocities higher than 400~\si{\micro\meter\per\second}. Therefore, on the basis of the assumptions of this model, only under severely impaired vascular conditions would hyperthermia get effective levels of cell death in a perfused tissue.
\par

\subsubsection{Effect of thermal dose on cell populations}
\begin{figure}[!b]
    \centering
    \includegraphics[width=1 \textwidth]{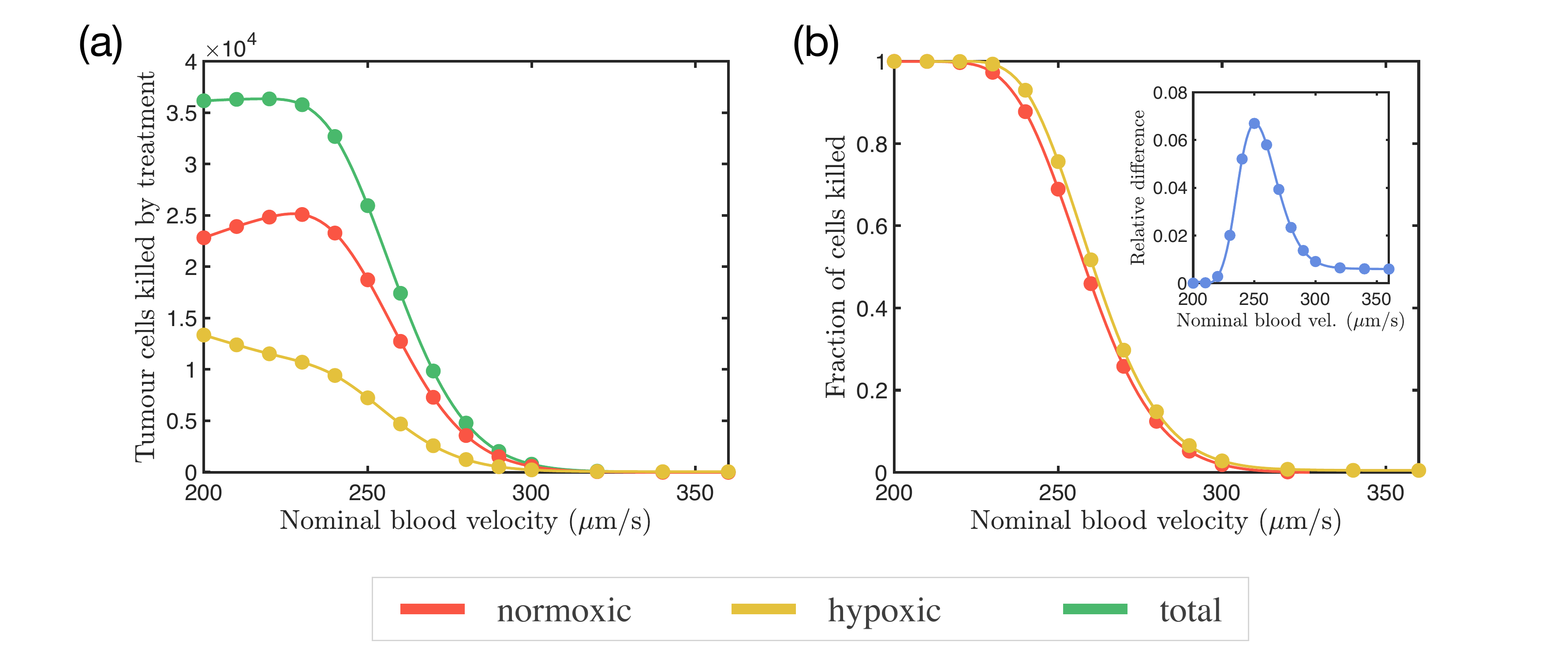}
    \caption{Exploration of the distribution of cell death in simulations of 30-minute hyperthermia treatment with a power of $P=40$~\si{\watt\per\liter} for different values of the nominal blood velocity ($v_{01}$, $v_{02}$) in the vessels irrigating the tissue. \textbf{(a)} Number of tumour cells killed in the normoxic and hypoxic populations and sum of both in different simulations varying on blood flow velocity. \textbf{(b)} Fraction of tumour cells killed by the treatment in the normoxic population and in the hypoxic population with respect to their number before the start. The inset illustrates the specific damage on hypoxic cells by showing the difference between the relative death in them and that on the normoxic cells.} 
    \label{Fig:fig11}
\end{figure}

We studied next what the reflection of this thermal dose was on cell populations. From the carried out simulations we took cases where there was a relevant accumulated thermal dose after the treatment, that is, samples where the blood velocity is under $v_{01}=v_{02}=360$ \si{\micro\meter\per\second}. From them we calculated the number of cells that got killed by the treatment in each population, i.e. the difference between the number of cells between the final point of the treatment and the beginning. In \cref{Fig:fig11}(a) the number of dead cells for each population are plotted against the value of the nominal velocity in the vessels. As the value of blood velocity decreases, the total number of cells eliminated by the treatment (green curve) gets higher. At the minimum studied value of blood velocity ($v_{01}=v_{02}=200$~\si{\micro\meter\per\second}), the number of cells killed by the treatment is the total number of initial tumour cells in the tissue, since for this lowest value of blood flow the cooling is not enough to compensate the external power and the accumulated thermal dose ($\overline{t_{43}}$=90.6 min) is enough to eliminate the population. The \cref{Fig:fig11}(a) also shows the number of normoxic (red) and hypoxic (yellow) tumour cells that got affected by the treatment: as the velocity decreases, the number of hypoxic cells present in the tissue before the beginning of the treatment gets higher due to the lower levels of oxygen. Since for those regimes the cell death is also higher, the favourable window of opportunity of hyperthermia treatment will be associated to the death of hypoxic tumour cells, which are also more resistant to radiotherapy---even though these do not make up the majority of dead cells during our simulations. This strengthens the idea of hyperthermia treatment as an adjuvant to radiotherapy. To get further insight on the effect of thermal induced cell death on this system, we calculated the proportion of tumour cells that died after the application of the treatment as a proportion of those present on their same population just before the beginning of the treatment (\cref{Fig:fig11}(b)). The span where cell death occurs is restricted to a small range of the velocities below $v_{01}=v_{02}=360$ \si{\micro\meter\per\second}. Again, for very low velocities, the cell death is complete, and for velocities between 230 \si{\micro\meter\per\second} and 320 \si{\micro\meter\per\second} there is a variation in the level of cell death that appears. In this range, the relative levels of cell death appearing in the hypoxic population is always higher than in the normoxic population, being the difference represented in the inset of \cref{Fig:fig11}(b). 
\par

\begin{figure}[!b]
    \centering
    \includegraphics[width=1 \textwidth]{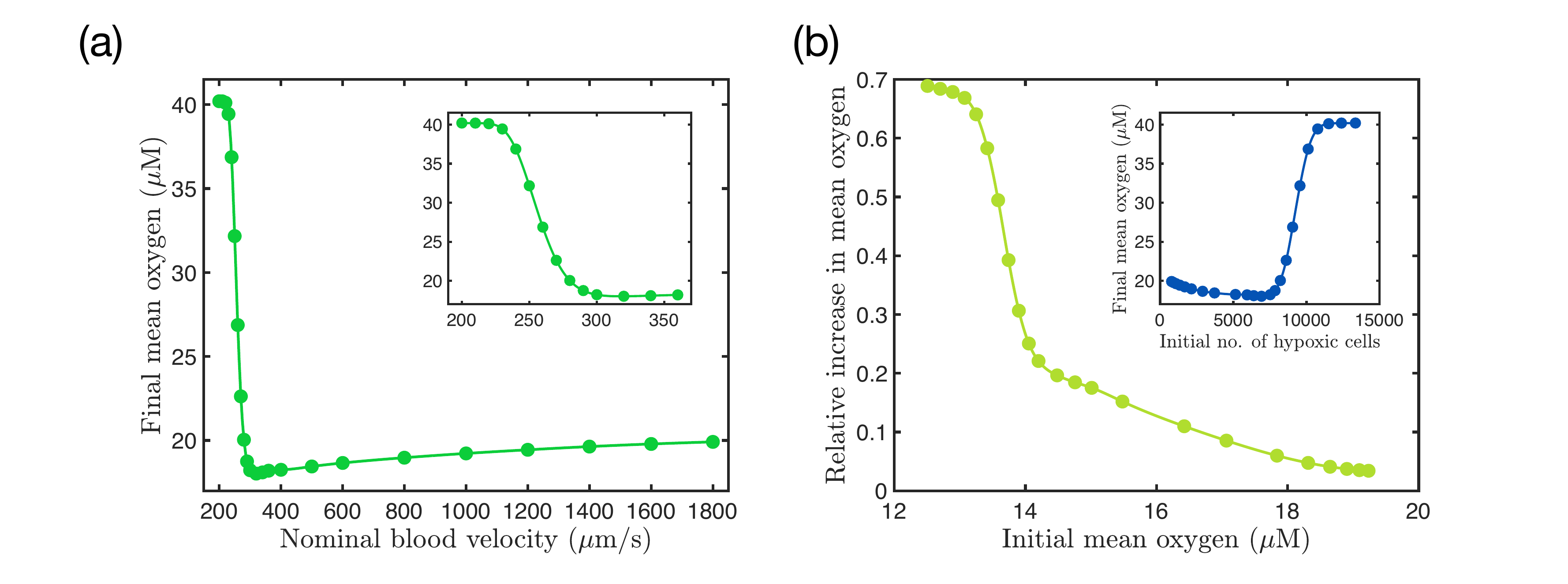}
    \caption{Reoxygenation in the tissue as a consequence of hyperthermia treatment according to our model's results. \textbf{(a)} Mean oxygen concentration in the tissue at the end of the thermal treatment ($\overline{T}(t_{\mathrm{end}})$) for the different simulation carried out with different levels or blood flow represented by different nominal blood velocities. The inset zooms in the range of low velocities where the rises are more pronounced. \textbf{(b)} Relative increase in the level of mean oxygen in the tissue between the end and the beginning of the treatment as a function of the initial mean oxygen in the tissue. The cases that were under a higher level of hypoxia prior to the treatment are those that got higher reoxygenation according to our model. Inset: mean oxygen concentration at the end of the treatment as a function of the hypoxic tumour cells present at the beginning of the treatment. In our model's study cases, a higher population of hypoxic cells before the treatment is indicative of a better reoxygenation due to hyperthermia.} 
    \label{Fig:fig12}
\end{figure}

\subsubsection{Thermal effect on oxygenation}
We have already shown the values of mean oxygen concentration that appear in the free evolution of the cancer population as it was reflected in \cref{Fig:fig6}. We are now interested in the effect that the therapy has on these levels at the end of the treatment and the possible improvements due to hyperthermia treatment. In \cref{Fig:fig12}(a) we have represented the mean oxygen concentrations from our simulations for varying velocities of blood in the vessels. It is apparent that the oxygen levels are in every case higher than the characteristic one of the free tumour evolution previous to therapy administration (\cref{Fig:fig6}). For a value of the nominal velocity of $v_{01}=v_{02}=320$ \si{\micro\meter\per\second} we find the minimum improvement in oxygenation. Velocities higher than that value produced an enhancement in oxygenation coming from the higher blood flow provided by the thermal dilation of the blood vessels. This effect increases with the original level of blood velocity and reaches the maximum at the higher considered value of blood velocity ($v_{01}=v_{02}=1800$ \si{\micro\meter\per\second}). Nevertheless, the improvements in tissue oxygenation that appear on the regime of higher velocities are modest compared to what happens to low perfused cases. When the velocity (see inset of \cref{Fig:fig12}) is smaller than $v_{01}=v_{02}=300$ \si{\micro\meter\per\second}, there appears a strong elevation in the mean oxygen level, which gets higher as the nominal velocity of blood---surrogate for blood flow---decreases. This improvement comes, in this case, from the reduction in oxygen consumers that appears as a consequence of cell death due to the thermal treatment, which is higher as blood flow decreases. We wished to know what the relative increase in mean oxygen concentration depending on the initial levels of oxygen was. In \cref{Fig:fig12}(b) we represent this information from the performed simulations, showing that the areas that were initially in the worst oxygenation scenarios are the most benefited from reoxygenation after hyperthermia treatment. Moreover, these areas are more densely populated by hypoxic tumour cells. In the inset of \cref{Fig:fig12}(b) we represent the final level of oxygen concentration after hyperthermia application as a function of the total number of hypoxic tumour cells existing previous to the application of the treatment. We found that those areas where there was a higher number of hypoxic tumour cells previous to the treatment were the ones that show higher levels of oxygen concentration when the treatment finishes. Therefore, these areas that are initially resistant to radiation therapy will be the main ones benefited from hyperthermia followed by subsequent radiotherapy.
\par

\section{Discussion}

Recent observations in both {\em in vivo} murine models and large cohorts of cancer patients of different histologies have found superlinear scaling laws relating proliferation and tumour size~\cite{perez2020universal}. The paradigm shift that this finding entails requires further analysis from a modelling perspective in order to shed light on the underlying mechanisms and the implications for the patient's progression and therapy response. At least three important attributes of cancer biology are in play to execute the changes that demand the acquisition of increasing energy needs. First of all, the presence of interacting populations with different phenotypes~\cite{perez2020universal}; secondly, profound changes in metabolic pathways, for instance, the appearance of the Warburg effect, by which tumour cells deviate glucose oxidation to other biochemical routes that prioritise cell proliferation at the expense of an efficient generation of ATP, even when other more energy-efficient strategies are available \cite{martinez2021cancer}; lastly, the involvement of the vasculature, whose transport properties––being key to the allocation of resources––explain other metabolic scaling laws present in living beings \cite{banavar2002supply,banavar2010general}. In this study we take into account simplified hallmarks of all these features as a way to gain insight in the complexity arising from them. The very same aspects ruling the metabolic uptake are key to a successful outcome from hyperthermia treatment of cancer. In this case, blood flow is the main player affecting tumour temperature, and the anomalous properties characterising the tumour vasculature determine the effect achieved by the therapy. Out of the multiple effects ascribed to non-ablative cancer thermal therapy, the centre of attention is put on its influence on oxygen concentration, and particularly the capability to impact radio-resistant hypoxic populations of cells. This implies phenotype-specific results that carry consequences to the multi-population tumour ecology and their evolutionary dynamics. Moreover, the alterations to the microenvironment are strongly non-linearly dependent on cell population changes, so the metabolic routes can also get very deregulated by the treatment~\cite{moon2010nadph}.
\par  

The biological and clinical implications of these aspects, both in the free cancer progression and under hyperthermia treatment, have being explored by mathematical models \cite{jimenez2021mesoscopic,bosque2021interplay}. However, some key aspects of the environmental conditions of the tumour, such as the presence of variable levels of oxygen or the presence of locally heterogeneous vasculature, have been neglected in those and other works. Here we open an avenue of research by using a first approach that investigates the microscopic conditions of populations evolving around tumour cords affected by different conditions of blood supply. As it has been shown, the diverse vascular conditions lead to radically different evolution of the various subpopulations, as well as a distinct response upon treatment. The results highlight the relevance of the local biophysical conditions to the outcome of the therapy. Due to the spatial heterogeneity characteristic of human cancers~\cite{jimenez2021evolutionary}, the particular state in each place of the tumour will be different from that encountered in other parts of the same tumour. However, the local conditions are not completely independent of the aggregate, and different configuration or patterns that emerge in the spatial distribution can be related to differences in the patients' outcome. Specifically, previous research on the distribution of fluorodeoxyglucose ($^{18}$F) uptake in breast cancer and non-small cell lung cancer has found that the location of the voxel of maximum uptake at the time of diagnosis is related to the overall survival of the patient \cite{jimenez2021evolutionary}. These patterns arise at a mesoscale emerging from the interactions existing at a microscopic level, therefore the present work is a first step to understand the lower scale, i.e. the sub-voxel scale. Future research should merge the local effects depending on individual vessels that are reported here with a wider scope that comprises an integrative representation of the overall tumour along with its vascular system. A modelling scenario describing such a system will allow for the \textit{in silico} testing of traditional therapies like radiotherapy and hyperthermia, but also novel approaches such as the internalisation of nanoparticles~\cite{Cortes2021} and immunotherapies~\cite{Moy2017}.
\par

Out of the many interrelating elements that influence cancer biophysical development, the role of the vasculature is paramount, both for the tumour natural development and for its treatment. In particular, its configuration and function is the most important element influencing the levels of oxygenation and hypoxia \cite{martin2019normalizing}, and also the temperature in the tissue \cite{rossmann2014review}. Therefore, oxygenation and temperature in a tumour are both interconnected by the important influence of the vascular tree that irrigates it. Even though much emphasis has been put on the effect of hyperthermia treatment on tumour oxygenation \cite{oei2020molecular} and also the relevance of blood flow to the former, reflecting the duality between both aspects, not many studies have stressed the role of blood vessels as an unifying thread for both phenomena---oxygenation and temperature. The results shown here demonstrate that indeed these two are two sides of the same coin, namely tumour vasculature; as a consequence, the results derived from experimental settings cannot be isolated from the specific vascular characteristics of the experimental model. In order to model different vascular settings, in this work we used blood velocity in the vessels as a proxy for blood flow and vessel functionality. Other options reflecting a variation in blood flow would have been equivalent and, in this case, we chose the simplest way to capture these variations without having to perform intricate analyses of the specific fluid mechanics in the vessels.
\par

The results of our model show a rise in tissue temperature as tumour cells progress and the metabolic requirements get bigger. As stated before, this is coupled to a decrease in oxygen levels (see~\cref{Fig:fig7}). In ecology, the key role of temperature in organism development and metabolism has long been known \cite{gillooly2001effects}, being the energy usage higher with higher temperature according to an Arrhenius relationship, what affects all the biological processes, e.g. developmental time \cite{gillooly2002effects}. Moreover, since oxygen is a requirement for aerobic organisms, temperature-driven variations in metabolism are linked to differences in oxygen consumption and, if the availability of oxygen is not enough to satisfy the metabolic demands, the hypoxic condition is induced; one of the implications of this is that the pair oxygen-temperature define the places where organisms can and cannot live \cite{deutsch2020metabolic}. The temperature variations that may happen within the tumour tissue, together with oxygen heterogeneity, might also have some relevance in cancer ecology, similarly to what occurs with pH levels \cite{alfarouk2011tumor,estrella2013acidity}, however this is an aspect that for the moment has remained unexplored. A possible reason is that, contrary to what is observed in organisms, temperature variations are not high and therefore the implications to evolutionary dynamics might be small. During thermal treatment the temperature of the tissue can go up by several degrees, thus inducing important differences with the natural evolution, even in the range of mild hyperthermia (below 43 \si{\degreeCelsius}). Strikingly, some effects of oxygen deprivation such as the induction of HIF-1 and the switch to a glycolytic metabolism are also produced by thermal therapy \cite{moon2010nadph}, highlighting this way the idea that an interdependence oxygen-temperature, similar to that of marine organisms, might operate in tumours during thermal therapy. Nevertheless, the reactions of catabolism gain importance as the 40~\si{\degreeCelsius} are surpassed, what modifies the relations usually employed to model that interdependence. Further mathematical modelling of the oxygen-temperature pair in cancer on the grounds of metabolic pathways might show significant results.
\par

Since very early in the history of hyperthermia treatment of cancer it was suggested that hypoxic cells (more difficult to kill by radiotherapy) are more sensitive to the action of heat~\cite{gerweck1974killing}. In this work we were interested in the differences that may appear between the hypoxic and normoxic populations sensitivity to heat in a physiological setting. For that we avoid explicitly modelling preferential sensitivity of any of the populations. Our results, however, show that hypoxic cells are indeed more prone to thermal death due to the fact that they tend to appear in places characterised by a low perfusion or located far from the blood vessels. Therefore, they tend to be exposed to relatively higher temperatures. This preferential death of hypoxic cells is an interesting feature of thermal therapy that suggests that it could be coupled with a subsequent application of radiation therapy. The former would eliminate the resistant hypoxic cells and elevate the levels of oxygen improving the outcome after administration of the latter. Moreover, since the reoxygenation arising in our model comes from the death of part of the tumour population, this effect would be long lasting and could widen the time window for the application of subsequent radiotherapy. There is current discussion as to whether radiotherapy has to be applied immediately after application of heat \cite{kroesen2019effect,crezee2019impact}. Our model suggests that there might be more than one route to heat-induced reoxygenation and, when cell death becomes relevant, radiotherapy may not need be applied immediately after. Nonetheless, other mechanisms such as vasodilation might only be present during the time that the temperature remains elevated~\cite{bosque2021interplay}. In the case that oxygenation after hyperthermia is due to metabolic changes in the cells~\cite{moon2010nadph}, the characteristic time to return to an oxidative metabolism should be evaluated. On the other hand, our model demonstrates that application of heat alone would not be sufficient for its success as a stand-alone therapy; it would partially affect only parts of the tumour environment, but those regions having an adequate perfusion are not expected to experience such thermally-induced cell death. It is interesting to note that the standard application of hyperthermia treatment lasts for about 60 minutes as it tries to maximise sensitisation to other combined therapies while avoiding risk to the healthy tissue~\cite{dobvsivcek2019quality,trefna2017quality}. This is based on the hypothesis that a longer time will optimise the absorption of the thermal dose by the tissue, therefore improving the outcome. In our model we have simulated the outcome during only 30 minutes of therapy since a longer duration had no impact on the results. We found that only a part of the cases simulated gave rise to a notable cell death. These cases, corresponding to low perfusion regimes, were rapidly affected by cell death and after a few minutes showed an important decrease on the cell population. In contrast, those cases where the blood vessels remained functional did not show an important cell death and therefore did not benefit from the therapy. Therefore, cell death and the time scales involved strongly depend on the specific vascular scenario used to model thermal therapy. It would be interesting to perform further research on more elaborate models of cell death in order to identify new therapeutic protocols better adapted to the vascular status of the patient. 
\par

\section{Conclusions}
To sum up, we have put forward a transport-based mathematical model to analyse the simultaneous role of temperature and oxygen in the progression of cancer cells at the microscale between two blood vessels, as well as their response to hyperthermia treatment. While the spatial scales were restricted to sub-voxel sizes, our framework was capable of capturing different physiological scenarios depending on the functionality of the local vasculature. We solved the system of partial differential equations numerically by means of the method of lines for a range of variable blood velocities, which we used as a proxy for vasculature performance. We found that temperature and oxygenation are highly influenced by the local status of the vasculature, encompassing both biophysical variables. According to our model, when hyperthermia treatment is applied, only regions with a prominently disrupted vasculature show a relevant level of cell death. These regions are more often populated by hypoxic cells and, as a consequence, these get especially affected by the treatment. Additionally, the reduction of consumption by cells gives rise to an effect of reoxygenation that is more pronounced in areas that were previously under a higher level of hypoxia. Our model also predicted the characteristic time scales during which all these processes occur. The framework presented herewith could be of use in understanding potential thermal feedback loops where accelerated tumour metabolism, as the one observed in~\cite{perez2020universal}, would induce a local heating capable of overcoming the damaged tumour vasculature' ability to pump out heat and thus elevate the tumour temperature. By this rationale, increasingly higher temperatures could lead to faster enzymatic reactions and thus to further temperature elevations.

\section*{Acknowledgement}
The authors thank Víctor M. Pérez-García and Rogelio Ortigosa for discussion. J.J.B. acknowledges a grant with reference 2018-CPUCLM-7798 funded by the University of Castilla-La Mancha with participation of the European Social Fund. G.F.C. is supported by the Spanish Ministerio de Ciencia e Innovación, MCIN/AEI/10.13039/501100011033 (grant PID2019-110895RB-I00) and by Junta de Comunidades de Castilla-La Mancha (SBPLY/19/180501/000211). M.C.N. is supported by the Ministerio de Ciencia e Innovación (grant PID2019-109652GB-I00).


\appendix
\setcounter{figure}{0} 
\setcounter{table}{0} 

\section{Phenotypic switch}
\label{Sec:App_PhenotypicSwitch}
\begin{figure}[ht]
    \centering
    \includegraphics[width=1 \textwidth]{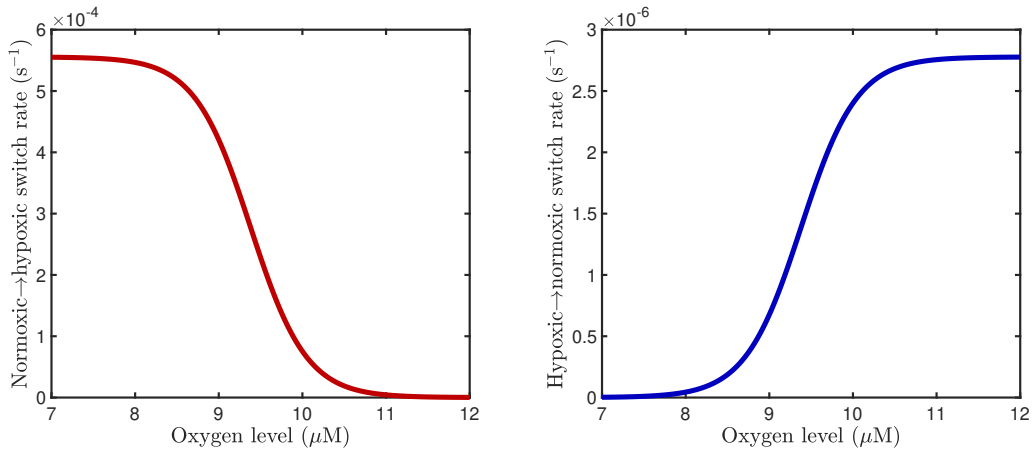}
    \caption{Illustration of the functions of the oxygen concentration $\sigma_{nh}(s)$ and $\sigma_{hn}(s)$ that model the switch rates between the normoxic phenotype to the hypoxic phenotype and vice versa.}
    \label{Fig:figA1}
\end{figure}

The tumour populations cells are subjected to a phenotypic switch between normoxic and hypoxic and vice versa, which depends on the level of oxygen $s$ that they are subjected to. The rates of switching between the two populations are modelled by means of a hyperbolic tangent with a sigmoidal shape where the parameters $\tau_{nh}$ and $\tau_{hn}$ are the characteristic times of change in each of the directions~\cite{Alicia2012}. The threshold limit that divides the regions of oxygen concentration where the rate halves is $s_{S}$, and the width of the window where the transition occurs is controlled by $\Delta s$. With these parameters, the terms modelling the switching read as
\begin{align}
	\sigma_{nh}(s) &= \frac{1}{2 \tau_{nh}} \left( 1 - \text{tanh}\left( \frac{s - s_{S}}{\Delta s} \right)   \right), &  \sigma_{hn}(s) &= \frac{1}{2 \tau_{hn}} \left( 1 + \text{tanh}\left( \frac{s - s_{S}}{\Delta s} \right) \right).  
\end{align}
\noindent Both functions are illustrated in \cref{Fig:figA1}, where each of the parameters take the values used for the simulations shown in this work (see \cref{Tab:Tabla}). The term $\sigma_{nh}(s)$, which quantifies the rate of transition from the normoxic phenotype to the hypoxic one, is also used to model the death of healthy cells due to the lack of oxygen, since these ones are much less resistant to harsh environments than their tumour counterparts due to the lack of flexibility of their metabolic pathways.

\section{Thermal dose and thermal cell death}
\label{Sec:App_ThermalDeath}

\setcounter{figure}{0} 
\setcounter{table}{0} 

The thermal dose received by cells under hyperthermia treatment is usually quantified by the ``equivalent time at 43\si{\degreeCelsius}'' that Sapareto and Dewey introduced in their seminal paper \cite{sapareto1984thermal}. This dose accumulates the time intervals $t_i$ that the cells expend under different temperatures $T_i$, and considers a weighting function to the temperature exposure as follows

\begin{equation}
\label{Eq:t43Sapareto}
	t_{43} = \sum_i t_i \cdot R^{43-T_i}.
\end{equation}

The weighting term $R$ has been experimentally studied and slightly different values have been employed for it. Here we follow the usual convention where $R$ is taken as 

\begin{equation}
\label{Eq:R_classic}
	R = 
	\begin{cases}
		0, & T \leq 40 \si{\degreeCelsius}, \\
    		0.25, & 40 \si{\degreeCelsius} < T \leq 43 \si{\degreeCelsius}, \\
    		0.5, & T  > 43 \si{\degreeCelsius}. \\
  	\end{cases}
\end{equation}

In order to use an equivalent expression to \cref{Eq:t43Sapareto} that allows for the calculation of the dose accumulated under temperatures that change in a continuous fashion with time (as opposed to discrete time intervals), we employ the following extension that gives the accumulated dose $t_{43}(t)$ at time $t$ of a tissue subjected to a temperature $T(t)$ (a varying function of time) between instants $t_\textrm{init}$ and $t$

\begin{equation}
\label{Eq_t43}
    t_{43}(t) = \int_{t_\textrm{init}}^t R^{43-T(\tau)} \D\tau. 
\end{equation}

Additionally, we approximate the discrete levels of the term $R$ from \cref{Eq:R_classic} with the following continuous function that assumes a progressive variation of the effect with temperature changes and avoids potential numerical instabilities

\begin{equation} 
	R = \frac{1}{4} + \frac{1}{4} \text{tanh} \left( \frac{T-42}{1.5} \right).
\end{equation}

\begin{figure}[H]
    \centering
    \includegraphics[width=1 \textwidth]{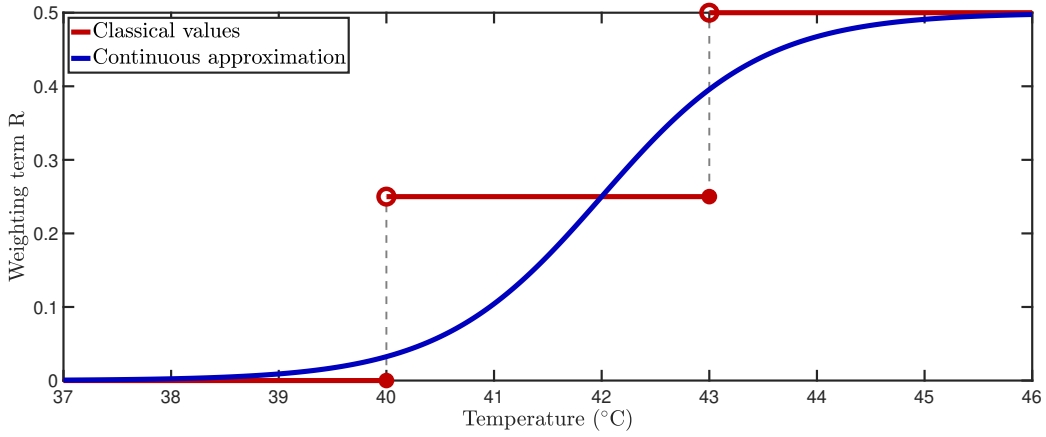}
    \caption{Functional form of the continuous approximation (blue) to the weighting term $R$ versus the discrete values that are typically used (red).}
    \label{Fig:figB1}
\end{figure}

Both functional forms for $R$ are depicted in \cref{Fig:figB1}.

For the thermal death of cells exposed to high temperatures we base our model on the approach conceived by Brüningk \textit{et al.} \cite{bruningk2018combining}, where the authors model the survival of cell populations under thermal insult with a linear-quadratic function of the thermal dose $t_{43}$ similar to that used in radiotherapy \cite{mcmahon2019linear}. After the application of the treatment, the survival fraction $S_{HT}$, that is, the number of remaining cells alive $P$ divided by the initial number $P_0$, is given by the expression
\begin{equation}
\label{Eq:Sht_Th}
    S_{HT} = \frac{P}{P_0} = \exp{\left( -(\alpha_{0,HT} - \alpha_{R,HT})t_{43} - \beta_{HT} t_{43}^2 \right)},
\end{equation}
where the parameters $\alpha_{0,HT}$ and $\alpha_{0R,HT}$ are equal since no radiotherapy is considered, and $\beta_{HT}=\frac{\alpha_0^2}{2}$ \si{\minute^{-2}} according to \cite{bruningk2018combining}. The potential influence of the cell cycle that is taken into account by the authors in their discrete model has been neglected here. Moreover, we have assumed that all damage is reparable, which entails a small error only in the high doses range. We have, then, that the only remaining free parameter after these assumptions, $\alpha_0$, has the following form as a function of temperature
\begin{equation}
\label{Eq:alpha0}
    \alpha_0 = \frac{1}{20} \exp\left( \frac{2}{3} \, \si{\degreeCelsius^{-1}} \; (T-43 \, \si{\degreeCelsius}) \right) \; \si{\minute^{-1}}.
\end{equation}
With the assumptions made, we have that, combining \cref{Eq:Sht_Th} and \cref{Eq:alpha0}, the survival function under hyperthermia treatment is given by
\begin{equation}
\label{Eq:Sht_Bru}
    S_{HT}=\exp \left( -\frac{1}{800} \exp\left(\frac{4}{3}(T-43)\right) t_{43}^2 \right),
\end{equation}
where $t_{43}$ in expressed in minutes. We then investigate the correct way to accommodate the survival term expressed by \cref{Eq:Sht_Bru} into our continuous equations for the evolution of populations. For that purpose, we take a generic form of \cref{Eq:normoxic,Eq:hypoxic,Eq:healthy}, where we model the evolution of a generic population with cell density $P$ and ignore all proliferation and spatial effects, considering only the cell death affecting an initial population $P_0(x,z)$. This equation has the form
\begin{equation}
    \frac{\partial P}{\partial t} = -\sigma_t(T,t_{43};x,z,t)P(x,z,t),
\end{equation}
which, due to the lack of spatial effects, is analogous to a separable first order ordinary differential equation and can be easily solved, what leads to the expression
\begin{equation}
    \ln \left( \frac{P(x,z,t)}{P_0(x,z)} \right) = \int_0^t - \sigma_t(\tau) \D\tau.
\end{equation}
Here we can introduce the definition of the survival fraction from \cref{Eq:Sht_Th} which gives us its value at each time point $t$
\begin{equation}
    \ln \left( S_{HT}(t) \right) = \int_0^t - \sigma_t(\tau) \D\tau.
\end{equation}
Applying the fundamental theorem of calculus to the right hand side, we solve for the term $\sigma_t$ that yields
\begin{equation}
\label{Eq:sigma_t_The}
    \sigma_t(t)=-\frac{\D}{\D t} \ln(S_{HT}(t)).
\end{equation}
Therefore, introducing the already known empirical expression for the survival fraction after hyperthermia (given by \cref{Eq:Sht_Bru}) into \cref{Eq:sigma_t_The}, and neglecting the spurious death terms related strictly to temperature variations, we get
\begin{equation}
\label{Eq:sigmaT}
    \sigma_t(T,t_{43}) = \frac{1}{400} \exp \left( \frac{4}{3} (T-43) \right) \cdot t_{43} \cdot \frac{\D t_{43}}{\D t}.
\end{equation}
Since the accumulated dose $t_{43}$ in a continuous setting is expressed by the integral from \cref{Eq_t43} we can apply the fundamental theorem of calculus to get rid of the derivative in \cref{Eq:sigmaT}. In this way we arrive at the definitive form of the thermal cell death term
\begin{equation}
    \sigma_t(T,t_{43}) = \frac{1}{400} \exp \left( \frac{4}{3} (T-43) \right) \cdot t_{43} \cdot \frac{R^{43-T}}{60}.
\end{equation}

\section{Spatial discretisation and grid}
\label{Sec:App_Grid}

\setcounter{figure}{0} 
\setcounter{table}{0}

\begin{figure}[ht]
    \centering
    \includegraphics[width=0.8 \textwidth]{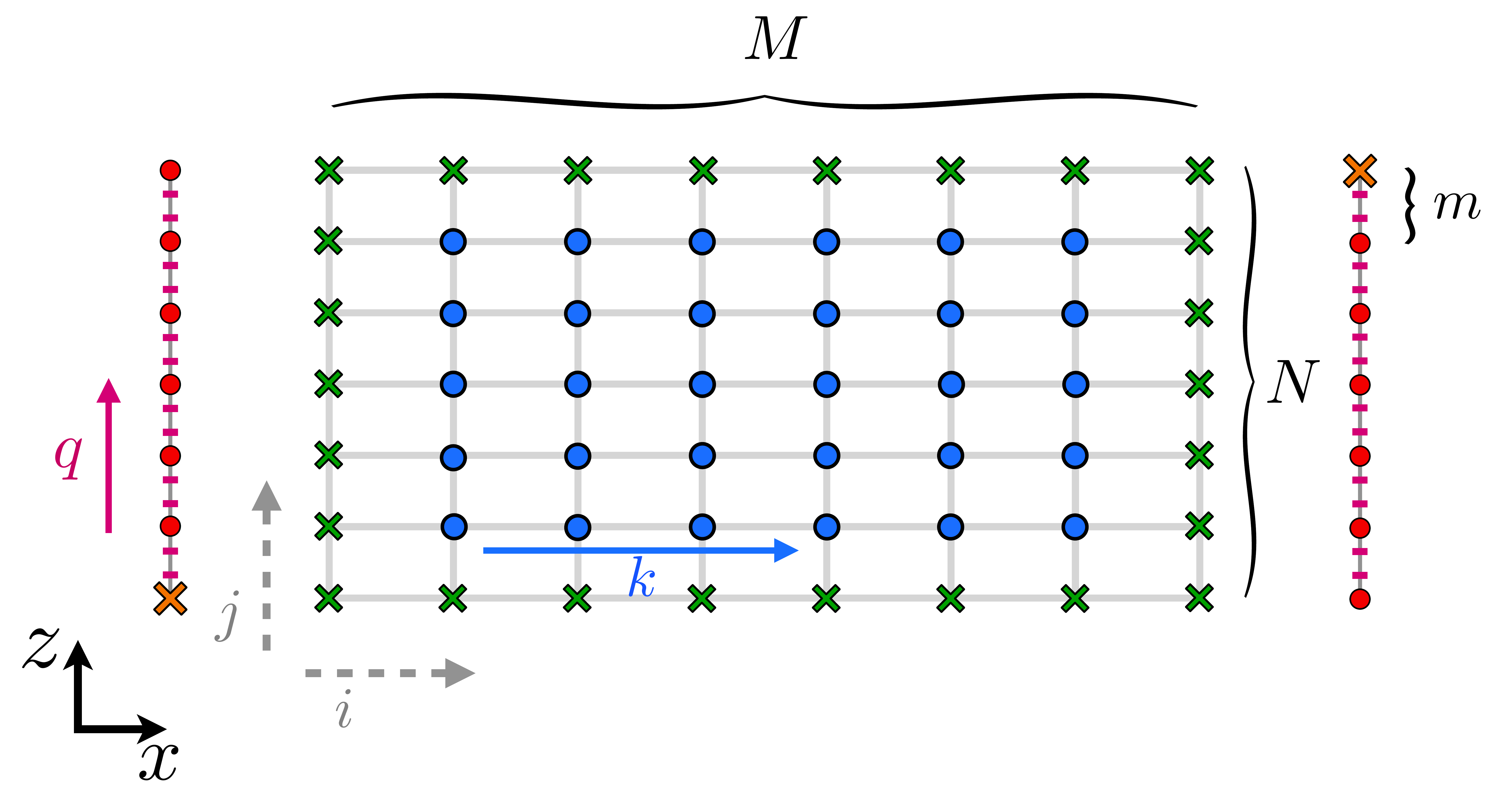}
    \caption{Sketch of the discretised grids used to handle the spatial relationships among the variables of the PDEs in \cref{Sec:MathModel}. The spatial domain for the tissue $[0,L_{x}]\times[0,L_{z}]$ is partitioned in $M$ and $N$ equispaced subintervals in the $x$ and $z$-directions, respectively. The interior points (blue dots) are associated to one ODE per unknown and the exterior points (green crosses) are calculated from the boundary conditions by means of algebraic relations. The interior points are indexed by an index $k$ following the direction of the arrow. The spatial domain for each vessel $[0,L_{z}]$ is divided in $N_V=N\times m$ partitions, with $m\in{\mathbb N}$, whose points are denoted by an index $q$ and marked with purple lines, or red dots in case that the point matches a point from the tissue. The points of the vessel where the values are imposed by boundary conditions are marked by orange crosses.}
    \label{Fig:figC1}
\end{figure}

In order to solve the system of PDEs developed in \cref{Sec:MathModel} we performed a discretisation of the spatial terms as explained in \cref{Sec:Calculation}. An schematic illustration of the discrete geometric grids that are used is shown by \cref{Fig:figC1}. The 2D tissue domain is partitioned in $M$ divisions in the $x$ direction and $N$ divisions in the $z$ direction. For each of the $N_i=(M-1)\times(N-1)$ interior points (blue dots) there is an ODE associated to the corresponding unknown. These points are numbered by an index $k$, as indicated in that figure, and each of the ODEs share the same numeration (see \crefrange{Eq:ODE1}{Eq:ODE7}). The spatial relationships enforced by the discretisation of the PDEs by a fourth order finite difference scheme relates the variables from different points yielding a coupled system. Additionally, the different boundary conditions imposed on each border point (green crosses) are functions of the interior points. In the case of the oxygen concentration and the temperature equations, the calculation of the points on the left and right boundary, which are in contact with the respective blood vessels, involves also matching the values from the blood vessel grids (red dots). Regarding the tissue grid as a $(M+1)\times(N+1)$ matrix with rows $i$ and columns $j$ facilitates the numerical implementation of the relations between the different points coming from boundary conditions. \Crefrange{Eq:P_i1}{Eq:P_M+1j} and \crefrange{Eq:U_1j}{Eq:U_M+1j} give these relations following this matrix notation. 
\par

Moreover, we model here the role of blood vessels that interchange oxygen and thermal energy with the tissue at its left and right boundaries, with the the blood flowing upwards for the left vessel and downwards for the right one. The evolution of energy and oxygen concentration in each section $z$ of the blood they carry is modelled by two 1D transport PDEs for each vessel. Each of the vessels has its own grid which is discretised in $N_V=N\times m$ equal partitions (therefore this grid is $m$ times denser than the grid of the tissue in the $z$ direction). Each of the discretised points is marked with a purple line, or a red dot when the specific point matches a point from the tissue. The discretisation of space maps each of the transport PDEs to $N_V$ coupled ODEs (one for each point of the grid with the exception of one which is given by the Dirichlet boundary conditions and is marked with an orange cross in the sketch). Each of the points, and therefore the ODEs, are numerated by an index $q$ following the direction of the $z$ axis. All the ODEs from the vessel are coupled to the points in the tissue boundary, but only some vessel points (marked with a red dot) match exactly with the points of the tissue boundary. In the other points, a linear interpolation of the two nearest points adjusted for the distance is used. On the other hand, the points of the vessels that match the point are the ones that are used for the calculation of the boundary values in the tissue. Since all the ODEs coming from the discretisation of the PDEs as well as the algebraic relations coming from the boundary conditions are coupled, the overall system has to be solved simultaneously. 
\par

\section{Differentiation matrices}
\label{Sec:App_DerMat}

\setcounter{figure}{0} 
\setcounter{table}{0} 

\noindent The terms $G_{kl}$ and $H_{kl}$ from \cref{Eq:ODE1,Eq:ODE2,Eq:ODE5,Eq:ODE6} are the components of two differentiation matrices $G$ and $H$ that perform a numerical second-order derivative along the $x$ and $z$ directions, respectively. Therefore, the summations applied on the arrays give the numerical Laplacian. Here, these operators are based on a finite difference scheme with fourth order of accuracy. The interior points of the grids use a central scheme, while the points adjacent to the borders use forward or backward schemes so only real points are used and there is no need to employ auxiliary points out of the grid. These two matrices are built in the following way

\begin{equation}
\widetilde{G} = \frac{1}{12 (\Delta x)^2}
\begin{pmatrix}
0 & 0 & 0 & 0 & 0 & 0 & 0 & \dots  & 0 \\
10 & -15 & -4 & 14 & -6 & 1 & 0 & \dots  & 0  \\
-1 & 16 & -30 & 16 & -1 & 0 & 0 & \dots  & 0 \\
0 & -1 & 16 & -30 & 16 & -1 & 0 & \dots  & 0 \\
   & \ddots & & & \ddots & & & \ddots \\
0 & \dots & 0 & -1 & 16 & -30 & 16 & -1 & 0 \\
0 & \dots & 0 & 0 & -1 & 16 & -30 & 16 & -1 \\
0 & \dots & 0 & 1 & -6 & 14 & -4 & -15 & 10 \\
0 & \dots & 0 & 0 & 0 & 0 & 0 & 0 & 0
\end{pmatrix}_{(M+1)\times(M+1)}
\end{equation}

\begin{equation}
\label{Eq:G}
    G = I_{N+1} \otimes \widetilde{G},
\end{equation} 

\begin{equation}
\widetilde{H} = \frac{1}{12 (\Delta z)^2}
\begin{pmatrix}
0 & 0 & 0 & 0 & 0 & 0 & 0 & \dots  & 0 \\
10 & -15 & -4 & 14 & -6 & 1 & 0 & \dots  & 0  \\
-1 & 16 & -30 & 16 & -1 & 0 & 0 & \dots  & 0 \\
0 & -1 & 16 & -30 & 16 & -1 & 0 & \dots  & 0 \\
   & \ddots & & & \ddots & & & \ddots \\
0 & \dots & 0 & -1 & 16 & -30 & 16 & -1 & 0 \\
0 & \dots & 0 & 0 & -1 & 16 & -30 & 16 & -1 \\
0 & \dots & 0 & 1 & -6 & 14 & -4 & -15 & 10 \\
0 & \dots & 0 & 0 & 0 & 0 & 0 & 0 & 0
\end{pmatrix}_{(N+1)\times(N+1)}
\end{equation}

\begin{equation}
\label{Eq:H}
    H = \widetilde{H} \otimes I_{M+1},
\end{equation} 

\noindent where $\Delta x=L_x/M$ and $\Delta z = L_z/N$ denote the spatial discretisation steps in the $x$ and $z$ directions, respectively, and $I_{M+1},\,I_{N+1}$ are square unitary matrices of the size indicated by the subindex. Here $\otimes$ represents the Kronecker product that expands the standard differentiation matrices to be ready to be applied over the column sub-array that contains the values of the given variable. No derivatives are actually calculated in those points belonging to the border since the solution at them comes from the boundary conditions applied to them, as we explain in \cref{Sec:Calculation}. However, the respective arrays of $k$ points are expanded to include also the values at the boundaries so that ODEs in \crefrange{Eq:ODE1}{Eq:ODE7} where these points intervene are calculated accurately. This explains the first and last rows of zeros in $\widetilde G$ and $\widetilde H$, which correspond to the calculation of the derivatives at the boundary points; their values do not explicitly enter into the set of ODEs, but are a necessary by-product of the boundary conditions. 
\par

The matrices $A$ and $B$ from \crefrange{Eq:ODE8}{Eq:ODE11}, whose components are $A_{qr}$ and $B_{qr}$, are the differentiation matrices for a first-order spatial derivative in the direction $z$ along the vessel path. To eliminate perturbations that may typically affect transport equations, we apply a five point biased upwind approximation in them, which enforces us to use two different derivatives, one ($A$) for the left vessel, where blood flows in the same direction of the axis, and another ($B$) for the right vessel, where blood flows opposite to the direction of the axis.

\begin{equation}
\label{Eq:A}
A = \frac{1}{12 \Delta z}
\begin{pmatrix}
0 & 0 & 0 & 0 & 0 & 0 & 0 & \dots  & 0\\
-3 & -10 & 18 & -6 & 1 & 0 & 0 & \dots  & 0  \\
1 & -8 & 0 & 8 & -1 & 0 & 0 & \dots  & 0 \\
-1 & 6 & -18 & 10 & 3 & 0 & 0 & \dots  & 0 \\
0 & -1 & 6 & -18 & 10 & 3 & 0 & \dots  & 0 \\
   & \ddots & & & \ddots & & & \ddots \\
0 & \dots & 0 & -1 & 6 & -18 & 10 & 3 & 0 \\
0 & \dots & 0 & 0 & -1 & 6 & -18 & 10 & 3 \\
0 & \dots & 0 & 0 & 3 & -16 & 36 & -48 & 25
\end{pmatrix}_{(N_V+1)\times(N_V+1)}
\end{equation}

\begin{equation}
\label{Eq:B}
B = \frac{1}{12 \Delta z}
\begin{pmatrix}
-25 & 48 & -36 & 16 & -3 & 0 & 0 &\dots  & 0\\
-3 & -10 & 18 & -6 & 1 & 0 & 0 &\dots  & 0  \\
0 & -3 & -10 & 18 & -6 & 1  & 0 &\dots & 0 \\
   & \ddots & & & \ddots & & & \ddots \\
0 & \dots & 0 & -3 & -10 & 18 & -6 & 1 & 0 \\
0 & \dots & 0 & 0& -3 & -10 & 18 & -6 & 1 \\
0 & \dots & 0 & 0 & 1 & -8 & 0 & 8 & -1 \\
0 & \dots & 0 & 0 & -1 & 6 & -18 & 10 & 3 \\
0 & \dots & 0 & 0 & 0 & 0 & 0 & 0 & 0
\end{pmatrix}_{(N_V+1)\times(N_V+1)}
\end{equation}

\FloatBarrier
\section{Tables of variables and parameters used in the simulations}
\label{Sec:App_Tabla}

\setcounter{figure}{0} 
\setcounter{table}{0} 

\begin{table}[H]
\centering
\caption{Variables and parameters of the mathematical model}
\resizebox{0.95\textwidth}{!}{

\begin{tabular}{lllll}
\noalign{\global\arrayrulewidth=1pt}
\hline
\textbf{Description} & \textbf{Symbol} & \textbf{Value} & \textbf{Unit} & \textbf{Reference} \\ 
\hline
Normoxic population cell density & $n$ & Variable & - &   \\ 
Hypoxic population cell density & $h$ & Variable & - &   \\ 
Healthy population cell density & $w$ & Variable & - &   \\ 
Necrotic population cell density & $c$ & Variable & - &   \\ 
Oxygen concentration in the tissue & $s$ & Variable & \si{\micro\Molar} &   \\ 
Tissue temperature & $T$ & Variable & \si{\degreeCelsius} &   \\ 
Accumulated thermal dose & $t_{43}$ & Variable & \si{\minute} &   \\ 
Blood temperature & $T_{b1},\;T_{b2}$ & Variable & \si{\degreeCelsius} &   \\
Oxygen concentration in blood & $s_{b1},\;s_{b2}$ & Variable & \si{\degreeCelsius} &   \\
\Xhline{0.5\arrayrulewidth}
Domain horizontal length & $L_x$ & 400 & \si{\micro\meter} & Model design \\
Domain vertical length & $L_z$ & 600 & \si{\micro\meter} & Model design \\
Vessel diameter & $d_1,\;d_2$ & 28 & \si{\micro\meter} & Model design \\
\Xhline{0.5\arrayrulewidth}
Carrying capacity & $K$ & \num{5e-4} & cell \si{\per\cubic\micro\meter} & Model design \\
Normoxic cells mobility & $D_n$ & \num{6.6e-4} & \si{\square\micro\meter\per\second} & \cite{wang2009prognostic} \\
Hypoxic cells mobility & $D_h$ & \num{6.6e-3} & \si{\square\micro\meter\per\second} & \cite{Alicia2012} \\
Normoxic cells proliferation rate & $\rho_n$ & \num{5.75e-7} & \si{\per\second} & \cite{ke2000relevance} \\
Hypoxic cells proliferation rate & $\rho_h$ & \num{3.35e-7} & \si{\per\second} & \cite{giese2003cost} \\
Healthy cells proliferation rate & $\rho_h$ & \num{1e-9} & \si{\per\second} & Estimated \\
Healthy cells loss rate due to tumour cells contact & $\lambda$ & \num{1.2e-6} & \si{\per\second} & Estimated \\
Occupancy of necrotic cells with respect to normal cells & $\xi$ & 0.5 & - & Estimated \\
Normoxic maximum initial density & $n_{00}$ & \num{5e-2} & - & Model design \\
Characteristic widths of initial normoxic distribution & $\sigma_{x},\;\sigma_{z}$ & 30 & \si{\micro\meter} & Model design \\
Healthy maximum initial density & $w_{00}$ & 0.4 & - & Model design \\
\Xhline{0.5\arrayrulewidth}
Oxygen diffusion coefficient & $D_s$ & 1500 & \si{\square\micro\meter\per\second} & \cite{dacsu2003theoretical} \\
Normoxic cells oxygen consumption rate & $\alpha_n$ & 4.65 & \si{\micro\Molar.s^{-1}} & Estimated \\
Hypoxic cells oxygen consumption rate & $\alpha_h$ & 0.93 & \si{\micro\Molar.s^{-1}} & Estimated \\
Healthy cells oxygen consumption rate & $\alpha_w$ & 4.65 & \si{\micro\Molar.s^{-1}} & Estimated \\
Michaelis-Menten saturation constant & $K_M$ & 3.35 & \si{\micro\Molar} & \cite{dacsu2003theoretical} \\
Vessel permeability for oxygen & $\gamma$ & 1000 & \si{\micro\meter\per\second} & \cite{owen2011mathematical} \\
Initial value for oxygen & $s_0$ & 20.1 & \si{\micro\Molar} & Model design \\
Concentration of oxygen in incoming blood & $s_{b0}$ & 40.2 & \si{\micro\Molar} & Model design \\
\Xhline{0.5\arrayrulewidth}
Tissue volumetric heat capacity & $\delta C$ & \num{3.86e-12} & \si{\joule\kelvin^{-1}\micro\meter^{-3}} & \cite{herman2016metabolism} \\
Blood heat capacity & $\delta_b C_b$ & \num{3.82e-12} & \si{\joule\kelvin^{-1}\micro\meter^{-3}} & \cite{herman2016metabolism} \\
Thermal conductivity & $\kappa$ & \num{5.1e-7} & \si{\watt\kelvin^{-1}\micro\meter^{-1}} & \cite{herman2016metabolism} \\
Normoxic cells metabolic heat production & $Q_n$ & \num{2.23e-13} & \si{\watt\micro\meter^{-3}} & Estimated \\
Hypoxic cells metabolic heat production & $Q_h$ & \num{3.47e-13} & \si{\watt\micro\meter^{-3}} & Estimated \\
Healthy cells metabolic heat production & $Q_w$ & \num{7.44e-14} & \si{\watt\micro\meter^{-3}} & Estimated \\
Heat transfer coefficient in blood-tissue interface & $\eta$ & \num{1.73e-8} & \si{\watt\kelvin^{-1}\micro\meter^{-2}} & Calculated \\
Vessel expansion coefficient & $\chi$ & \num{0.03} & \si{\degreeCelsius^{-1}} & Model design \\
Externally applied power & $P$ & 40 & \si{\watt\per\liter} & \cite{wust2002hyperthermia} \\
Tissue temperature initial value & $T_{0}$ & 36.6 & \si{\degreeCelsius} & Model design \\
Temperature of incoming blood & $T_{b0}$ & 36.5 & \si{\degreeCelsius} & Model design \\
\Xhline{0.5\arrayrulewidth}
Characteristic time for $n\rightarrow h$ phenotypic switch & $\tau_{nh}$ & 1800 & \si{\second} & \cite{jewell2001induction} \\
Characteristic time for $h\rightarrow n$ phenotypic switch & $\tau_{hn}$ & \num{3.6e5} & \si{\second} & Estimated \\
Characteristic time for $h\rightarrow c$ death & $\tau_{hc}$ & 1800 & \si{\second} & Estimated \\
Phenotypic switch oxygen threshold & $s_{S}$ & 9.38 & \si{\micro\Molar} & \cite{vaupel2004role} \\
Hypoxic cells oxygen death threshold & $s_{C}$ & 0.938 & \si{\micro\Molar} & Estimated \\
Sensitivity around transition threshold & $\Delta s$ & 0.67 & \si{\micro\Molar} & Estimated \\
\hline
\label{Tab:Tabla}
\end{tabular}
}
\end{table}

\section{Jacobian sparcity pattern}
\label{Sec:App_Jacobian}

\setcounter{figure}{0} 
\setcounter{table}{0}

\begin{figure}[ht]
    \centering
    \includegraphics[width=0.6 \textwidth]{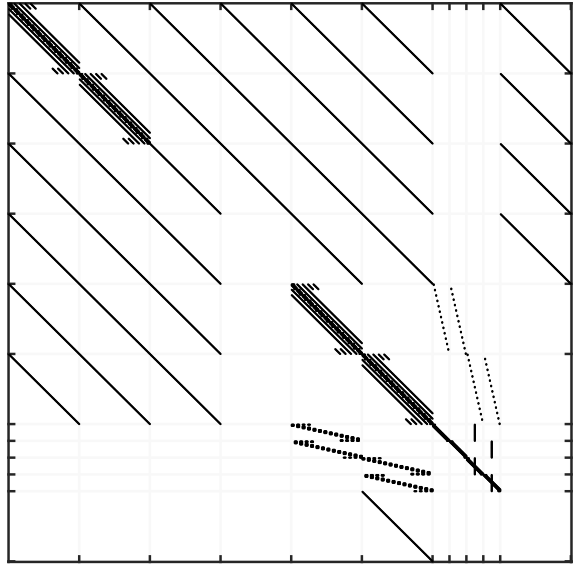}
    \caption{Visualisation of the sparcity matrix for the jacobian of the ODE system \crefrange{Eq:ODE1}{Eq:ODE7} and \crefrange{Eq:ODE8}{Eq:ODE11} that discretises our PDEs model. For illustration purposes, the matrix is generated for a simplified mesh with $M=10$, $N=14$ and $m=2$ so the patterns are apparent; however the mesh used for the simulations are much denser.}
    \label{Fig:figF1}
\end{figure}

In order to efficiently use a Runge-Kutta solver for the integration of large systems of ODEs (in the case of this work {\fontfamily{cmtt}\selectfont ode15s} from MATLAB) it is important to provide the Jacobian sparcity pattern of the system which is a matrix of 0s and 1s where the 1s indicate the positions where there might be nonzero elements of the Jacobian. For the system of $7N_{\textrm{int}}+4N_V$ ODEs comprising \crefrange{Eq:ODE1}{Eq:ODE7} and \crefrange{Eq:ODE8}{Eq:ODE11}, where $N_{\textrm{int}}$ is the number of internal points in the tissue $N_{\textrm{int}}=(M-1)\times(N-1)$, and $N_V$, the number of divisions in the vessel $N_V=N\times m$ (which is also the number of points that enter the calculation in the vessels, the first point being given by the boundary condition at the entry), we have the following general form of the Jacobian sparcity pattern, where the empty places indicate blocks of zeros:

\begin{equation}
J=
\NiceMatrixOptions{code-for-first-row = \lightgray,
                   code-for-last-col = \lightgray}
\begin{pNiceArray}{cccc|cc|cccc|c}[first-row,first-col,last-col,nullify-dots]
  &  n  &  h  &  w  &  c  &  s  &  T  & s_{b1} & s_{b2} & T_{b1} & T_{b2} & t_{43}\\
  &  K  &  I  &  I  &  I  &  I  &  I  &        &        &        &        &   I     &  n \\
  &  I  &  K  &  I  &  I  &  I  &  I  &        &        &        &        &   I     &  h \\
  &  I  &  I  &  I  &  I  &  I  &  I  &        &        &        &        &   I     &  w \\
  &  I  &  I  &  I  &     &  I  &  I  &        &        &        &        &   I     &  c \\
\hline      
  &  I  &  I  &  I  &     &  K  &     &   V    &   W    &        &        &         &  s \\  
  &  I  &  I  &  I  &     &     &  K  &        &        &   V    &   W    &         &  T \\ 
\hline
  &     &     &     &     &  E  &     &   K_v  &        &  L_1   &        &         & s_{b1} \\
  &     &     &     &     &  F  &     &        &   K_w  &        &  L_2   &         & s_{b2} \\
  &     &     &     &     &     &  E  &        &        &  K_V   &        &         & T_{b1} \\
  &     &     &     &     &     &  F  &        &        &        &   K_W  &         & T_{b2}\\
\hline
  &     &     &     &     &     &  I  &        &        &        &        &         & t_{43}
\end{pNiceArray}
\end{equation}

\noindent Each of the submatrices belonging to $J$ express the relationship of the elements of the array discretising the variable, indicated in the row, against the elements of the array lists in the column. For instance, the submatrix $I$ corresponds to a square identity matrix of size $N_\textrm{int}$. The submatrix $K$ comes from the relationships among points imposed by the discretised Laplacian and is simply the logical counterpart (0 or 1) of the sum $G+H$ from the matrices resulting from \cref{Eq:G,Eq:H}. Likewise, the submatrices $V$ and $W$ are the logical counterparts of the upwind differentiation matrices $A$ and $B$ used in the vessels and given by \cref{Eq:A,Eq:B}.
\par

The sparcity pattern of the Jacobian is graphically illustrated in \cref{Fig:figF1} for a simplified grid with $M=10$, $N=14$ and $m=2$.


\section{Geometrical interpretation of the problem}
\label{Sec:App_Geometrical}
Notice that even though we are interested in the 3D biological setting of cells interacting with two nearby blood vessels from which they obtain oxygen and nutrients and exchange energy---traditionally known as \textit{tumour cords}---we only consider a 2D slice of cells between these vessels for the sake of simplicity in our mathematical model. In view of the fact that in the real 3D biological configuration all the cells surrounding the vessels contribute to the consumption of oxygen and the transfer of heat, we must use a correction factor that links the results of our 2D model to the 3D biological situation. This factor enters into the calculation of our 2D model to account for the number of cells that surround the vessel in a cylindrical geometry and add up to the transport from and to the vessel. First of all, we assume that the slice from our 2D model has the thickness of one cell. The maximum number of cells that the tissue can accommodate per unit volume is represented by the carrying capacity $K$. Thus, its inverse $\frac{1}{K}$ gives the volume of one cell in a fully packed tissue. Assuming that the cells have the same characteristic length in every direction, then the thickness of a one-celled slice is $\frac{1}{\sqrt[3]{K}}$. Since we can approximately regard that each vessel (separated by a distance $L_x$ from their centres) feeds half of the considered slice, the service volume for each vessel in our model is $\frac{1}{\sqrt[3]{K}} \cdot \frac{L_x}{2} \cdot L_z$. On the other hand, the hollow cylinder (excluding the blood volume inside the vessel) that surrounds each blood vessel has a volume $\left( \pi\left( \frac{L_x}{2} \right)^2 - \pi R^2 \right) L_z$. Therefore, we have that the cylinder is $\phi'$ times bigger than the slice, with
\begin{equation}
\label{Eq:phiprima}
    \phi' = \frac{\pi\left( \frac{L_x}{2} \right)^2 - \pi R^2}{\frac{L_x}{2} \frac{1}{\sqrt[3]{K}}} = \pi \left( \frac{L_x}{2} \sqrt[3]{K} - \frac{2 R^2 \sqrt[3]{K}}{L_x}  \right) = \pi \sqrt[3]{K} \left( \frac{L_x}{2} - \frac{2R^2}{L_x} \right)
\end{equation}

To account for the total number of cells of a population with a density $P(x,z,t)$, which has to be integrated in the 2D domain, and as the density varies between 0 and 1, we multiply this integral by the carrying capacity $K$ and by the slice thickness to transform the result to cell number. Additionally, since we want to take all the cylinder into account we introduce $\phi'$ from \cref{Eq:phiprima}. Therefore the result for the number of cells is
\begin{equation}
    P_{\mathrm{num}}(t) = \phi' \cdot K \cdot \frac{1}{\sqrt[3]{K}} \cdot \int_0^{L_x} \!\!\! \int_0^{L_z} P(x,z,t) \D x \D z.
\end{equation}
We can combine the multiplying term in a single factor $\phi$
\begin{equation}
    P_{\mathrm{num}}(t) = \phi \int_0^{L_x} \!\!\! \int_0^{L_z} P(x,z,t) \D x \D z,
\end{equation}
with
\begin{equation}
    \phi = \pi \sqrt[3]{K} \left( \frac{L_x}{2} - \frac{2R^2}{L_x} \right) \cdot K \cdot \frac{1}{\sqrt[3]{K}} = \pi K \left( \frac{L_x}{2} - \frac{2R^2}{L_x} \right),
\end{equation}
which is the form that we use to express the number of cells in \crefrange{Eq:n_num}{Eq:c_num}.


\section{Temperature and oxygen levels in the blood vessels}

\setcounter{figure}{0} 
\setcounter{table}{0} 

\begin{figure}[t]
    \centering
    \includegraphics[width=1 \textwidth]{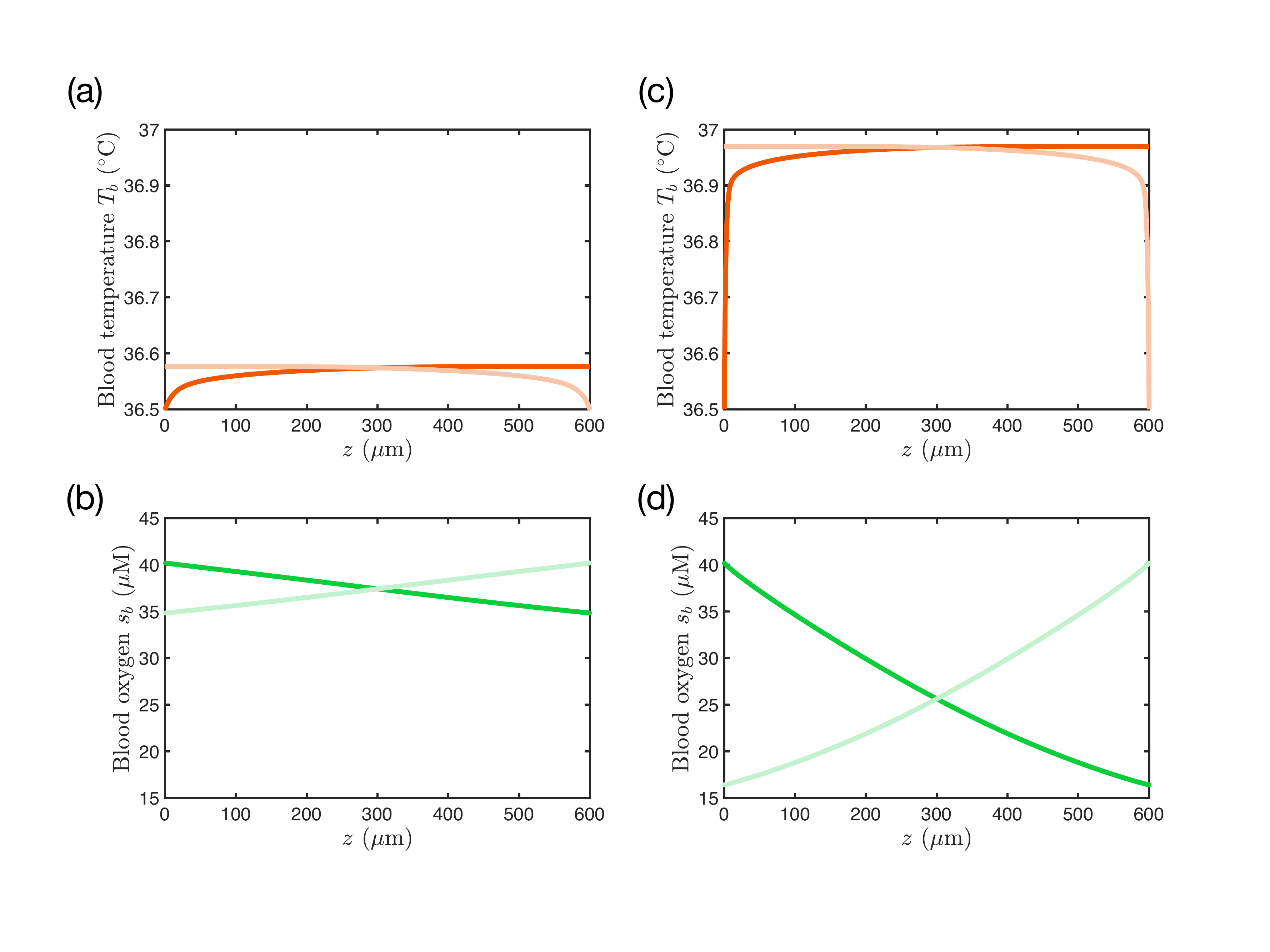}
    \caption{Results of the simulations performed in \cref{SubSec:Results_FREE} for the variables of the blood vessels at the final instant of the simulation ($t=245$ days). \textbf{(a)} Temperature in each section $z$ of the left (dark orange) and right (light orange) blood vessels ($T_{b1}(z)$ and $T_{b2}(z)$) for a simulated case where the velocity in the vessels is $v_1=v_2=1600$~\si{\micro\meter\per\second}, corresponding to a functional vasculature. \textbf{(b)} Oxygen drop in the blood vessels for that same case, with the left vessel in dark green ($s_{b1}(z)$) and the right vessel ($s_{b2}(z)$) in light green. \textbf{(c)} Temperature in the vessels ($T_{b1}(z)$, $T_{b2}(z)$) for the simulations with $v_1=v_2=300$ \si{\micro\meter\per\second} (impaired vasculature) and \textbf{(d)} oxygen concentration ($s_{b1}(z)$, $s_{b2}(z)$) following the same colour convention as the left panels.}
    \label{Fig:figH1}
\end{figure}

\Cref{Fig:figH1} depicts the spatial distribution of blood temperatures along the blood vessels path $z$ at the final instant of the simulations detailed in \cref{SubSec:Results_FREE} whose tissue variables are shown in \cref{Fig:fig3} (corresponding to \cref{Fig:figH1}(a) and \cref{Fig:figH1}(b)) and \cref{Fig:fig5} (corresponding to \cref{Fig:figH1}(c) and \cref{Fig:figH1}(d)).

\FloatBarrier

\bibliographystyle{elsarticle-num.bst}
\bibliography{Bosque_2021.bib}

\begin{thebibliography}{10}
\expandafter\ifx\csname url\endcsname\relax
  \def\url#1{\texttt{#1}}\fi
\expandafter\ifx\csname urlprefix\endcsname\relax\def\urlprefix{URL }\fi
\expandafter\ifx\csname href\endcsname\relax
  \def\href#1#2{#2} \def\path#1{#1}\fi

\bibitem{wilson2011targeting}
W.~R. Wilson, M.~P. Hay, Targeting hypoxia in cancer therapy, Nature Reviews
  Cancer 11~(6) (2011) 393--410.
\newblock \href {http://dx.doi.org/10.1038/nrc3064}
  {\path{doi:10.1038/nrc3064}}.

\bibitem{Bhandari2019}
V.~Bhandari, C.~Hoey, L.~Y. Liu, E.~Lalonde, J.~Ray, J.~Livingstone, R.~Lesurf,
  Y.-J. Shiah, T.~Vujcic, X.~Huang, S.~M.~G. Espiritu, L.~E. Heisler,
  F.~Yousif, V.~Huang, T.~N. Yamaguchi, C.~Q. Yao, V.~Y. Sabelnykova,
  M.~Fraser, M.~L.~K. Chua, T.~van~der Kwast, S.~K. Liu, P.~C. Boutros, R.~G.
  Bristow, Molecular landmarks of tumor hypoxia across cancer types, Nature
  Genetics 51~(2) (2019) 308--318.
\newblock \href {http://dx.doi.org/10.1038/s41588-018-0318-2}
  {\path{doi:10.1038/s41588-018-0318-2}}.

\bibitem{michiels2016cycling}
C.~Michiels, C.~Tellier, O.~Feron, Cycling hypoxia: A key feature of the tumor
  microenvironment, Biochimica et Biophysica Acta (BBA)-Reviews on Cancer
  1866~(1) (2016) 76--86.
\newblock \href {http://dx.doi.org/10.1016/j.bbcan.2016.06.004}
  {\path{doi:10.1016/j.bbcan.2016.06.004}}.

\bibitem{McDougall2006}
S.~McDougall, A.~Anderson, M.~Chaplain, Mathematical modelling of dynamic
  adaptive tumour-induced angiogenesis: Clinical implications and therapeutic
  targeting strategies, Journal of Theoretical Biology 241~(3) (2006) 564--589.
\newblock \href {http://dx.doi.org/10.1016/j.jtbi.2005.12.022}
  {\path{doi:10.1016/j.jtbi.2005.12.022}}.

\bibitem{Owen2009}
M.~Owen, T.~Alarcon, P.~Maini, H.~Byrne, Angiogenesis and vascular remodelling
  in normal and cancerous tissues, Journal of Mathematical Biology 58 (2009)
  689--721.
\newblock \href {http://dx.doi.org/10.1007/s00285-008-0213-z}
  {\path{doi:10.1007/s00285-008-0213-z}}.

\bibitem{pries2009structural}
A.~R. Pries, A.~J. Cornelissen, A.~A. Sloot, M.~Hinkeldey, M.~R. Dreher,
  M.~H{\"o}pfner, M.~W. Dewhirst, T.~W. Secomb, Structural adaptation and
  heterogeneity of normal and tumor microvascular networks, PLoS Computational
  Biology 5~(5) (2009) e1000394.
\newblock \href {http://dx.doi.org/10.1371/journal.pcbi.1000394}
  {\path{doi:10.1371/journal.pcbi.1000394}}.

\bibitem{bernabeu2020abnormal}
M.~O. Bernabeu, J.~K{\"o}ry, J.~A. Grogan, B.~Markelc, A.~Beardo,
  M.~d’Avezac, R.~Enjalbert, J.~Kaeppler, N.~Daly, J.~Hetherington, et~al.,
  Abnormal morphology biases hematocrit distribution in tumor vasculature and
  contributes to heterogeneity in tissue oxygenation, Proceedings of the
  National Academy of Sciences 117~(45) (2020) 27811--27819.
\newblock \href {http://dx.doi.org/10.1073/pnas.2007770117}
  {\path{doi:10.1073/pnas.2007770117}}.

\bibitem{al2019hypoxia}
W.~Al~Tameemi, T.~P. Dale, R.~M.~K. Al-Jumaily, N.~R. Forsyth, Hypoxia-modified
  cancer cell metabolism, Frontiers in Cell and Developmental Biology 7 (2019)
  4.
\newblock \href {http://dx.doi.org/10.3389/fcell.2019.00004}
  {\path{doi:10.3389/fcell.2019.00004}}.

\bibitem{dzyubak2021multi}
L.~Dzyubak, O.~Dzyubak, J.~Awrejcewicz, Multi-parametric evolution of
  conditions leading to cancer invasion in biological systems, Applied
  Mathematical Modelling 90 (2021) 46--60.
\newblock \href {http://dx.doi.org/10.1016/j.apm.2020.08.079}
  {\path{doi:10.1016/j.apm.2020.08.079}}.

\bibitem{giese2003cost}
A.~Giese, R.~Bjerkvig, M.~Berens, M.~Westphal, Cost of migration: invasion of
  malignant gliomas and implications for treatment, Journal of Clinical
  Oncology 21~(8) (2003) 1624--1636.
\newblock \href {http://dx.doi.org/10.1200/JCO.2003.05.063}
  {\path{doi:10.1200/JCO.2003.05.063}}.

\bibitem{lewis2016intratumoral}
D.~M. Lewis, K.~M. Park, V.~Tang, Y.~Xu, K.~Pak, T.~K. Eisinger-Mathason, M.~C.
  Simon, S.~Gerecht, Intratumoral oxygen gradients mediate sarcoma cell
  invasion, Proceedings of the National Academy of Sciences 113~(33) (2016)
  9292--9297.
\newblock \href {http://dx.doi.org/10.1073/pnas.1605317113}
  {\path{doi:10.1073/pnas.1605317113}}.

\bibitem{DArcy2019}
M.~D'Arcy, Cell death: a review of the major forms of apoptosis, necrosis and
  autophagy, Cell Biology International 43~(6) (2019) 582--592.
\newblock \href {http://dx.doi.org/10.1002/cbin.11137}
  {\path{doi:10.1002/cbin.11137}}.

\bibitem{Rong2006}
Y.~Rong, D.~Durden, E.~Van~Meir, D.~Brat, 'pseudopalisading' necrosis in
  glioblastoma: A familiar morphologic feature that links vascular pathology,
  hypoxia, and angiogenesis, Journal of Neuropathology and Experimental
  Neurology 65~(6) (2006) 529--539.
\newblock \href {http://dx.doi.org/10.1097/00005072-200606000-00001}
  {\path{doi:10.1097/00005072-200606000-00001}}.

\bibitem{dewhirst2017transport}
M.~W. Dewhirst, T.~W. Secomb, Transport of drugs from blood vessels to tumour
  tissue, Nature Reviews Cancer 17~(12) (2017) 738--750.
\newblock \href {http://dx.doi.org/10.1038/nrc.2017.93}
  {\path{doi:10.1038/nrc.2017.93}}.

\bibitem{wouters2007implications}
A.~Wouters, B.~Pauwels, F.~Lardon, J.~B. Vermorken, Implications of in vitro
  research on the effect of radiotherapy and chemotherapy under hypoxic
  conditions, The Oncologist 12~(6) (2007) 690--712.
\newblock \href {http://dx.doi.org/10.1634/theoncologist.12-6-690}
  {\path{doi:10.1634/theoncologist.12-6-690}}.

\bibitem{hall2018radiobiology}
E.~J. Hall, A.~J. Giaccia, Radiobiology for the Radiologist, Vol.~8, Wolters
  Kluwer Health, 2018.
\newblock \href {http://dx.doi.org/10.1007/s13246-018-0684-1}
  {\path{doi:10.1007/s13246-018-0684-1}}.

\bibitem{joiner2019basic}
M.~C. Joiner, A.~J. van~der Kogel, Basic clinical radiobiology, CRC press,
  2019.
\newblock \href {http://dx.doi.org/10.1201/9780429490606}
  {\path{doi:10.1201/9780429490606}}.

\bibitem{valle2021chemoimmunotherapy}
P.~A. Valle, L.~N. Coria, K.~D. Carballo, Chemoimmunotherapy for the treatment
  of prostate cancer: Insights from mathematical modelling, Applied
  Mathematical Modelling 90 (2021) 682--702.
\newblock \href {http://dx.doi.org/10.1016/j.apm.2020.09.021}
  {\path{doi:10.1016/j.apm.2020.09.021}}.

\bibitem{pang2021mathematical}
L.~Pang, S.~Liu, F.~Liu, X.~Zhang, T.~Tian, Mathematical modeling and analysis
  of tumor-volume variation during radiotherapy, Applied Mathematical Modelling
  89 (2021) 1074--1089.
\newblock \href {http://dx.doi.org/10.1016/j.apm.2020.07.028}
  {\path{doi:10.1016/j.apm.2020.07.028}}.

\bibitem{wigerup2016therapeutic}
C.~Wigerup, S.~P{\aa}hlman, D.~Bexell, Therapeutic targeting of hypoxia and
  hypoxia-inducible factors in cancer, Pharmacology \& Therapeutics 164 (2016)
  152--169.
\newblock \href {http://dx.doi.org/10.1016/j.pharmthera.2016.04.009}
  {\path{doi:10.1016/j.pharmthera.2016.04.009}}.

\bibitem{Li2021}
Y.~Li, L.~Zhao, X.-F. Li, Targeting hypoxia: Hypoxia-activated prodrugs in
  cancer therapy, Frontiers in Oncology 11 (2021) 700407.
\newblock \href {http://dx.doi.org/10.3389/fonc.2021.700407}
  {\path{doi:10.3389/fonc.2021.700407}}.

\bibitem{mattoso2021pointwise}
R.~Mattoso, A.~A. Novotny, Pointwise antennas design in hyperthermia therapy,
  Applied Mathematical Modelling 89 (2021) 89--104.
\newblock \href {http://dx.doi.org/10.1016/j.apm.2020.07.046}
  {\path{doi:10.1016/j.apm.2020.07.046}}.

\bibitem{ghasemi2020computational}
M.~Ghasemi, S.~Sivaloganathan, {A computational study of combination
  HIFU--chemotherapy as a potential means of overcoming cancer drug
  resistance}, Mathematical Biosciences 329 (2020) 108456.
\newblock \href {http://dx.doi.org/10.1016/j.mbs.2020.108456}
  {\path{doi:10.1016/j.mbs.2020.108456}}.

\bibitem{calvo2020modelling}
G.~F. Calvo, B.~Cort{\'e}s-Llanos, J.~Belmonte-Beitia, G.~Salas,
  {\'A}.~Ayuso-Sacido, Modelling the role of flux density and coating on
  nanoparticle internalization by tumor cells under centrifugation, Applied
  Mathematical Modelling 78 (2020) 98--116.
\newblock \href {http://dx.doi.org/10.1016/j.apm.2019.10.005}
  {\path{doi:10.1016/j.apm.2019.10.005}}.

\bibitem{li2020thermo}
X.~Li, Q.-H. Qin, X.~Tian, Thermo-viscoelastic analysis of biological tissue
  during hyperthermia treatment, Applied Mathematical Modelling 79 (2020)
  881--895.
\newblock \href {http://dx.doi.org/10.1016/j.apm.2019.11.007}
  {\path{doi:10.1016/j.apm.2019.11.007}}.

\bibitem{Elming2019}
P.~Elming, B.~Sorensen, A.~Oei, N.~Franken, J.~Crezee, J.~Overgaard,
  M.~Horsman, Hyperthermia: The optimal treatment to overcome radiation
  resistant hypoxia, Cancers 11~(1) (2019) 60.
\newblock \href {http://dx.doi.org/10.3390/cancers11010060}
  {\path{doi:10.3390/cancers11010060}}.

\bibitem{ahmed2015hyperthermia}
K.~Ahmed, Y.~Tabuchi, T.~Kondo, Hyperthermia: an effective strategy to induce
  apoptosis in cancer cells, Apoptosis 20~(11) (2015) 1411--1419.
\newblock \href {http://dx.doi.org/10.1007/s10495-015-1168-3}
  {\path{doi:10.1007/s10495-015-1168-3}}.

\bibitem{van2016cem43}
G.~C. van Rhoon, {Is CEM43 still a relevant thermal dose parameter for
  hyperthermia treatment monitoring?}, International Journal of Hyperthermia
  32~(1) (2016) 50--62.
\newblock \href {http://dx.doi.org/10.3109/02656736.2015.1114153}
  {\path{doi:10.3109/02656736.2015.1114153}}.

\bibitem{gupta2013numerical}
P.~K. Gupta, J.~Singh, K.~N. Rai, A numerical study on heat transfer in tissues
  during hyperthermia, Mathematical and Computer Modelling 57~(5-6) (2013)
  1018--1037.
\newblock \href {http://dx.doi.org/10.1016/j.mcm.2011.12.050}
  {\path{doi:10.1016/j.mcm.2011.12.050}}.

\bibitem{suleman2020mathematical}
M.~Suleman, S.~Riaz, R.~Jalil, A mathematical modeling approach toward magnetic
  fluid hyperthermia of cancer and unfolding heating mechanism, Journal of
  Thermal Analysis and Calorimetry (2020) 1--27\href
  {http://dx.doi.org/10.1007/s10973-020-10080-8}
  {\path{doi:10.1007/s10973-020-10080-8}}.

\bibitem{damaghi2021harsh}
M.~Damaghi, J.~West, M.~Robertson-Tessi, L.~Xu, M.~C. Ferrall-Fairbanks, P.~A.
  Stewart, E.~Persi, B.~L. Fridley, P.~M. Altrock, R.~A. Gatenby, et~al., {The
  harsh microenvironment in early breast cancer selects for a Warburg
  phenotype}, Proceedings of the National Academy of Sciences 118~(3).
\newblock \href {http://dx.doi.org/10.1073/pnas.2011342118}
  {\path{doi:10.1073/pnas.2011342118}}.

\bibitem{mendonsa2018cadherin}
A.~M. Mendonsa, T.-Y. Na, B.~M. Gumbiner, E-cadherin in contact inhibition and
  cancer, Oncogene 37~(35) (2018) 4769--4780.
\newblock \href {http://dx.doi.org/10.1038/s41388-018-0304-2}
  {\path{doi:10.1038/s41388-018-0304-2}}.

\bibitem{pavel2018contact}
M.~Pavel, M.~Renna, S.~J. Park, F.~M. Menzies, T.~Ricketts, J.~F{\"u}llgrabe,
  A.~Ashkenazi, R.~A. Frake, A.~C. Lombarte, C.~F. Bento, et~al., {Contact
  inhibition controls cell survival and proliferation via YAP/TAZ-autophagy
  axis}, Nature communications 9~(1) (2018) 1--18.
\newblock \href {http://dx.doi.org/10.1038/s41467-018-05388-x}
  {\path{doi:10.1038/s41467-018-05388-x}}.

\bibitem{kumari2018reactive}
S.~Kumari, A.~K. Badana, R.~Malla, Reactive oxygen species: a key constituent
  in cancer survival, Biomarker insights 13 (2018) 1177271918755391.
\newblock \href {http://dx.doi.org/10.1177/1177271918755391}
  {\path{doi:10.1177/1177271918755391}}.

\bibitem{Enderling2014}
H.~Enderling, M.~Chaplain, Mathematical modeling of tumor growth and treatment,
  Current Pharmaceutical Design 20 (2014) 4934--4940.
\newblock \href {http://dx.doi.org/10.2174/1381612819666131125150434}
  {\path{doi:10.2174/1381612819666131125150434}}.

\bibitem{Michor2015}
P.~Altrock, L.~Liu, F.~Michor, The mathematics of cancer: integrating
  quantitative models, Nature Rev. Cancer 15 (2015) 730--745.
\newblock \href {http://dx.doi.org/10.1038/nrc4029}
  {\path{doi:10.1038/nrc4029}}.

\bibitem{Swanson2017}
J.~Alfonso, K.~Talkenberger, M.~Seifert, B.~Klink, A.~Hawkins-Daarud,
  K.~Swanson, H.~Hatzikirou, A.~Deutsch, The biology and mathematical modelling
  of glioma invasion: a review, Journal of the Royal Society Interface 14~(136)
  (2017) 20170490.
\newblock \href {http://dx.doi.org/10.1098/rsif.2017.0490}
  {\path{doi:10.1098/rsif.2017.0490}}.

\bibitem{Rockne2019}
R.~Rockne, A.~Hawkins-Daarud, K.~Swanson, J.~Sluka, J.~Glazier, P.~Macklin,
  D.~Hormuth, A.~Jarrett, E.~Lima, J.~Oden, G.~Biros, T.~Yankeelov, K.~Curtius,
  I.~Al~Bakir, W.~Dominik, N.~Komarova, L.~Aparicio, M.~Bordyuh, R.~Rabadan,
  S.~Finley, H.~Enderling, J.~Caudell, E.~G. Moros, A.~Anderson, R.~Gatenby,
  A.~Kaznatcheev, P.~Jeavons, N.~Krishnan, J.~Pelesko, R.~Wadhwa, N.~Yoon,
  D.~Nichol, A.~Marusyk, M.~Hinczewski, J.~Scott, The 2019 mathematical
  oncology roadmap, Physical Biology 16~(4) (2019) 041005.
\newblock \href {http://dx.doi.org/10.1088/1478-3975/ab1a09}
  {\path{doi:10.1088/1478-3975/ab1a09}}.

\bibitem{Belmonte2014}
J.~Belmonte-Beitia, G.~F. Calvo, V.~M. P\'erez-Garc\'{\i}a, {Effective particle
  methods for Fisher-Kolmogorov equations: theory and applications to brain
  tumor dynamics}, Commun. Nonlinear Sci. Numer. Simulat. 19 (2014) 3267--3283.
\newblock \href {http://dx.doi.org/10.1016/j.cnsns.2014.02.004}
  {\path{doi:10.1016/j.cnsns.2014.02.004}}.

\bibitem{el2019revisiting}
M.~El-Hachem, S.~W. McCue, W.~Jin, Y.~Du, M.~J. Simpson, {Revisiting the
  Fisher--Kolmogorov--Petrovsky--Piskunov equation to interpret the
  spreading--extinction dichotomy}, Proceedings of the Royal Society A
  475~(2229) (2019) 20190378.
\newblock \href {http://dx.doi.org/10.1098/rspa.2019.0378}
  {\path{doi:10.1098/rspa.2019.0378}}.

\bibitem{elazab2018macroscopic}
A.~Elazab, Y.~M. Abdulazeem, A.~M. Anter, Q.~Hu, T.~Wang, B.~Lei, Macroscopic
  cerebral tumor growth modeling from medical images: A review, IEEE Access 6
  (2018) 30663--30679.
\newblock \href {http://dx.doi.org/10.1109/ACCESS.2018.2839681}
  {\path{doi:10.1109/ACCESS.2018.2839681}}.

\bibitem{Badoual2021}
L.~Adenis, S.~Plaszczynski, B.~Grammaticos, J.~Pallud, M.~Badoual, The effect
  of radiotherapy on diffuse low-grade gliomas evolution: confronting theory
  with clinical data, J. Pers. Med. 11 (2021) 818.
\newblock \href {http://dx.doi.org/10.3390/jpm11080818}
  {\path{doi:10.3390/jpm11080818}}.

\bibitem{perez2011bright}
V.~M. P\'erez-Garc\'{\i}a, G.~F. Calvo, J.~Belmonte-Beitia, D.~Diego,
  L.~P\'erez-Romasanta, Bright solitary waves in malignant gliomas, Physical
  Review E 84 (2011) 021921.
\newblock \href {http://dx.doi.org/10.1103/PhysRevE.84.021921}
  {\path{doi:10.1103/PhysRevE.84.021921}}.

\bibitem{perez2017glioblastoma}
J.~P{\'e}rez-Beteta, A.~Mart{\'\i}nez-Gonz{\'a}lez, D.~Molina, M.~Amo-Salas,
  B.~Luque, E.~Arregui, M.~Calvo, J.~M. Borr{\'a}s, C.~L{\'o}pez,
  M.~Claramonte, et~al., {Glioblastoma: does the pre-treatment geometry matter?
  A postcontrast T1 MRI-based study}, European Radiology 27~(3) (2017)
  1096--1104.
\newblock \href {http://dx.doi.org/10.1007/s00330-016-4453-9}
  {\path{doi:10.1007/s00330-016-4453-9}}.

\bibitem{Alicia2012}
A.~Mart\'{\i}nez-Gonz\'{a}lez, G.~F. Calvo, L.~A. P\'erez-Romasanta, V.~M.
  P\'erez-Garc\'{\i}a, Hypoxic cell waves around necrotic cores in
  glioblastoma: A biomathematical model and its therapeutic implications,
  Bulletin of Mathematical Biology 74 (2012) 2875--2896.
\newblock \href {http://dx.doi.org/10.1007/s11538-012-9786-1}
  {\path{doi:10.1007/s11538-012-9786-1}}.

\bibitem{Berezhnoi2018}
A.~Berezhnoi, M.~Schwarz, A.~Buehler, S.~Ovsepian, J.~Aguirre,
  V.~Ntziachristos, Assessing hyperthermia-induced vasodilation in human skin
  in vivo using optoacoustic mesoscopy, J. Biophotonics 11~(11) (2018)
  201700359.
\newblock \href {http://dx.doi.org/10.1002/jbio.201700359}
  {\path{doi:10.1002/jbio.201700359}}.

\bibitem{kingsley2021bridging}
J.~L. Kingsley, J.~R. Costello, N.~Raghunand, K.~A. Rejniak, Bridging
  cell-scale simulations and radiologic images to explain short-time
  intratumoral oxygen fluctuations, PLoS Computational Biology 17~(7) (2021)
  e1009206.
\newblock \href {http://dx.doi.org/10.1371/journal.pcbi.1009206}
  {\path{doi:10.1371/journal.pcbi.1009206}}.

\bibitem{schiesser2012partial}
W.~E. Schiesser, Partial differential equation analysis in biomedical
  engineering: case studies with MATLAB, Cambridge University Press, 2012.
\newblock \href {http://dx.doi.org/10.1017/cbo9781139137096}
  {\path{doi:10.1017/cbo9781139137096}}.

\bibitem{perez2020universal}
V.~M. P{\'e}rez-Garc{\'\i}a, G.~F. Calvo, J.~J. Bosque, O.~Le{\'o}n-Triana,
  J.~Jim{\'e}nez, J.~Perez-Beteta, J.~Belmonte-Beitia, M.~Valiente, L.~Zhu,
  P.~Garc{\'\i}a-G{\'o}mez, et~al., Universal scaling laws rule explosive
  growth in human cancers, Nature Physics 16~(12) (2020) 1232--1237.
\newblock \href {http://dx.doi.org/10.1038/s41567-020-0978-6}
  {\path{doi:10.1038/s41567-020-0978-6}}.

\bibitem{martinez2021cancer}
I.~Mart{\'\i}nez-Reyes, N.~S. Chandel, Cancer metabolism: looking forward,
  Nature Reviews Cancer 21 (2021) 669--680.
\newblock \href {http://dx.doi.org/10.1038/s41568-021-00378-6}
  {\path{doi:10.1038/s41568-021-00378-6}}.

\bibitem{banavar2002supply}
J.~R. Banavar, J.~Damuth, A.~Maritan, A.~Rinaldo, Supply--demand balance and
  metabolic scaling, Proceedings of the National Academy of Sciences 99~(16)
  (2002) 10506--10509.
\newblock \href {http://dx.doi.org/10.1073/pnas.162216899}
  {\path{doi:10.1073/pnas.162216899}}.

\bibitem{banavar2010general}
J.~R. Banavar, M.~E. Moses, J.~H. Brown, J.~Damuth, A.~Rinaldo, R.~M. Sibly,
  A.~Maritan, A general basis for quarter-power scaling in animals, Proceedings
  of the National Academy of Sciences 107~(36) (2010) 15816--15820.
\newblock \href {http://dx.doi.org/10.1073/pnas.1009974107}
  {\path{doi:10.1073/pnas.1009974107}}.

\bibitem{moon2010nadph}
E.~J. Moon, P.~Sonveaux, P.~E. Porporato, P.~Danhier, B.~Gallez,
  I.~Batinic-Haberle, Y.-C. Nien, T.~Schroeder, M.~W. Dewhirst, {NADPH
  oxidase-mediated reactive oxygen species production activates
  hypoxia-inducible factor-1 (HIF-1) via the ERK pathway after hyperthermia
  treatment}, Proceedings of the National Academy of Sciences 107~(47) (2010)
  20477--20482.
\newblock \href {http://dx.doi.org/10.1073/pnas.1006646107}
  {\path{doi:10.1073/pnas.1006646107}}.

\bibitem{jimenez2021mesoscopic}
J.~Jim{\'e}nez-S{\'a}nchez, {\'A}.~Mart{\'\i}nez-Rubio, A.~Popov,
  J.~P{\'e}rez-Beteta, Y.~Azimzade, D.~Molina-Garc{\'\i}a, J.~Belmonte-Beitia,
  G.~F. Calvo, V.~M. P{\'e}rez-Garc{\'\i}a, A mesoscopic simulator to uncover
  heterogeneity and evolutionary dynamics in tumors, PLoS computational biology
  17~(2) (2021) e1008266.
\newblock \href {http://dx.doi.org/10.1371/journal.pcbi.1008266}
  {\path{doi:10.1371/journal.pcbi.1008266}}.

\bibitem{bosque2021interplay}
J.~J. Bosque, G.~F. Calvo, V.~M. P{\'e}rez-Garc{\'\i}a, M.~C. Navarro, The
  interplay of blood flow and temperature in regional hyperthermia: a
  mathematical approach, Royal Society Open Science 8~(1) (2021) 201234.
\newblock \href {http://dx.doi.org/10.1098/rsos.201234}
  {\path{doi:10.1098/rsos.201234}}.

\bibitem{jimenez2021evolutionary}
J.~Jim{\'e}nez-S{\'a}nchez, J.~J. Bosque, G.~A.~J. Londo{\~n}o,
  D.~Molina-Garc{\'\i}a, {\'A}.~Mart{\'\i}nez, J.~P{\'e}rez-Beteta,
  C.~Ortega-Sabater, A.~F.~H. Mart{\'\i}nez, A.~M.~G. Vicente, G.~F. Calvo,
  et~al., Evolutionary dynamics at the tumor edge reveal metabolic imaging
  biomarkers, Proceedings of the National Academy of Sciences 118~(6).
\newblock \href {http://dx.doi.org/10.1073/pnas.2018110118}
  {\path{doi:10.1073/pnas.2018110118}}.

\bibitem{Cortes2021}
B.~Cort{\'e}s-Llanos, S.~M. Ocampo, L.~de~la Cueva, G.~F. Calvo, ,
  J.~Belmonte-Beitia, L.~P{\'e}rez, G.~Salas, {\'A}.~Ayuso-Sacido, {Influence
  of coating and size of magnetic nanoparticles on cellular uptake for in vitro
  MRI}, Nanomaterials 11 (2021) 2888.
\newblock \href {http://dx.doi.org/10.3390/nano11112888}
  {\path{doi:10.3390/nano11112888}}.

\bibitem{Moy2017}
A.~Moy, J.~Tunnell, Combinatorial immunotherapy and nanoparticle mediated
  hyperthermia, Advanced Drug Delivery Reviews 8 (2017) 175--184.
\newblock \href {http://dx.doi.org/10.1016/j.addr.2017.06.008}
  {\path{doi:10.1016/j.addr.2017.06.008}}.

\bibitem{martin2019normalizing}
J.~D. Martin, G.~Seano, R.~K. Jain, Normalizing function of tumor vessels:
  progress, opportunities, and challenges, Annual review of physiology 81
  (2019) 505--534.
\newblock \href {http://dx.doi.org/10.1146/annurev-physiol-020518-114700}
  {\path{doi:10.1146/annurev-physiol-020518-114700}}.

\bibitem{rossmann2014review}
C.~Rossmann, D.~Haemmerich, Review of temperature dependence of thermal
  properties, dielectric properties, and perfusion of biological tissues at
  hyperthermic and ablation temperatures, Critical Reviews in Biomedical
  Engineering 42~(6) (2014) 467--–492.
\newblock \href {http://dx.doi.org/10.1615/CritRevBiomedEng.2015012486}
  {\path{doi:10.1615/CritRevBiomedEng.2015012486}}.

\bibitem{oei2020molecular}
A.~Oei, H.~Kok, S.~Oei, M.~Horsman, L.~Stalpers, N.~Franken, J.~Crezee,
  Molecular and biological rationale of hyperthermia as radio-and
  chemosensitizer, Advanced drug delivery reviews 163 (2020) 84--97.
\newblock \href {http://dx.doi.org/10.1016/j.addr.2020.01.003}
  {\path{doi:10.1016/j.addr.2020.01.003}}.

\bibitem{gillooly2001effects}
J.~F. Gillooly, J.~H. Brown, G.~B. West, V.~M. Savage, E.~L. Charnov, Effects
  of size and temperature on metabolic rate, Science 293~(5538) (2001)
  2248--2251.
\newblock \href {http://dx.doi.org/10.1126/science.1061967}
  {\path{doi:10.1126/science.1061967}}.

\bibitem{gillooly2002effects}
J.~F. Gillooly, E.~L. Charnov, G.~B. West, V.~M. Savage, J.~H. Brown, Effects
  of size and temperature on developmental time, Nature 417~(6884) (2002)
  70--73.
\newblock \href {http://dx.doi.org/10.1038/417070a}
  {\path{doi:10.1038/417070a}}.

\bibitem{deutsch2020metabolic}
C.~Deutsch, J.~L. Penn, B.~Seibel, Metabolic trait diversity shapes marine
  biogeography, Nature 585~(7826) (2020) 557--562.
\newblock \href {http://dx.doi.org/10.1038/s41586-020-2721-y}
  {\path{doi:10.1038/s41586-020-2721-y}}.

\bibitem{alfarouk2011tumor}
K.~O. Alfarouk, A.~K. Muddathir, M.~E. Shayoub, Tumor acidity as evolutionary
  spite, Cancers 3~(1) (2011) 408--414.
\newblock \href {http://dx.doi.org/10.3390/cancers3010408}
  {\path{doi:10.3390/cancers3010408}}.

\bibitem{estrella2013acidity}
V.~Estrella, T.~Chen, M.~Lloyd, J.~Wojtkowiak, H.~H. Cornnell,
  A.~Ibrahim-Hashim, K.~Bailey, Y.~Balagurunathan, J.~M. Rothberg, B.~F.
  Sloane, et~al., Acidity generated by the tumor microenvironment drives local
  invasion, Cancer research 73~(5) (2013) 1524--1535.
\newblock \href {http://dx.doi.org/10.1158/0008-5472.CAN-12-2796}
  {\path{doi:10.1158/0008-5472.CAN-12-2796}}.

\bibitem{gerweck1974killing}
L.~E. Gerweck, E.~L. Gillette, W.~C. Dewey, Killing of chinese hamster cells in
  vitro by heating under hypoxic or aerobic conditions, European journal of
  cancer 10~(10) (1974) 691--693.
\newblock \href {http://dx.doi.org/10.1016/0014-2964(74)90009-7}
  {\path{doi:10.1016/0014-2964(74)90009-7}}.

\bibitem{kroesen2019effect}
M.~Kroesen, H.~T. Mulder, J.~Van~Holthe, A.~Aangeenbrug, J.~Mens, H.~Van~Doorn,
  M.~M. Paulides, E.~Oomen-de Hoop, R.~Vernhout, L.~Lutgens, et~al., The effect
  of the time interval between radiation and hyperthermia on clinical outcome
  in 400 locally advanced cervical carcinoma patients, Frontiers in oncology 9
  (2019) 134.
\newblock \href {http://dx.doi.org/10.3389/fonc.2019.00134}
  {\path{doi:10.3389/fonc.2019.00134}}.

\bibitem{crezee2019impact}
H.~Crezee, H.~Kok, A.~L. Oei, N.~A. Franken, L.~J. Stalpers, The impact of the
  time interval between radiation and hyperthermia on clinical outcome in
  patients with locally advanced cervical cancer, Frontiers in oncology 9
  (2019) 412.
\newblock \href {http://dx.doi.org/10.3389/fonc.2019.00412}
  {\path{doi:10.3389/fonc.2019.00412}}.

\bibitem{dobvsivcek2019quality}
H.~Dob{\v{s}}{\'\i}{\v{c}}ek~Trefn{\'a}, M.~Schmidt, G.~Van~Rhoon, H.~Kok,
  S.~Gordeyev, U.~Lamprecht, D.~Marder, J.~Nadobny, P.~Ghadjar,
  S.~Abdel-Rahman, et~al., Quality assurance guidelines for interstitial
  hyperthermia, International Journal of Hyperthermia 36~(1) (2019) 276--293.
\newblock \href {http://dx.doi.org/10.1080/02656736.2018.1564155}
  {\path{doi:10.1080/02656736.2018.1564155}}.

\bibitem{trefna2017quality}
H.~D. Trefn{\'a}, H.~Crezee, M.~Schmidt, D.~Marder, U.~Lamprecht, M.~Ehmann,
  J.~Hartmann, J.~Nadobny, J.~Gellermann, N.~van Holthe, et~al., {Quality
  assurance guidelines for superficial hyperthermia clinical trials: I.
  Clinical requirements}, International Journal of Hyperthermia 33~(4) (2017)
  471--482.
\newblock \href {http://dx.doi.org/10.1080/02656736.2016.1277791}
  {\path{doi:10.1080/02656736.2016.1277791}}.

\bibitem{sapareto1984thermal}
S.~A. Sapareto, W.~C. Dewey, Thermal dose determination in cancer therapy,
  International Journal of Radiation Oncology* Biology* Physics 10~(6) (1984)
  787--800.
\newblock \href {http://dx.doi.org/10.1016/0360-3016(84)90379-1}
  {\path{doi:10.1016/0360-3016(84)90379-1}}.

\bibitem{bruningk2018combining}
S.~Br{\"u}ningk, G.~Powathil, P.~Ziegenhein, J.~Ijaz, I.~Rivens, S.~Nill,
  M.~Chaplain, U.~Oelfke, G.~ter Haar, Combining radiation with hyperthermia: a
  multiscale model informed by in vitro experiments, Journal of the Royal
  Society Interface 15~(138) (2018) 20170681.
\newblock \href {http://dx.doi.org/10.1098/rsif.2017.0681}
  {\path{doi:10.1098/rsif.2017.0681}}.

\bibitem{mcmahon2019linear}
S.~J. McMahon, The linear quadratic model: usage, interpretation and
  challenges, Physics in Medicine \& Biology 64~(1) (2019) 01TR01.
\newblock \href {http://dx.doi.org/10.1088/1361-6560/aaf26a}
  {\path{doi:10.1088/1361-6560/aaf26a}}.

\bibitem{wang2009prognostic}
C.~H. Wang, J.~K. Rockhill, M.~Mrugala, D.~L. Peacock, A.~Lai, K.~Jusenius,
  J.~M. Wardlaw, T.~Cloughesy, A.~M. Spence, R.~Rockne, et~al., Prognostic
  significance of growth kinetics in newly diagnosed glioblastomas revealed by
  combining serial imaging with a novel biomathematical model, Cancer research
  69~(23) (2009) 9133--9140.
\newblock \href {http://dx.doi.org/10.1158/0008-5472.CAN-08-3863}
  {\path{doi:10.1158/0008-5472.CAN-08-3863}}.

\bibitem{ke2000relevance}
L.~D. Ke, Y.-X. Shi, S.-A. Im, X.~Chen, W.~A. Yung, The relevance of cell
  proliferation, vascular endothelial growth factor, and basic fibroblast
  growth factor production to angiogenesis and tumorigenicity in human glioma
  cell lines, Clinical Cancer Research 6~(6) (2000) 2562--2572.

\bibitem{dacsu2003theoretical}
A.~Da{\c{s}}u, I.~Toma-Da{\c{s}}u, M.~Karlsson, Theoretical simulation of
  tumour oxygenation and results from acute and chronic hypoxia, Physics in
  Medicine \& Biology 48~(17) (2003) 2829.
\newblock \href {http://dx.doi.org/10.1088/0031-9155/48/17/307}
  {\path{doi:10.1088/0031-9155/48/17/307}}.

\bibitem{owen2011mathematical}
M.~R. Owen, I.~J. Stamper, M.~Muthana, G.~W. Richardson, J.~Dobson, C.~E.
  Lewis, H.~M. Byrne, Mathematical modeling predicts synergistic antitumor
  effects of combining a macrophage-based, hypoxia-targeted gene therapy with
  chemotherapy, Cancer Research 71~(8) (2011) 2826--2837.
\newblock \href {http://dx.doi.org/10.1158/0008-5472.CAN-10-2834}
  {\path{doi:10.1158/0008-5472.CAN-10-2834}}.

\bibitem{herman2016metabolism}
I.~P. Herman, {Metabolism: Energy, Heat, Work, and Power of the Body}, in:
  Physics of the Human Body, Springer, 2016, pp. 393--489.
\newblock \href {http://dx.doi.org/10.1007/978-3-319-23932-3}
  {\path{doi:10.1007/978-3-319-23932-3}}.

\bibitem{wust2002hyperthermia}
P.~Wust, B.~Hildebrandt, G.~Sreenivasa, B.~Rau, J.~Gellermann, H.~Riess,
  R.~Felix, P.~Schlag, Hyperthermia in combined treatment of cancer, The Lancet
  Oncology 3~(8) (2002) 487--497.
\newblock \href {http://dx.doi.org/10.1016/S1470-2045(02)00818-5}
  {\path{doi:10.1016/S1470-2045(02)00818-5}}.

\bibitem{jewell2001induction}
U.~R. Jewell, I.~Kvietikova, A.~Scheid, C.~Bauer, R.~H. Wenger, M.~Gassmann,
  {Induction of HIF--1$\alpha$ in response to hypoxia is instantaneous}, The
  FASEB Journal 15~(7) (2001) 1312--1314.
\newblock \href {http://dx.doi.org/10.1096/fj.00-0732fje}
  {\path{doi:10.1096/fj.00-0732fje}}.

\bibitem{vaupel2004role}
P.~Vaupel, The role of hypoxia-induced factors in tumor progression, The
  Oncologist 9 (2004) 10--17.
\newblock \href {http://dx.doi.org/10.1634/theoncologist.9-90005-10}
  {\path{doi:10.1634/theoncologist.9-90005-10}}.

\end{thebibliography}

\end{document}